\documentclass[a4paper,11pt,twoside,fancyhdr]{article}
\usepackage{epsfig,amsmath,amssymb,graphicx,fancyheadings,indentfirst,%
color,multirow,slashed}
\usepackage[english]{babel}
\usepackage[colorlinks=true,urlcolor=black]{hyperref}

\usepackage{feynmf} 

\usepackage[nottoc,numbib]{tocbibind}

\hypersetup{backref = true,pagebackref = true,hyperindex = true, colorlinks = true,
  breaklinks = true, urlcolor = blue, linkcolor = blue, bookmarks = true,
  bookmarksopen = true, citecolor=red}

\graphicspath{{figures/}}
\newcommand{\ep}{\varepsilon}

\renewcommand{\theequation}{\thesection.\arabic{equation}}

\definecolor{lightgrey}{gray}{0.9}

\newcommand{\N}{\mathbb{N}}

\def\al{\alpha}
\def\eps{\epsilon}

\def\veps{\varepsilon}
\def\ep{\varepsilon}

\def\be{\begin{equation}}
\def\ee{\end{equation}}
\def\bea{\begin{eqnarray}}
\def\eea{\end{eqnarray}}
\def\bse{\begin{subequations}}
\def\ese{\end{subequations}}
\def\bc{\begin{center}}
\def\ec{\end{center}}

\def\ms{\medskip}
\def\bs{\bigskip}

\def\ni{\noindent}
\def\ra{\rightarrow}
\def\Ra{\Rightarrow}
\def\lra{\leftrightarrow}

\def\nonum{\nonumber}

\newcommand{\fleq}[1]{\begin{flalign}#1\end{flalign}}

\def\I{{\rm i}}

\def\D{{\rm d}}
\def\Ord{{\rm O}}

\newcommand{\comment}[1]{}

\newcommand{\ie}{{\it i.e.}}
\newcommand{\eg}{{\it e.g.}}

\newtheorem{theorem}{Theorem}

\catcode`,\active

\catcode`\,12

\allowdisplaybreaks

\begin{document}
\pagestyle{empty}

\pagenumbering{arabic}

\begin{titlepage}
\begin{flushright}
\begin{tabular}{l}
\end{tabular}
\end{flushright}

\vskip 2cm

\begin{center}
 {\Large Multi-loop techniques for massless Feynman diagram calculations}
\end{center}

\vspace{1cm}

\centerline{\sc A.~V.~Kotikov$^1$ and S.~Teber$^{2}$}

\vspace{10mm}

\centerline{\it $^1$Bogoliubov Laboratory of Theoretical Physics, Joint Institute for Nuclear Research, 141980 Dubna, Russia.}
\centerline{\it $^2$Sorbonne Universit\'e, CNRS, Laboratoire de Physique Th\'eorique et Hautes Energies, LPTHE, F-75005 Paris, France.}

\vspace{1cm}

\centerline{\bf Abstract}

\vspace{5mm}
We review several multi-loop techniques for analytical massless Feynman diagram calculations in relativistic quantum field theories: integration by parts,
the method of uniqueness, functional equations and the Gegenbauer polynomial technique. A brief, historically oriented, overview of 
some of the results obtained over the decades for the massless 2-loop propagator-type diagram is given. 
Concrete examples of up to $5$-loop diagram calculations are also provided.

\vspace{1cm}

\tableofcontents

\end{titlepage}



\pagestyle{fancy}






\section{Introduction}

In relativistic quantum field theories, the first perturbative techniques using covariance and gauge invariance 
were independently developed in the 1940s by Tomonaga \cite{Tomonaga:1946zz}, Schwinger \cite{Schwinger:1948yk+Schwinger:1948yj+Schwinger:1949ra}
and Feynman \cite{Feynman:1948fi+Feynman:1949zx+Feynman:1950ir} and unified by Dyson \cite{PhysRev.75.486}.
This led to the discovery of the concept of renormalization as an attempt to give a meaning to divergent integrals appearing in the perturbative series.
Specific ideas of the field theoretic renormalization group (RG) were then developed starting from the 50s
\cite{GellMann:1954fq}-\cite{Bogolyubov:1980nc}.
Early success came
from their application to quantum electrodynamics (QED) with unprecedented agreement between high precision experiments and 
high precision theoretical computations of measurable quantities (anomalous magnetic moment, Lamb's shift, etc...).
In the 50s, non-abelian gauge theories were discovered 
\cite{Yang:1954ek} and, in the 60s, the weak interaction was unified with electromagnetism (electro-weak interaction) \cite{Glashow61+Salam:1964ry+Weinberg:1967tq}.
In the 70s, 't Hooft and Veltman \cite{'tHooft:1971rn,'tHooft:1972fi} proved the renormalizability of non-abelian gauge field theories and, to this purpose, invented
the dimensional regularization (DR) technique which was also independently discovered in
\cite{Bollini:1972ui,Cicuta:1972jf,Ashmore:1972uj}.
Combined with the minimal subtraction (MS) scheme \cite{'tHooft:1973mm} within an RG framework, this regularization technique is particularly well suited to
compute radiative corrections and we shall use it extensively in this review.
In the wake of these far-reaching developments, asymptotic freedom has then been discovered \cite{Politzer:1973fx,Gross:1973id}.
It eliminates the problem of the Landau poles and allows for the existence of theories which are well defined at arbitrarily low energies.
The first example of such a theory is quantum chromodynamics (QCD) which describes quarks as well as their interactions (strong interaction).
Contrarily to QED, a perturbative approach to QCD is possible only at high energy.
At low energies, a strong coupling regime takes place corresponding to quark confinement. 
At this point, let us recall that, back in the 70s, 't Hooft discovered a simplification of U$(N)$ gauge theories in the limit of large $N$ \cite{'tHooft:1973jz}.
This work initiated the large-$N$ study of gauge theories, see the monograph \cite{Brezin:1994eb} for a review.
Recently, it played an important role in examining the conjecture relating string theories in anti-de Sitter spaces to (the 't Hooft limit of) superconformal field theories
in one less space-time dimension (AdS/CFT correspondence) \cite{Maldacena:1997re+Gubser:1998bc+Witten:1998qj}. The most prominent example of such a field theory is the
$3+1$-dimensional $\mathcal{N}=4$ super-Yang-Mills theory at the conformal point which now serves as a kind of theoretical laboratory to gain insight on the beautiful and complex
structure of quantum field theories. 

The fundamental developments which led to the elaboration of such theories in particle/high-energy physics have had profound consequences
in other fields such as statistical mechanics. Formally, a $D$-dimensional quantum system is equivalent to a $D+1$-dimensional (in Euclidean space)
statistical mechanics one. Physically, complex systems composed of a large number of interacting particles are subject to emergent phenomena such as
phase transitions. It is Wilson that first realized that the vicinity of a second order phase transition may be described by a
continuous quantum field theory (QFT)  and formulated the so-called momentum-shell RG \cite{Wilson:1971bg,Wilson:1971dh}.
This led to the development of the small-$\veps$ expansion technique to compute critical exponents by Wilson and Fisher \cite{Wilson:1971dc}
and brought up the important notion of (non-trivial) infra-red (IR) fixed points.\footnote{According to \cite{Brezin:1994eb}, the large-$N$ expansion also first appeared
in the context of statistical mechanics through the work of Stanley on spin systems \cite{Stanley:1968gx}.} The works of the Saclay group, see, \eg, \cite{Brezin:1974eb}, led to the development of the field-theoretic Wilsonian RG
culminating in the book \cite{zinn2002quantum}.

The ability to access high order corrections, and therefore achieve high precision calculations, arose in the 80s with the developments
of powerful methods to compute renormalization group functions, {\it i.e.}, $\beta$-functions and anomalous dimensions of fields, with wide applications ranging from particle physics,
to statistical mechanics and condensed matter physics. These methods include, \eg, the method of infra-red rea-rangement (IRR) \cite{Vladimirov:1979zm} and the
$R^*$-operation \cite{Chetyrkin:1982nn}-\cite{Chetyrkin:2017ppe},
the method of uniqueness  \cite{D'Eramo:1971zz}-\cite{Kotikov:1987mw}
and (Vasil'ev's) conformal bootstrap technique \cite{Vasiliev:1982dc},
integration by parts (IBP)
\cite{Vasiliev:1981dg,Tkachov:1981wb,Chetyrkin:1981qh},
the Gegenbauer polynomial technique \cite{Chetyrkin:1980pr}
and the combination of these methods with symmetry arguments \cite{Broadhurst:1986bx,Barfoot:1987kg}. These techniques were applied to four-dimensional relativistic theories (and their classical analogues)
using RG in DR within the MS scheme which is incomparably more efficient computationally  than the more physically
appealing momentum-shell RG, see the classic monographs: \cite{Collins:1984xc,kleinert2001critical} and especially \cite{Vasil'evbook} for a beautiful historical introduction; see also the
more recent books \cite{grozin2007lectures,smirnov2013analytic} devoted to Feynman diagram calculations.
Among their greatest early achievements, let us mention, \eg, the computation of the $3$-loop $\beta$-function of QCD \cite{Tarasov:1980au}
and the computation of the $5$-loop $\beta$-function of $\Phi^4$-theory
  \cite{Gorishnii:1983gp,Kazakov:1984km,Kazakov:1983pk}.
  The following decades witnessed further ground-breaking developments, \eg,
  in discovering dimensional recurrence relations for Feynman diagrams \cite{Tarasov:1996br,Lee:2009dh},
  in the exact evaluation of massless Feynman diagram, \eg,
    \cite{Gracey:1992ew,Kotikov:1995cw,Broadhurst:1996ur},
  in developing and applying new techniques to deal
  with massive ones, \eg,  \cite{Boos:1990rg}-\cite{Kotikov:1991pm},
in performing high order $\veps$-expansions together with understanding some intriguing relations with number theoretical issues,
\eg, \cite{Broadhurst:1996ur}-\cite{Brown:2009ta},
in applying such techniques to numerous models, \eg, \cite{Kleinert:1991rg}-\cite{Gracey:1993iu},
in the Hopf-algebraic interpretation of the renormalization group \cite{Kreimer:1997dp,Connes:1998qv}, in the notion of a cosmic Galois group \cite{Cartier:2001,Kontsevich:1999,Connes:2004zi}...

Throughout the years, there has been a dramatic increase in the complexity of the calculations.
Modern challenges often require the manipulation of thousands of diagrams each one of them eventually breaking into hundreds of integrals.
Some of the most complicated integrals, the so called ``master integrals'' \cite{Broadhurst:1987ei}, require human assistance for their analytic evaluation and will be the subject of this review. 
Other tasks, on the other hand, are rather systematic in nature as well as highly symbolic: from generating the diagrams, to performing eventual gamma matrix algebra, to reducing large numbers of
diagrams to a few master integrals... This led to the automation of such tasks via the developments of powerful computer algebra systems,
\eg, from SCHOONSHIP \cite{Veltman}, to REDUCE \cite{Hearn}, FORM \cite{Vermaseren:2000nd}, GiNac \cite{Bauer:2000cp} and the commercial MATHEMATICA \cite{wolfram1991mathematica}, see \cite{Weinzierl:2002cg} for a short review.
Specific algorithms were developed to generate Feynman diagrams, \eg, QGRAF \cite{Nogueira:1991ex} and EXP \cite{Seidensticker:1999bb}.
Others to deal with the reduction problem such as Laporta's algorithm \cite{Laporta:2001dd}, Baikov's method \cite{Baikov:1996rk}
as well as computer codes combining several algorithms to achieve this task such as REDUZE \cite{Studerus:2009ye,vonManteuffel:2012np}, 
FIRE \cite{Smirnov:2008iw}, KIRA \cite{Maierhoefer:2017hyi} and LiteRed \cite{Lee:2013mka}. 
Some algorithms are devoted to the subtle problems of dealing with subdivergences and generating the Laurent expansion
such as, \eg, the sector decomposition technique \cite{Binoth:2000ps,Bogner:2007cr}, see also the dissertation of Bogner \cite{Bogner2009},
parametric integration using hyperlogarithms \cite{Panzer:2014caa}, see also the dissertation of Panzer \cite{Panzer:2015ida} and the nice recent paper \cite{Georgoudis:2018olj}, the recently discovered
method of graphical functions \cite{Schnetz:2013hqa} and its combination with parametric
integration \cite{Golz:2015rea} together with the automation of the $R$ and $R^*$ operators by Batkovich and Kompaniets \cite{Batkovich:2014rka}, see also the very recent \cite{Herzog:2017bjx}.
Though computer assisted, all these remarkable developments
may be considered in some sense as analytical as opposed to the numerical methods only used at the final stage of the procedure in order to extract a numerical value for the coefficients of the Laurent series associated with a given diagram.
Often, they involve advanced mathematical concepts from, \eg, graph theory, algebraic geometry and number theory.
Nowadays, computer algebra systems combined with appropriate algorithms are an integral part of the field of multi-loop calculations.
For the years 2016/2017 alone, they allowed breakthrough achievements such as, {\it e.g.}, the 4-loop $\beta$-function calculation for the Gross-Neveu \cite{Gracey:2016mio}~\footnote{Notice that the lower number of loops
presently achieved for the Gross-Neveu model with respect to other models is related to the loss of multiplicative renormalizability of 4-fermion operators in dimensional regularization and the generation of
evanescent operators; so calculations for this model are less straightforward than in other models.}, see also \cite{Gracey:2018qba} for very recent large-$N$ calculations, and Gross-Neveu-Yukawa models \cite{Mihaila:2017ble,Zerf:2017zqi},
the 5-loop $\beta$-function calculation for QCD \cite{Baikov:2016tgj}, its generalization to an arbitrary gauge group \cite{Luthe:2016xec,Luthe:2017ttc,Herzog:2017ohr} and
gauge fixing parameter \cite{Chetyrkin:2017bjc,Luthe:2017ttg},
the $6$-loop \cite{Batkovich:2016jus,Kompaniets:2016hct,Kompaniets:2017yct} and $7$-loop \cite{Schnetz:2016fhy} calculations 
of the $\Phi^4$-model renormalization group functions and the 7-loop anomalous dimension calculation of twist-2 operators in planar $\mathcal{N}=4$ super-Yang-Mills theory with the help of integrability arguments \cite{Marboe:2016igj}.

Having set the general framework, the focus of this review will be on some of the methods that we consider as being the most efficient from the point of 
view of the purely {\it analytic} computation of ``master integrals''. These integrals may be viewed as basic building blocs at the {\it core} of multi-loop calculations. Their evaluation is therefore a fundamental 
task prior to any automation. Their importance is witnessed by the recent appearance of ``Loopedia'' \cite{Bogner:2017xhp} which attempts at providing
a database for all known loop integrals. Many of such integrals related to four-dimensional models were known (but the integrals were scattered throughout various papers)
before the enormous developments mentioned above 
concerning particle physics, statistical mechanics and (supersymmetric) gauge field theories. 
In other cases such as, \eg, odd dimensional theories relevant to condensed matter physics systems, non-standard master integrals appear the systematic evaluation of which is more recent,
see, \eg, \cite{Teber:2012de}-\cite{Kotikov:2016rgs}.

In the following, we will restrict ourselves to the case of {\it massless} Feynman integrals focusing in particular on the two-loop massless propagator-type 
integrals~\footnote{The integrals with many legs are essentially more complicated (see the recent paper \cite{Chawdhry:2018awn} and references therein) and their
consideration is beyond the scope of this review.} which are 
the most elementary ones yet highly non-trivial in some peculiar cases as we will see. The main emphasis will be on the standard rules of perturbation theory for 
massless Feynman diagrams as described in the pioneering work of Kazakov \cite{Kazakov:1984km}, see also the beautiful lectures \cite{Kazakov:1984bw}. 
As will be discussed in details in the review, such rules allow, in principle, the computation of complicated Feynman diagrams using sequences of
simple transformations (without any explicit integration). A diagram is straightforwardly integrated once the appropriate sequence is found. 
In a sense, the method greatly simplifies multi-loop calculations. For a given diagram, the task of finding the sequence of transformations is, however, highly non-trivial. For the most complicated integrals, these
rules can be supplemented with the help of the Gegenbauer $x$-space technique \cite{Chetyrkin:1980pr} and, in particular, rules for integrating massless diagrams with Heaviside theta functions
and traceless products \cite{Kotikov:1995cw}. These rules provide series representation for the diagram which may be a convenient starting point for their $\veps$-expansion.

The manuscript is organized as follows. In Sec.~\ref{sec:basics}, we present the notations as well as the basics of Feynman diagram calculations at one-loop. In
Sec.~\ref{2-loop}, the two-loop massless propagator-type Feynman diagram is presented and a brief historical review of the results concerning this diagram is provided.
In Sec.~\ref{sec:meth}, we review some methods of calculations: parametric integration, integration by parts, the method of uniqueness, Fourier transform and duality and functional equations.
Excepting for the technique of parametric integration, all the mentioned methods allow to compute Feynman integrals without any explicit integration.
  In Sec.~\ref{sec:examples}, some concrete examples of $5$-loop calculations are provided. The conclusion is given in Sec.~\ref{sec:conclusion}. 
Finally, App.~\ref{App:A} shortly reviews the Gegenbauer polynomial technique including loop integration with Heaviside theta
functions into subintegral expressions.

\begin{fmffile}{fmf-review1}

\section{Basics of Feynman diagrams}
\label{sec:basics}

\subsection{Notations}

We consider an Euclidean space-time of dimensionality $D$. Throughout this manuscript we shall use dimensional regularization, 
see the textbook \cite{Collins:1984xc} as well as the early reviews \cite{Leibbrandt:1975dj,Narison:1980ti} for a more complete account on dimensional
regularization and the more recent \cite{Kilgore:2011ta} for a very instructive review on
{\it conventional} dimensional regularization that we actually use, and set, {\it e.g.}, $D=4-2\veps$ in the case of $(3+1)$-dimensional theories, where $\veps \ra 0$ is the
regularization parameter. Then, the infinitesimal volume element, {\it e.g.}, in momentum space, can be written as:
\be
\D^4 k = \mu^{2\eps}\,\D^D k\, ,
\label{d4k}
\ee
where $\mu$ is the so-called renormalization scale in the minimal subtraction ($\text{MS}$) scheme 
which is related to the corresponding parameter $\overline{\mu}$ in the modified minimal subtraction ($\overline{\text{MS}}$) scheme with the help of:
\be
\overline{\mu}^{\,2} = 4\pi e^{-\gamma_E} \mu^2\, ,
\label{muMSbar}
\ee
where $\gamma_E$ is Euler's constant. 
Our main focus will be on {\it massless Feynman diagrams of the propagator type} (the so called p-integrals \cite{Chetyrkin:1981qh}).
Diagrams will be analyzed mainly in momentum space but position space will also be considered in some cases. 
In momentum space, Feynman diagrams are defined as integrals over dummy momentum variables or loop momenta.
The dependence of these integrals on the external momentum follows from dimensional analysis and is power-like. The goal
of the calculations is then to compute the dimensionless coefficient function, $C_D$, associated with a given diagram (see below).
In some peculiar cases this function can be computed exactly. In most cases, only an approximate solution can be found. Often,
it takes the form of a Laurent series in $\veps$ and of great interest are the coefficients of negative power of $\veps$
which may be related to $\beta$-functions and anomalous dimensions of fields.

In the following, for simplicity, we shall assume that all algebraic manipulations related to gamma matrices 
such as, \eg, contraction of Lorentz indices, calculations of traces, etc... have been done.~\footnote{In some cases, \eg, for $n$-point functions, a
	tensorial reduction, the so-called Passarino-Veltman reduction scheme \cite{Passarino:1978jh}, see also \cite{Denner:1991kt} for a review,
	allows to express a tensor integral in terms of scalar ones with tensor coefficients depending on the external
        kinematic variables and eventually the metric tensor. We assume that such a reduction has been performed and essentially focus on the computation of the scalar integrals. Notice that, at
        one-loop, the Passarino-Veltman reduction has been automated in packages such as
	 FeynCalc \cite{Shtabovenko:2016sxi,Mertig:1990an}, LoopTools \cite{Hahn:1998yk} and (combined with FeynArts \cite{Hahn:2000kx}) FormCalc \cite{Hahn:2004rf}.}
The diagrams we shall be mainly interested in are therefore expressed in terms of scalar integrals; for completeness, and because they are of practical interest in concrete calculations,
we shall also consider diagrams with simple numerators such as traceless products. 
In reciprocal space, momentum is conserved at each vertex and integrations are over all internal
momenta. This has to be contrasted with calculations in real space where integrations are over the coordinates of all vertices.
In both spaces, the lines of such diagrams correspond to scalar propagators and are simple power laws. In momentum space, they take the form: $1/k^{2\al}$ where $\al$ is the so-called {\it index} of the line.
A line with an arbitrary index $\al$ can be represented graphically as:
\be
\frac{1}{k^{2\al}}: \quad \parbox{10mm}{
  \begin{fmfgraph*}(10,10)
    \fmfleft{i}
    \fmfright{o}
	  \fmf{plain,label=\small{$\al$},l.s=left}{i,o}
  \end{fmfgraph*}
} \quad , \qquad
\frac{k^\mu}{k^{2\al}}: \quad \parbox{10mm}{
  \begin{fmfgraph*}(10,10)
    \fmfleft{i}
    \fmfright{o}
	  \fmf{fermion,label=\small{$\al^\mu$},l.s=left}{i,o}
  \end{fmfgraph*}
} \quad  \, , \qquad
\frac{k^{\mu_1 \cdots \mu_n}}{k^{2\al}}: \quad \parbox{10mm}{
  \begin{fmfgraph*}(10,10)
    \fmfleft{i}
    \fmfright{o}
	  \fmf{fermion,label=\small{$\al^{\mu_1 \cdots \mu_n}$},l.s=left}{i,o}
  \end{fmfgraph*}
} \quad  \, ,
\label{def:lines}
\ee
where the absence of arrow implies a scalar propagator while arrows indicate the presence of a non-trivial numerator the tensorial structure of which is displayed on the index for clarity.
In momentum space, ordinary lines have index $1$. The link with coordinate space is given by the Fourier transform:
\be
\int [\D^D p]\,\frac{e^{\I p x}}{[p^2]^\al} = \frac{2^{D-2\al}}{(4\pi)^{D/2}}\,\frac{a(\al)}{[\,x^2\,]^{D/2-\al}}, \qquad a(\al) = \frac{\Gamma(D/2-\al)}{\Gamma(\al)} \, ,
\label{FT-line}
\ee
where  $\al \not= D/2, D/2+1, ...$, $\Gamma(x)$ is Euler's gamma function and, in the following, we shall often use the notation 
\be
[\D^D p] = \frac{\D^D p}{(2 \pi)^D}\, .
\ee
From Eq.~(\ref{FT-line}), we see that, in coordinate space, ordinary lines have dimension $D/2-1$.
For a line of arbitrary index $\al$, the indices $\al$ and $D/2-\al$ are said to be {\it dual} to each other in the sense of Fourier transform.
The inverse Fourier transform reads:
\be
\frac{1}{[\,p^2\,]^\al} = \frac{a(\al)}{\pi^{D/2}\,2^{2\al}}\,\int \D^D x \,\frac{e^{-\I p x}}{[\,x^2\,]^{D/2-\al}}\, .
\label{FT-inv-line}
\ee
In $p$-space, a zero index means that the corresponding line should shrink to a point while in 
$x$-space it means that the line has to be deleted. 

Let us then consider a general $L$-loop propagator-type Feynman diagram with $N$-internal lines.
Schematically, this diagram can be represented as:
\bea
\parbox{15mm}{
  \begin{fmfgraph*}(15,13)
      \fmfleft{i}
      \fmfright{o}
      \fmf{plain}{i,v}
      \fmfblob{.75w}{v}
      \fmf{plain}{v,o}
  \end{fmfgraph*}
}
\quad = \quad 
\frac{(p^2)^{\frac{LD}{2} - \sum_{i=1}^N\,\al_i}}{(4\pi)^{\frac{LD}{2}}}\,\text{C}_D(\vec{\al})\, ,
\label{def:p-integral}
\eea
where $\vec{\al} = (\al_1,\,\al_2,\, \cdots,\,\al_N)$ and the index of the diagram corresponds to the sum of the indices of its constituent lines: 
$\sum_{i=1}^N\,\al_i$. 
Eq.~(\ref{def:p-integral}) defines the dimensionless coefficient function, $\text{C}_D(\vec{\al})$, 
of the propagator-type diagram. This function depends on the indices, $\vec \al$, and the dimension of space-time, $D$. 
In the following, we shall extensively study the one-loop and two-loop p-integrals. At this point, we briefly consider vacuum-type diagrams (the so-called v-integrals).
A general multi-loop v-integral can be represented as:
\bea
\parbox{15mm}{
  \begin{fmfgraph*}(15,13)
      \fmfleft{v}
      \fmfblob{.75w}{v} \qquad \, '
  \end{fmfgraph*}
} 
\eea
and can be obtained from the p-integral by putting the external momenta to zero. Such diagrams are therefore scaleless.

\subsection{Massless vacuum diagrams (and Mellin-Barnes transformation)}

In the frame of dimensional regularization, the one-loop massless vacuum diagram obeys the following identity \cite{Gorishnii:1984te}:
\be
\parbox{10mm}{
  \begin{fmfgraph*}(10,10)
      \fmfleft{i}
      \fmfright{o}
	  \fmf{plain,left,label=\small{$\al$},l.s=right}{i,o}
      \fmf{plain,left}{o,i}
  \end{fmfgraph*}
} 
\quad = \quad
\int \frac{\D^D k}{(k^2)^\al} \quad = \quad i \pi \, \Omega_D \, \delta(\al-D/2)\, ,
\label{one-loop-v-int}
\ee
where $\Omega_D = 2 \pi^{D/2} / \Gamma(D/2)$ is the surface of the unit hypersphere in $D$-dimensional Euclidean space-time.
One way to check the consistency of this relation is to consider the Mellin-Barnes transformation of a massive scalar propagator \cite{Usyukina:1975yg,Boos:1990rg}:
\be
\frac{1}{(k^2+m^2)^\al} = \frac{1}{2\I \pi \Gamma(\al)}\,\int_{-\I \infty}^{\I \infty} \D s \,\frac{(m^2)^s}{(k^2)^{s+\al}}\,\Gamma(-s) \Gamma(\al+s)\, .
\label{Mellin-Barnes}
\ee
Interestingly, this transformation allows to express a massive propagator in terms of a contour integral involving a massless one.
As noticed by Boos and Davydychev \cite{Boos:1990rg}, this suggests that techniques for computing massless Feynman diagrams may be of importance to compute massive ones.
For our present purpose, only the expression of the massive one-loop tadpole integral will be useful.~\footnote{See, \eg, Refs.~\cite{Usyukina:1975yg}-\cite{Davydychev:1992mt}
for more examples of the use of the Mellin-Barnes transformation in Feynman diagram calculations.} It reads:
\be
\int \frac{[\D^D k]}{(k^2 + m^2)^\al} = \frac{(m^2)^{D/2-\al}}{(4\pi)^{D/2}}\,\frac{\Gamma(\al - D/2)}{\Gamma(\al)}\, ,
\label{massive-one-loop-tadpole}
\ee
where the mass dependence follows from dimensional analysis and the dimensionless factor $\Gamma(\al - D/2)/\Gamma(\al)$ corresponds to the coefficient function of this simple diagram. 
Eq.~(\ref{massive-one-loop-tadpole}) can be straightforwardly obtained using the standard parametric integration technique, see Sec.~\ref{sec:meth:param-int}. 
It can also be obtained using the Mellin-Barnes transformation (\ref{Mellin-Barnes}) upon assuming that
Eq.~(\ref{one-loop-v-int}) holds. This proves the consistency of Eq.~(\ref{one-loop-v-int}).

There is a connection between v-integrals and p-integrals, see Ref.~\cite{Gorishnii:1984te}. The latter may be derived by turning a p-integral
into a v-integral upon multiplying it by $(p^2)^{-\sigma}$ and integrating over $p$.~\footnote{This is also known as gluing, see Ref.~\cite{Chetyrkin:1981qh} and \cite{Baikov:2010hf} for a recent review.} 
From Eq.~(\ref{def:p-integral}), such a procedure yields:
\begin{subequations}
\bea
\parbox{15mm}{
  \begin{fmfgraph*}(15,13)
      \fmfleft{i}
      \fmfright{o}
      \fmf{plain}{i,v}
      \fmfblob{.75w}{v}
      \fmf{plain}{v,o}
	  \fmf{plain,right,label=\small{$\sigma$},l.s=right}{i,o}
  \end{fmfgraph*}
}
\quad & = & \quad
\frac{C_D(\vec{\al})}{(4\pi)^{\frac{LD}{2}}}\,\int \frac{\D^D p}{(p^2)^{\sigma + \sum_{i=1}^N\, \al_i - \frac{LD}{2}}} 
\nonum \\
& = & \quad 
\frac{i \pi}{(4\pi)^{\frac{LD}{2}}} \, \Omega_D \, C_D(\vec{\al})\,\delta(\sigma + \sum_{i=1}^N\, \al_i - \frac{LD}{2}) 
\\
& = & \quad 
\frac{i \pi}{(4\pi)^{\frac{LD}{2}}} \, \Omega_D \, X_D(\vec{\al},\sigma)\,\delta(\sigma + \sum_{i=1}^N\, \al_i - \frac{LD}{2}) \, .
\eea
\end{subequations}
Hence, the coefficient functions of the v-type diagram, $X_D$, and the p-type diagram, $C_D$, are related by:
\be
C_D(\vec{\al}) = X_D(\vec{\al},\sigma)|_{\sigma = \frac{LD}{2} - \sum_{i=1}^N\, \al_i}\, .
\label{v-vs-p}
\ee

As can be noticed from Eq.~(\ref{one-loop-v-int}), vacuum diagrams are rather ambiguously defined within dimensional regularization. Their scaleless
nature does not provide any clue of what their actual value might be a priori: it can be zero, infinity or even some finite number, 
see also the review \cite{Leibbrandt:1975dj}. This results from a subtle interplay between infrared and ultraviolet divergences of massless
diagrams. Following t'Hooft and Veltman, we shall often assume that the continuous dimension regularizes such highly divergent integrals to {\it zero}. 
Similarly, whenever a diagram contains a scaleless subdiagram, {\it e.g.}, such as the massless tadpole diagram,
its value will be set to zero. In principle, however, care must be taken in the case where $D=2\al$. 
On the other hand, the consequence of Eq.~(\ref{one-loop-v-int}) is unambiguous for integrals over polynomials corresponding to the case where $\al<0$ in (\ref{one-loop-v-int});
within dimensional regularization such integrals vanish identically. Summarizing, in the following, we shall always assume that:~\footnote{In the case where $\al=D/2$ is encountered, it is
also possible to use the following trick: introduce a regulator $\delta \ra 0$ shifting the index $\al$, \eg, $\al \ra \al+\delta$. The limit $\delta \ra 0$ is taken at the end of the calculation. 
See Ref.~\cite{Kotikov:2013kcl} for an example.}
\be
\int \frac{\D^D k}{(k^2)^\al} = 0 \qquad (\al \not= D/2)\, .
\label{one-loop-v-int2}
\ee

\subsection{Massless one-loop propagator-type diagram}
\label{sec:one-loop-p-diag}

The one-loop (scalar) p-type massless integral is defined as:
\be
J(D,p,\al,\beta) = \int [\D^D k] \frac{1}{k^{2\al} (p-k)^{2\beta}}\, ,
\label{def:one-loop-p-int}
\ee
where $p$ is the external momentum and $\al$ and $\beta$ are arbitrary indices.
In graphical form it is represented as:
\bea
J(D,p,\al,\beta) =
\qquad
\parbox{13mm}{
    \begin{fmfgraph*}(15,13)
      \fmfleft{i}
      \fmfright{o}
      \fmfleft{ve}
      \fmfright{vo}
      \fmffreeze
      \fmfforce{(-0.3w,0.5h)}{i}
      \fmfforce{(1.3w,0.5h)}{o}
      \fmfforce{(0w,0.5h)}{ve}
      \fmfforce{(1.0w,0.5h)}{vo}
      \fmffreeze
	    \fmf{fermion,label=\footnotesize{$p$}}{i,ve}
	    \fmf{plain,left=0.6,label=\footnotesize{$\al$},l.d=0.1h}{ve,vo}
	    \fmf{plain,left=0.6,label=\footnotesize{$\beta$},l.d=0.05h}{vo,ve}
      \fmf{plain}{vo,o}
      \fmffreeze
      \fmfdot{ve,vo}
    \end{fmfgraph*}
} \quad \qquad .
\eea
In Eq.~(\ref{def:one-loop-p-int}), the momentum dependence is easily extracted from dimensional analysis which allows to write the diagram in the following form:
\be
J(D,p,\al,\beta) =  \frac{(p^2)^{D/2 - \al - \beta}}{(4\pi)^{D/2}}\,G(D,\al,\beta)\, ,
\label{def:one-loop-G-func}
\ee
where $G(D,\al,\beta)$ is the (dimensionless) coefficient function of the diagram. In graphical form, the latter is represented by a diagram similar to the one for $J(D,p,\al,\beta)$ but with amputated external legs:
\be
G(D,\al,\beta) = \text{C}_D\bigg[\int \frac{[\D^D k]}{k^{2\al} (p-k)^{2\beta}}\bigg] \quad =
\quad
  \parbox{13mm}{
  \begin{fmfgraph*}(15,13)
      \fmfleft{i}
      \fmfright{o}
	  \fmf{plain,right=0.6,label=\footnotesize{$\al$},l.d=0.1h}{i,o}
	  \fmf{plain,right=0.6,label=\footnotesize{$\beta$},l.d=0.05h}{o,i}
      \fmfdot{i,o}
  \end{fmfgraph*}
}\qquad  .
\ee
In the one-loop case, the so-called $G$-function is known exactly and reads:
\be
G(D,\al,\beta) = \frac{a(\al) a(\beta)}{a(\al + \beta -D/2)}, \qquad a(\al) = \frac{\Gamma(D/2 - \al)}{\Gamma(\al)}\, .
\label{one-loop-G}
\ee
All these results may be generalized to integrands with numerators. In particular:
\be
J^{\mu_1 \cdots  \mu_n}(D,p,\al,\beta) = \int [\D^D k] \frac{k^{\mu_1 \cdots  \mu_n}}{k^{2\al} (p-k)^{2\beta}} = \frac{(p^2)^{D/2 - \al - \beta}}{(4\pi)^{D/2}}\,p^{\mu_1 \cdots \mu_n}\,G^{(n,0)}(D,\al,\beta)\, ,
\label{def:one-loop-p-int+tp}
\ee
where $k^{\mu_1 \cdots  \mu_n}$ denotes the traceless product and
\be
G^{(n,0)}(D,\al,\beta) = \frac{a_n(\al) a_0(\beta)}{a_n(\al + \beta -D/2)}, \qquad a_n(\al) = \frac{\Gamma(n+D/2 - \al)}{\Gamma(\al)}\, ,
\label{one-loop-Gn}
\ee
where $G(D,\al,\beta) \equiv G^{(0,0)}(D,\al,\beta)$ and, for simplicity, the argument $D$ is also sometimes dropped unless an ambiguity may arise. Graphically, a one-loop p-type 
diagram with numerator can be represented as:
%
%
\be
G^{(1,0)}(D,\al,\beta) = \text{C}_D\bigg[\int \frac{[\D^D k]~~k^\mu}{k^{2\al} (p-k)^{2\beta}}\bigg] \quad =
\quad
  \parbox{13mm}{
  \begin{fmfgraph*}(15,13)
      \fmfleft{i}
      \fmfright{o}
	  \fmf{fermion,left=0.6,label=\footnotesize{$\al^\mu$},l.d=0.2h}{i,o}
	  \fmf{plain,left=0.6,label=\footnotesize{$\beta$},l.d=0.05h}{o,i}
      \fmfdot{i,o}
  \end{fmfgraph*}
} \quad = \quad
-\quad 
  \parbox{13mm}{
  \begin{fmfgraph*}(15,13)
      \fmfleft{i}
      \fmfright{o}
	  \fmf{fermion,right=0.6,label=\footnotesize{$\al^\mu$},l.d=0.2h}{o,i}
	  \fmf{plain,right=0.6,label=\footnotesize{$\beta$},l.d=0.05h}{i,o}
      \fmfdot{i,o}
  \end{fmfgraph*}
} \quad\,~~ .
\ee
Whenever the integrand contains a scalar product, the corresponding lines are arrowed. As an example:
\begin{flalign}
\text{C}_D\bigg[\int \frac{[\D^D k]~~(k,p-k)}{k^{2\al} (p-k)^{2\beta}}\bigg] \quad = \quad
  \parbox{13mm}{
  \begin{fmfgraph*}(15,13)
      \fmfleft{i}
      \fmfright{o}
	  \fmf{fermion,left=0.6,label=\footnotesize{$\al^\mu$},l.d=0.15h}{i,o}
	  \fmf{fermion,right=0.6,label=\footnotesize{$\beta_\mu$},l.d=0.1h}{i,o}
      \fmfdot{i,o}
  \end{fmfgraph*}
} \quad = \quad
  \parbox{13mm}{
  \begin{fmfgraph*}(15,13)
      \fmfleft{i}
      \fmfright{o}
	  \fmf{fermion,right=0.6,label=\footnotesize{$\al^\mu$},l.d=0.15h}{o,i}
	  \fmf{fermion,left=0.6,label=\footnotesize{$\beta_\mu$},l.d=0.1h}{o,i}
      \fmfdot{i,o}
  \end{fmfgraph*}
} \quad = \quad - \quad
  \parbox{13mm}{
  \begin{fmfgraph*}(15,13)
      \fmfleft{i}
      \fmfright{o}
	  \fmf{fermion,left=0.6,label=\footnotesize{$\al^\mu$},l.d=0.15h}{i,o}
	  \fmf{fermion,left=0.6,label=\footnotesize{$\beta_\mu$},l.d=0.1h}{o,i}
      \fmfdot{i,o}
  \end{fmfgraph*}
} \qquad ,
\end{flalign}
\vskip 2mm

\ni where the notation $(k,p)=k^\mu p_\mu$ denotes the scalar product of the $D$-dimensional momenta $k$ and $p$.
When such a diagram is encountered, it can be evaluated as:
\vspace{2mm}
\begin{flalign}
  \parbox{13mm}{
  \begin{fmfgraph*}(15,13)
      \fmfleft{i}
      \fmfright{o}
	  \fmf{fermion,left=0.6,label=\footnotesize{$\al^\mu$},l.d=0.15h}{i,o}
	  \fmf{fermion,right=0.6,label=\footnotesize{$\beta_\mu$},l.d=0.1h}{i,o}
      \fmfdot{i,o}
  \end{fmfgraph*}
} \quad = \quad - \quad
  \parbox{13mm}{
  \begin{fmfgraph*}(15,13)
      \fmfleft{i}
      \fmfright{o}
	  \fmf{plain,left=0.6,label=\footnotesize{$\al-1$},l.d=0.15h}{i,o}
	  \fmf{plain,right=0.6,label=\footnotesize{$\beta$},l.d=0.1h}{i,o}
      \fmfdot{i,o}
  \end{fmfgraph*}
} \quad + \qquad
\parbox{13mm}{
    \begin{fmfgraph*}(15,13)
      \fmfleft{i}
      \fmfleft{ve}
      \fmfright{vo}
      \fmffreeze
      \fmfforce{(-0.3w,0.5h)}{i}
      \fmfforce{(0w,0.5h)}{ve}
      \fmfforce{(1.0w,0.5h)}{vo}
      \fmffreeze
	    \fmf{fermion,label=\footnotesize{$p_\mu$}}{i,ve}
	    \fmf{fermion,left=0.6,label=\footnotesize{$\al^\mu$},l.d=0.15h}{ve,vo}
	    \fmf{plain,left=0.6,label=\footnotesize{$\beta$},l.d=0.1h}{vo,ve}
      \fmffreeze
      \fmfdot{ve,vo}
    \end{fmfgraph*}
} \qquad = \quad -G(D,\al-1,\beta) + G^{(1,0)}(D,\al,\beta)\, ,
\label{oneloop+arrows}
\end{flalign}
\vskip 2mm

\ni where, in the third diagram, the internal momentum is dotted with an external one. Hence:
\be
\int \frac{[\D^D k]~~(k,p-k)}{k^{2\al} (p-k)^{2\beta}} = \frac{(p^2)^{D/2+1-\al-\beta}}{(4\pi)^{D/2}}\,
\bigg( G^{(1,0)}(D,\al,\beta) - G(D,\al-1,\beta)\bigg)\, .
\label{oneloop+arrows:res}
\ee

The expression of $G(D,\al,\beta)$ given above may be derived by using parametric integration, see Sec.~\ref{sec:meth:param-int}.
Following Ref.~\cite{Kazakov:1984bw}, an alternative simple derivation consists in first going to $x$-space using Eq.~(\ref{FT-inv-line})
and then going back to $p$-space:
\bea
\int \frac{\D^D k}{k^{2\al} (p-k)^{2\beta}}
&=&  \frac{a(\al) a(\beta)}{\pi^D\,2^{2(\al + \beta)}}\,\int \frac{\D^D x\, \D^D y\,\D^D k\, e^{-\I k x-\I (p-k)y}}{[\,x^2\,]^{D/2-\al}\,[\,y^2\,]^{D/2-\beta}}
\nonum \\
&=&  \frac{a(\al) a(\beta)}{2^{2(\al + \beta)-D}}\,\int \frac{\D^D x\, e^{-\I p x}}{[\,x^2\,]^{D-\al-\beta}}
\nonum \\
&=&  \pi^{D/2}\,\frac{a(\al) a(\beta)}{a(\al + \beta - D/2)}\,\frac{1}{[\,p^2\,]^{\al + \beta - D/2}} \, .
\label{def:loop_calc}
\nonum
\eea
This leads to Eq.~(\ref{one-loop-G}).

An important property of the $G$-function is that it vanishes whenever one (or more) of the indices is zero or a negative {\it integer}:
\be
G(D,n,m) = 0, \qquad n \leq 0, \quad m \leq 0\, .
\ee
This property follows from the fact that a massless one-loop p-integral with a zero or negative integer index corresponds to a massless vacuum
diagram (possibly with a polynomial numerator) which vanishes according to Eqs.~(\ref{one-loop-v-int2}) 
(provided the case $D=2\al$ is not encountered). 

Finally, the $G$-function informs about the singularities of Eq.~(\ref{def:one-loop-p-int}). The latter can be either ultraviolet or infrared in nature. In both cases, they will appear 
as $1/\veps$ poles in the expression of the $G$-function as dimensional regularization treats both types of singularities on an equal footing. 
In order to see that, let's note that from dimensional analysis Eq.~(\ref{def:one-loop-p-int}) has an ultraviolet singularity ($k \ra \infty$)
for $\al + \beta \leq D/2$; on the other hand, it has an infrared singularity ($k \ra 0$) for $\al \geq D/2$ and/or $\beta \geq D/2$. Then, from the explicit expression of $G(D,\al, \beta)$ 
in terms of $\Gamma$-functions:
\be
G(D,\al,\beta) = \frac{\Gamma(D/2-\al) \Gamma(D/2-\beta)\Gamma(\al + \beta -D/2)}{\Gamma(\al)\Gamma(\beta)\Gamma(D-\al-\beta)}\, ,
\label{G:IR+UV}
\ee
we see that poles coming from either one of the first two $\Gamma$-functions in the numerator are IR poles while those coming from the last $\Gamma$-function in the numerator are UV poles.

\section{Massless two-loop propagator-type diagram}
\label{2-loop}

\subsection{Basic definition}

Central to this manuscript is the massless two-loop propagator-type diagram which is 
defined as:
\be
J(D,p,\al_1,\al_2,\al_3,\al_4,\al_5) = \int \frac{[\D^Dk][\D^Dq]}{(k-p)^{2\al_1}\,(q-p)^{2\al_2}\,q^{2\al_3}\,k^{2\al_4}\,(k-q)^{2\al_5}}\, ,
\label{def:two-loop-p-int}
\ee
where $p$ is the external momentum and the $\al_i$, $i=1-5$, are five arbitrary indices.
In graphical form, it is represented as:
\be
J(D,p,\al_1,\al_2,\al_3,\al_4,\al_5) 
\quad = \qquad
\parbox{16mm}{
    \begin{fmfgraph*}(16,14)
      \fmfleft{i}
      \fmfright{o}
      \fmfleft{ve}
      \fmfright{vo}
      \fmftop{vn}
      \fmftop{vs}
      \fmffreeze
      \fmfforce{(-0.3w,0.5h)}{i}
      \fmfforce{(1.3w,0.5h)}{o}
      \fmfforce{(0w,0.5h)}{ve}
      \fmfforce{(1.0w,0.5h)}{vo}
      \fmfforce{(.5w,0.95h)}{vn}
      \fmfforce{(.5w,0.05h)}{vs}
      \fmffreeze
	    \fmf{fermion,label=\footnotesize{$p$}}{i,ve}
      \fmf{plain,left=0.8}{ve,vo}
	    \fmf{phantom,left=0.7,label=\footnotesize{$\al_1$},l.d=-0.1w}{ve,vn}
	    \fmf{phantom,right=0.7,label=\footnotesize{$\al_2$},l.d=-0.1w}{vo,vn}
      \fmf{plain,left=0.8}{vo,ve}
	    \fmf{phantom,left=0.7,label=\footnotesize{$\al_3$},l.d=-0.1w}{vo,vs}
	    \fmf{phantom,right=0.7,label=\footnotesize{$\al_4$},l.d=-0.1w}{ve,vs}
	    \fmf{plain,label=\footnotesize{$\al_5$},l.d=0.05w}{vs,vn}
      \fmf{plain}{vo,o}
      \fmffreeze
      \fmfdot{ve,vn,vo,vs}
    \end{fmfgraph*}
} \qquad \, .
\label{graph:J}
\ee
Similarly to the one-loop case, the momentum dependence of Eq.~(\ref{def:two-loop-p-int}) follows from dimensional analysis which allows to write this diagram in the form:
\be
J(D,p,\al_1,\al_2,\al_3,\al_4,\al_5) = \frac{(p^2)^{D-\sum_{i=1}^5 \al_i}}{(4\pi)^D}\, G(D,\al_1,\al_2,\al_3,\al_4,\al_5)\, ,
\label{def:two-loop-G-func}
\ee
where the (dimensionless) coefficient function of the diagram, $G(D,\{\al_i\})$, has been defined according to the general case Eq.~(\ref{def:p-integral}).
Graphically, the latter is represented as:
\be
G(D,\al_1,\al_2,\al_3,\al_4,\al_5) = \text{C}_D[J(D,p,\al_1,\al_2,\al_3,\al_4,\al_5)]
\quad = \quad
\parbox{16mm}{
    \begin{fmfgraph*}(16,14)
      \fmfleft{ve}
      \fmfright{vo}
      \fmftop{vn}
      \fmftop{vs}
      \fmffreeze
      \fmfforce{(0w,0.5h)}{ve}
      \fmfforce{(1.0w,0.5h)}{vo}
      \fmfforce{(.5w,0.95h)}{vn}
      \fmfforce{(.5w,0.05h)}{vs}
      \fmffreeze
      \fmf{plain,left=0.8}{ve,vo}
	    \fmf{phantom,left=0.7,label=\footnotesize{$\al_1$},l.d=-0.1w}{ve,vn}
	    \fmf{phantom,right=0.7,label=\footnotesize{$\al_2$},l.d=-0.1w}{vo,vn}
      \fmf{plain,left=0.8}{vo,ve}
	    \fmf{phantom,left=0.7,label=\footnotesize{$\al_3$},l.d=-0.1w}{vo,vs}
	    \fmf{phantom,right=0.7,label=\footnotesize{$\al_4$},l.d=-0.1w}{ve,vs}
	    \fmf{plain,label=\footnotesize{$\al_5$},l.d=0.05w}{vs,vn}
      \fmffreeze
      \fmfdot{ve,vn,vo,vs}
    \end{fmfgraph*}
}\quad \, .
\label{graph:G2loop}
\ee
\begin{figure}
  \begin{center}
  \includegraphics{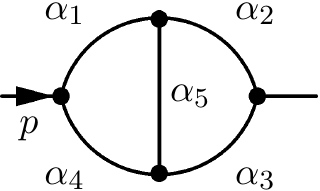}
  \caption{\label{fig:J}
  Two-loop scalar massless propagator-type diagram.}
  \end{center}
\end{figure}

As in the one-loop case, whenever the integrand contains a scalar product, the corresponding lines are arrowed. As an example:
\be
\text{C}_D\bigg[\int \frac{[\D^Dk][\D^Dq]\,(k-q,q-p)}{(k-p)^{2\al_1}\,(q-p)^{2\al_2}\,q^{2\al_3}\,k^{2\al_4}\,(k-q)^{2\al_5}} \bigg]
\quad = \quad
\parbox{16mm}{
    \begin{fmfgraph*}(16,14)
      \fmfleft{ve}
      \fmfright{vo}
      \fmftop{vn}
      \fmftop{vs}
      \fmffreeze
      \fmfforce{(0w,0.5h)}{ve}
      \fmfforce{(1.0w,0.5h)}{vo}
      \fmfforce{(.5w,0.95h)}{vn}
      \fmfforce{(.5w,0.05h)}{vs}
      \fmffreeze
      \fmf{plain,left=0.8}{ve,vo}
      \fmf{phantom_arrow,right=0.4}{vo,vn}
	    \fmf{phantom,left=0.7,label=\footnotesize{$\al_1$},l.d=-0.01w}{ve,vn}
	    \fmf{phantom,right=0.7,label=\footnotesize{$\al_{2 \mu}$},l.d=-0.01w}{vo,vn}
      \fmf{plain,left=0.8}{vo,ve}
	    \fmf{phantom,left=0.7,label=\footnotesize{$\al_3$},l.d=-0.01w}{vo,vs}
	    \fmf{phantom,right=0.7,label=\footnotesize{$\al_4$},l.d=-0.01w}{ve,vs}
	    \fmf{fermion,label=\footnotesize{$\al_5^\mu$},l.d=0.05w}{vs,vn}
      \fmffreeze
      \fmfdot{ve,vn,vo,vs}
    \end{fmfgraph*}
} \quad \, .
\label{2loop-arrows-1}
\ee
As another example, we may consider the case where the scalar product involves the external momentum:
\be
\text{C}_D\bigg[\int \frac{[\D^Dk][\D^Dq]\,(p,q)}{(k-p)^{2\al_1}\,(q-p)^{2\al_2}\,q^{2\al_3}\,k^{2\al_4}\,(k-q)^{2\al_5}} \bigg]
\qquad = \qquad
\parbox{16mm}{
    \begin{fmfgraph*}(16,14)
      \fmfleft{i}
      \fmfleft{ve}
      \fmfright{vo}
      \fmftop{vn}
      \fmftop{vs}
      \fmffreeze
      \fmfforce{(-0.3w,0.5h)}{i}
      \fmfforce{(0w,0.5h)}{ve}
      \fmfforce{(1.0w,0.5h)}{vo}
      \fmfforce{(.5w,0.95h)}{vn}
      \fmfforce{(.5w,0.05h)}{vs}
      \fmffreeze
	    \fmf{fermion,label=\footnotesize{$p^\mu$}}{i,ve}
      \fmf{plain,left=0.8}{ve,vo}
      \fmf{phantom_arrow,right=0.4}{vs,vo}
	    \fmf{phantom,left=0.7,label=\footnotesize{$\al_1$},l.d=-0.01w}{ve,vn}
	    \fmf{phantom,right=0.7,label=\footnotesize{$\al_2$},l.d=-0.01w}{vo,vn}
      \fmf{plain,left=0.8}{vo,ve}
	    \fmf{phantom,left=0.7,label=\footnotesize{$\al_{3 \mu}$},l.d=-0.01w}{vo,vs}
	    \fmf{phantom,right=0.7,label=\footnotesize{$\al_{4}$},l.d=-0.01w}{ve,vs}
	    \fmf{plain,label=\footnotesize{$\al_5$},l.d=0.05w}{vs,vn}
      \fmffreeze
      \fmfdot{ve,vn,vo,vs}
    \end{fmfgraph*}
}\quad \, .
\label{2loop-arrows-2}
\ee

\begin{figure}
\begin{center}
  \includegraphics{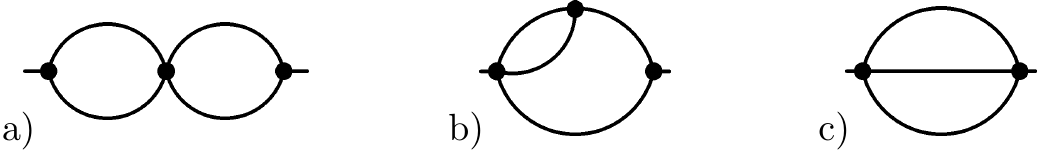}
  \caption{\label{fig:primitive}
  Two-loop primitive, or recursively one-loop, diagrams. Diagram a) corresponds to $\al_5=0$. Diagram b) corresponds to $\al_4=0$. Diagram c) corresponds to $\al_1=\al_3=0$.}
\end{center}
\end{figure}

Notice that that there is a single topological class of two-loop propagator-type diagrams represented in Fig.~\ref{fig:J}.
In some cases, two-loop diagrams may be reduced to products of one-loop diagrams (and hence products of $\Gamma$-functions) and are said to be {\it primitively one-loop}, or {\it recursively one-loop}, 
diagrams \cite{Chetyrkin:1981qh};  some examples of the later are given in Fig.~\ref{fig:primitive}. In general, however, no simple expression is known for the diagram of Fig.~\ref{fig:J} with
five arbitrary indices.

\subsection{Symmetries}
\label{sec:symmetries}

Symmetries are important because they yield identities among the coefficient functions with changed indices. 
We shall introduce a number of other such identities which follow from non-trivial transformations 
in the following sections. An appropriate use of identities among diagrams with different indices is central to multi-loop calculations 
and very often significantly reduces the amount of computations which has to be done. As a matter of fact,
these identities, when used in an appropriate way, may reduce a considerable number of two-loop diagrams to primitively
one-loop ones leaving only a small set of truly two-loop diagrams. As already mentioned in the Introduction, following Broadhurst, the set of irreducible integrals (at $1$, $2$ or higher loop order) 
is refereed to as the {\it master integrals} \cite{Broadhurst:1987ei}.

We start with some basic symmetries of the diagram which follow from changing the integration variables in Eq.~(\ref{def:two-loop-p-int}):
\begin{itemize}
\item the invariance of the integral upon changing $k \lra q$ implies that the diagram is invariant under the change $1 \leftrightarrow 2$
and $3 \leftrightarrow 4$. Geometrically, this can be viewed as an invariance of the diagram in a reflection 
through the plane perpendicular to the diagram and
containing the line of index $\al_5$. At the level of the coefficient function, Eq.~(\ref{def:two-loop-G-func}), this yields the following trivial identity:
\be
G(D,\al_1,\al_2,\al_3,\al_4,\al_5) = G(D,\al_2,\al_1,\al_4,\al_3,\al_5).
\label{transf1}
\ee
\item the invariance of the integral upon changing $k \lra k-p$ and $q \lra q-p$ implies that the diagram is 
invariant under the change $1 \leftrightarrow 4$ and $2 \leftrightarrow 3$. 
Geometrically, this can be viewed as an invariance of the diagram in a reflection through the plane perpendicular to the diagram and
to the line of index $\al_5$. This yields the following trivial identity among the coefficient functions with changed indices:
\be
G(D,\al_1,\al_2,\al_3,\al_4,\al_5) = G(D,\al_4,\al_3,\al_2,\al_1,\al_5).
\label{transf2}
\ee
\end{itemize}

It turns out that the 2-loop diagram is invariant under a very large group: $Z_2 \times S_6$ which has $1,440$ elements \cite{Broadhurst:1986bx,Barfoot:1987kg}.
Historically, some of the symmetry properties of the diagram were observed by the St Petersburg group \cite{Vasiliev:1981dg} (see also the textbook \cite{Vasil'evbook}). A few years later, 
the study of Gorishny and Isaev~\cite{Gorishnii:1984te} clearly revealed that the diagram has a full tetrahedral symmetry. 
To demonstrate this, they used the relation between the coefficient functions of the 2-loop p-integral 
and the related 3-loop v-integral. From Eq.~(\ref{v-vs-p}), 
this relation reads:\footnote{This is again the gluing (or glue and cut) method,
see Ref.~\cite{Chetyrkin:1981qh} and \cite{Baikov:2010hf} for a recent review.}
\be
G(D,\al_1,\al_2,\al_3,\al_4,\al_5) = X_D(\vec{\al},\sigma)|_{\sigma = D - \sum_{i=1}^5\, \al_i} \, .
\label{pv}
\ee
The 3-loop v-integral has the full tetrahedral symmetry. 
The coefficient function $X_D$ is therefore invariant under the permutations of the indices corresponding to the elements of the full
tetrahedral group (rotations and reflections) which is isomorphic to $S_4$, the symmetric group of 4 elements (the vertices of the tetrahedron).
This group has $4!=24$ elements and 3 generators;~\footnote{There are two possible sets of generators for the symmetric group $S_n$:
\begin{itemize}
\item $n-1$ generators formed by the transpositions $(1\,2),\,\,(2\,3),\,\, \cdots (n\,n-1)$,
\item 2 generators formed by a transposition $(1\,2)$ and an n-cycle: $(1\,2\,\cdots\,n)$.
\end{itemize}} 
for example: rotations around 2 axes passing by some vertex and the center
of the opposite side and one reflection. The generating elements can, for example, be chosen as: the rotation axes passing through the vertices $(\al_1,\,\al_5,\,\al_2)$
and $(\al_1,\,\sigma,\,\al_4)$ and a reflection corresponding to the permutation of the two vertices $(\al_1,\,\al_2,\,\al_5)$ and $(\al_3,\,\al_4,\,\al_5)$ (similar
to the transformation $(1 \leftrightarrow 4,\,\,2 \leftrightarrow 3)$).
For the v-integral, this yields:
\begin{subequations}
\label{S4generators}
\bea
X_D(\al_1,\al_2,\al_3,\al_4,\al_5,\sigma) & = & X_D(\al_4,\sigma,\al_2,\al_5,\al_1,\al_3)
\label{S4generators-1} \\
& = & X_D(\al_4,\al_5,\al_2,\sigma,\al_3,\al_1),
\label{S4generators-2} \\
& = & X_D(\al_4,\al_3,\al_2,\al_1,\al_5,\sigma).
\label{S4generators-3}
\eea
\end{subequations}
Then, cutting one line of the v-integral to transform it into a p-integral yields:
\begin{subequations}
\label{transf3}
\bea
G(D,\al_1,\al_2,\al_3,\al_4,\al_5) & = & G(D,\al_2,\al_5,\al_4,\frac{3D}{2}-\sum_i \al_i,\al_1),
\label{transf3-1} \\
& = & G(D,\al_4,\al_5,\al_2,\frac{3D}{2}-\sum_i \al_i,\al_3),
\label{transf3-2} \\
& = & G(D,\al_4,\al_3,\al_2,\al_1,\al_5).
\label{transf3-3}
\eea
\end{subequations}
The last transformation in Eq.~(\ref{transf3}) is the same as the one of Eq.~(\ref{transf2}).
Combining the 3 transformations of Eq.~(\ref{transf3}) one generates all the possible transformations of indices
of the 2-loop diagram compatible with the tetrahedral symmetry.

Gorishny and Isaev further noted that the 2-loop diagram is invariant under another transformation which follows
from the uniqueness relation (see Sec.~\ref{sec:meth:uniqueness} for more on uniqueness):
\bea
&&C_D[J(D,p,\al_1,\al_2,\al_3,\al_4,\al_5)] 
\nonum \\
&&= a(\al_2)a(\al_3)a(\al_5)a(D-t_2)\,C_D[J(D,p,\al_1,\tilde{\al}_3,\tilde{\al}_2,\al_4,t_2-D/2)],
\label{transf4}
\eea
where $\tilde{\al} = D/2-\al$ and $t_2=\al_2+\al_3+\al_5$.
The existence of this additional transformation suggests that the symmetry group of the 2-loop diagram is larger than $S_4$.
As a matter of fact, instead of the 3 generators of Eq.~(\ref{S4generators}) we could have chosen one transposition and a $6$-cycle:
\begin{subequations}
\label{S6generators}
\bea
X_D(\al_1,\al_2,\al_3,\al_4,\al_5,\sigma) & = & X_D(\al_2,\al_1,\al_3,\al_4,\al_5,\sigma),
\label{S6generators-1} \\
& = & X_D(\sigma,\al_1,\al_2,\al_3,\al_4,\al_5).
\label{S6generators-2}
\eea
\end{subequations}
The transformations of Eq.~(\ref{S6generators}) generate the group $S_6$. Furthermore, the transformation of Eq.~(\ref{transf4})
involves dual indices $\al \rightarrow D/2-\al$ suggesting that there might be an additional $Z_2$ symmetry.
That the actual group is indeed $Z_2 \times S_6$, which has $2 \times 6!=1,440$ elements, was realized by Broadhurst~\cite{Broadhurst:1986bx} 
soon after the work of Gorishny and Isaev. 
The 3 generators of the $Z_2 \times S_6$ group are: a transposition and a $6$-cycle which generate $S_6$; 
and the dual transformation $\al_i \ra D/2-\al_i$ which generates $Z_2$.
Broadhurst~\cite{Broadhurst:1986bx} and then Barfoot and Broadhurst~\cite{Barfoot:1987kg} not only defined the symmetry 
group but also gave the set of 10 group invariants. 
As we shall review in the next paragraph, this allowed a more
accurate expansion of the massless 2-loop propagator-type diagram. 

\subsection{Brief historical overview of some results}
\label{Sec:hist}

The importance of the 2-loop massless propagator-type diagram of Fig.~\ref{fig:J} comes from the fact that it is a basic building block of many 
multi-loop calculations. As such, it has been extensively studied during the last three decades, see Ref.~\cite{Grozin:2012xi,Bierenbaum:2003ud} for historical reviews.
In this section, we shall review some results obtained for this diagram over the years. 

\begin{figure}
  \begin{center}
  \includegraphics{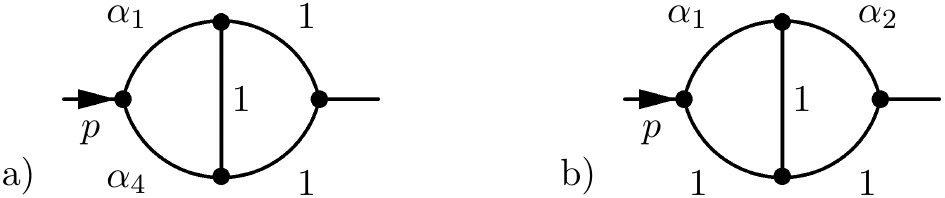}  
  \caption{\label{fig:triangle}
  Some simple two-loop massless propagator diagrams which satisfy the rule of triangle and can be computed exactly using IBP identities.}
  \end{center}
\end{figure}

Generally speaking, when all indices are integers the 2-loop massless propagator-type diagram is easily computed. 
One of the earliest and most well-known result is the one due to Chetyrkin, Kataev and Tkachov \cite{Chetyrkin:1980pr} who found an exact expression
for $G(D,1,1,1,1,1)$ with the help of the Gegenbauer polynomial technique (see App.~\ref{sec:meth:Gegenbauer} for an introduction to the latter).
Soon after, the exact result for $G(D,1,1,1,1,1)$ could be re-obtained in a much more simple and straightforward way
by Vasiliev et al.\ \cite{Vasiliev:1981dg}, Tkachov \cite{Tkachov:1981wb} as well as Chetyrkin and Tkachov \cite{Chetyrkin:1981qh} using integration by parts (see Sec.~\ref{sec:meth:IBP} for more on IBP).
The result reads:
\be
G(D,1,1,1,1,1) =  \frac{2}{D-4}\, \left[ \,\, G(D,1,1) G(D,1,2) - G(D,1,1) G(D,2,3-D/2) \right ]\, ,
\label{G(1,1,1,1,1)}
\ee
where $G(D,\al,\beta)$ is the coefficient function of the one-loop p-type integral, Eq.~(\ref{one-loop-G}).  The fact that
$G(D,1,1,1,1,1)$ reduces to products of $G$-functions implies that this peculiar 2-loop diagram  is actually primitively one-loop.
It can therefore be expressed in terms of $\Gamma$-functions:
\be
G(4-2\veps,1,1,1,1,1) = \frac{\Gamma(1+2\veps)}{\veps^3\,(1-2\veps)}\, \left[  \frac{\Gamma^4(1-\veps)\Gamma^2(1+\veps)}{\Gamma^2(1-2\veps)\Gamma(1+2\veps)}
- \frac{\Gamma^3(1-\veps)}{\Gamma(1-3\veps)} \right ]\, ,
\label{G(1,1,1,1,1)-G4d}
\ee
where the case $D=4-2\veps$ has been considered. From dimensional analysis, $G(4-2\veps,1,1,1,1,1)$ is expected to be UV finite with no $1/\veps$ poles.
This can be checked explicitly by writing it in expanded form and keeping only the first few terms for short:~\footnote{Notice that in Eq.~(\ref{G(1,1,1,1,1)-exp4d}), we have used
a scheme in which $\gamma_E$ and $\zeta_2$ were subtracted from the remaining $\veps$-expansion. There are several other such schemes, \eg, the $G$-scheme \cite{Chetyrkin:1980pr}, see Eq.~(\ref{G-scheme}), where
a factor of $G^l(\veps)$ is extracted from every $l$-loop diagram and may be absorbed in a redefinition of the renormalization scale $\mu$. 
As they resum part of the $\veps$-expansion, these schemes appear to converge faster than the $\overline{\text{MS}}$ scheme.}
%
\bea
G(4-2\veps,1,1,1,1,1)
&=&
\frac{e^{-2\gamma_E \veps - \zeta_2 \veps^2}}{1-2\veps}\,
\biggl(6 \zeta_3 + 9 \zeta_4 \veps + 42 \zeta_5 \veps^2 + (90 \zeta_6 - 46 \zeta_3^2)\,\veps^3
\nonumber \\
&&+ (294 \zeta_7 - 135 \zeta_3\zeta_4)\,\veps^4 + \Ord(\veps^5)
\biggr)\, ,
\label{G(1,1,1,1,1)-exp4d}
\eea
where the expansion formula for Gamma functions has been used:
\be
\Gamma(1+x) = \exp \left(-\gamma_E x + \sum_{n=2}^\infty\,\frac{(-1)^n}{n}\,\zeta_n\,x^n \right)\, ,
\label{gamma-expansion1}
\ee
and $\zeta_n$ is the Riemann zeta function which is defined as:
\be
\zeta_s = \zeta(s) = \sum_{n=1}^{\infty}\frac{1}{n^s} \qquad (\Re{s}>1)\, .
\label{def:zeta}
\ee
In Eq.~(\ref{G(1,1,1,1,1)-exp4d}), the $\Ord(1)$ term reduces to the celebrated $\zeta_3$ and all coefficients of higher order $\veps$ terms can be expressed in terms
of zeta functions.  The authors of \cite{Chetyrkin:1981qh} also noticed that the functions $G(D,\al_1,1,1,\al_4,1)$ as well as
$G(D,\al_1,\al_2,1,1,1)$, where $\al_1$, $\al_2$ and $\al_4$ are arbitrary indices, can also be computed exactly using IBP.
This follows from the so-called {\it rule of triangles} \cite{Chetyrkin:1981qh} whereby the 2-loop diagram can be exactly computed with the help of IBP
whenever three adjacent lines have integer indices, see Fig.~\ref{fig:triangle}.
For completeness, we give their expressions \cite{Chetyrkin:1981qh} (see Sec.~\ref{sec:meth:uniquenessApp} for a derivation of the first identity):
\begin{subequations}
\label{IBP:other-applications}
	\fleq{
&G(D,\al_1,1,1,\al_4,1)  =  \frac{G(D,1,1)}{D-\al_1-\al_4-2}\,\bigg[ \al_1 G(D,1+\al_1,\al_4) - \al_1 G(D,1+\al_1,2+\al_4-D/2) \bigg .
\nonum \\
	&\qquad \bigg. + \al_4 G(D,\al_1,1+\al_4) - \al_4 G(D,2+\al_1-D/2,1+\al_4) \bigg]\, ,
\label{sec:IBP:application1}
\\
&G(D,\al_1,\al_2,1,1,1) =  \frac{2(1+\al_1+\al_2-D/2)}{(D-3)(D-4)}\, \bigg[ \al_1 \, G(D,1,1+\al_1)G(D,1,\al_1+\al_2+2-D/2)
\nonum \\&\qquad
  + \{ \al_2 \leftrightarrow \al_1\}  \bigg]
- \frac{2\al_1 \al_2}{(D-3)(D-4)}\,G(D,1,1+\al_1)G(D,1,1+\al_2) \, .
\label{sec:IBP:application2}
}
\end{subequations}
These functions can be expanded for arbitrary indices $\al_1$, $\al_2$ and $\al_4$ in $D=n-2\veps$ ($n \in \N^*$) with the help of:
\be
\Gamma(x+\veps) = \Gamma(x)\,\exp \Big[ \,\,\sum_{k=1}^{\infty} \psi^{(k-1)}(x) \frac{\veps^k}{k!} \,\,\Big]\, ,
\label{gamma-expansion2}
\ee
where $\psi^{(k)}$ is the polygamma function of order $k$:
\be
\psi(x) = \psi^{(0)}(x) = \frac{\Gamma'(x)}{\Gamma(x)}, \quad  \psi^{(k)}(x) = \frac{\D^k}{\D x^k} \,\psi(x)\, ,
\label{def:polygamma}
\ee
$\psi(x)$ being the digamma function and $\psi'(x) = \psi^{(1)}(x)$ the trigamma function.

On the other hand, for arbitrary (non-integer) values of all the indices, the evaluation of the massless 2-loop p-type integral is highly non-trivial 
(even for the lowest order coefficients of the $\veps$-expansion) and peculiar cases have to be considered, see, {\it e.g.},
Refs.~\cite{Chetyrkin:1980pr}-\cite{Bierenbaum:2003ud}.
In the case where all the indices take the form:\footnote{Indices of this kind appear when considering multi-loop Feynman diagrams with integer indices. 
Upon integrating some of the subdiagrams using, \eg, IBP or another technique, the diagram transforms into a diagram with less loops but having lines where the integer indices are shifted by $\veps$ quantities.} 
$\al_i = n_i + a_i \veps$, where the $n_i$ are positive integers and the $a_i$ are non-negative real numbers,
the diagram is known only in the form of an $\veps$-expansion. For $n_i=1$, arbitrary $a_i$ and $D=4-2\veps$, an expansion to order $\veps^3$ could be 
carried out in the seminal paper of Kazakov using the method of uniqueness \cite{Kazakov:1983ns}, see also Sec.~\ref{sec:meth}. 
Using functional identities among complicated diagrams, 
Kazakov further managed to extend his computation to order $\veps^4$ \cite{Kazakov:1984km,Kazakov:1983pk}. 
Then, using the symmetry arguments outlined in Sec.~\ref{sec:symmetries}, Broadhurst~\cite{Broadhurst:1986bx} and then
Barfoot and Broadhurst~\cite{Barfoot:1987kg} managed to extend the computation to order $\veps^5$ and then $\veps^6$, respectively. 
%
%
%
Subsequently, the orders $\veps^7$ and $\veps^8$ were computed by Broadhurst, Gracey and Kreimer~\cite{Broadhurst:1996ur}. After two decades of calculations, 
an expansion to order $\veps^9$ was achieved in Ref.~\cite{Broadhurst:2002gb}. A this point, number theoretical issues were raised, see, \eg, \cite{Broadhurst:2002gb} and references therein. It was known 
since the early days of quantum field theory that the Riemann zeta function, Eq.~(\ref{def:zeta}), often arises in Feynman diagram computations. More complicated diagrams
depending on an additional scale, such as massive propagator-type Feynman diagrams, were also known to be expressed in terms of the polylogarithm. This
function is a generalization of the zeta function and is defined as:
\be
\mbox{Li}_s(z) = \sum_{n=1}^{\infty} \frac{z^n}{n^s}\, ,
\label{def:polylog}
\ee
where $\mbox{Li}_1(z) = - \log(1-z)$ and $\mbox{Li}_s(1) = \zeta_s$ with $s \geq 2$. 
Other generalizations relevant to multi-loop massless Feynman diagrams include multiple zeta functions, also known as multiple zeta values (MZV) or Euler sums; they are defined as:
\be
\zeta_{s_1,s_2, \cdots, s_l} = \zeta(s_1,s_2, \cdots, s_l) = \sum_{n_1>n_2> \cdots n_l>0} \frac{1}{n_1^{s_1} n_2^{s_2} \cdots n_l^{s_l}} \qquad (s_1>1,~~s_2,\cdots,s_l \in \N)\, ,
\label{def:MZV}
\ee
where the integer $l$ is referred to as the {\it length} of the multiple zeta value and $s = \sum_{i=1}^l s_i$ to its {\it weight}. In general, they reduce
to zeta functions, \eg, $\zeta(2,2) = (3/4)\zeta(4)$, $\zeta(3,2) = 3\zeta(2) \zeta(3) - (11/2) \zeta(5)$, etc... 
In some cases they are irreducible, \eg, at length $2$ and weight $8$, $\zeta(6,2)$ cannot be written in terms of zeta functions.
Multiple zeta functions are themselves a peculiar case of the multiple polylogarithm:
\be
\mbox{Li}_{s_1,s_2, \cdots, s_l}(z_1, z_2, \cdots, z_l) = \sum_{n_1>n_2> \cdots n_l>0} \frac{z_1^{n_1} z_2^{n_2} \cdots z_l^{n_l}}{n_1^{s_1} n_2^{s_2} \cdots n_l^{s_l}} \, ,
\label{def:mpl}
\ee
which appears in multi-scale Feynman diagrams. The important question that is then asked is whether Feynman diagrams (and in particular the coefficients of the $\veps$ expansion) 
can be fully expressed in terms of the zeta functions, Eq.~(\ref{def:zeta}), or generalization of these functions such as the multiple zeta functions and multiple polylogarithms?
In the general case, this question is very difficult to answer. At this point, let's note that in the case of massive propagator-type diagrams, it is well known that, starting from two-loops, 
there are integrals which cannot be expressed in terms of multiple polylogarithms. The simplest examples are given by the two-loop sunrise and kite integrals,
which can be expressed in terms of Elliptic polylogarithms, see the recent review \cite{Adams:2017xsu} and references therein.
Even more complicated integrals containing structures beyond multiple polylogarithms have been considered recently in \cite{Hidding:2017jkk}.

Returning to the massless two-loop p-type integral $J(D,p,\al_1,\al_2,\al_3,\al_4,\al_5)$ of Eq.~(\ref{def:two-loop-p-int}), shown in Fig.~\ref{fig:J}, we would like
  to note that progress appeared in 2003 with the work of Bierenbaum and Weinzierl \cite{Bierenbaum:2003ud}. 
 For $\al_i = n_i + a_i \veps$ and $D=2m-2\veps$ ($m \in \N$), they managed to automate its $\veps$-expansion; in principle, their computer assisted method allows an expansion to arbitrary order in $\veps$ 
the only restrictions arising from hardware constraints. Furthermore, they proved the following important theorem \cite{Bierenbaum:2003ud}:\footnote{Some generalization of the theorem appeared in 
Refs.~\cite{Brown:2008um,Brown:2009ta}.} 

\begin{theorem}[Bierenbaum and Weinzierl (2003)]
Multiple zeta values are sufficient for the Laurent expansion of the two-loop integral, $G(D,\al_1,\al_2,\al_3,\al_4,\al_5)$, with $D=2m-2\veps$ ($m \in \N$) if all powers of the propagators are of the form
$\al_i = n_i + a_i \veps$ where the $n_i$ are positive integers and the $a_i$ are non-negative real numbers.
\end{theorem}

On a more ``phenomenological'' level, a principle of ``maximum weight'' or ``maximal transcendentality''
was discovered in Refs.~\cite{Broadhurst:1996ur}, \cite{Fleischer:1998nb}-\cite{Kotikov:2001sc}
  \footnote{We were informed by David Broadhurst that this principle
appears to be first due to John Gracey in an example of supersymmetric nonlinear sigma model preceding Ref.~\cite{Broadhurst:1996ur}.}
(some reviews can be found in \cite{Kotikov:2010gf}).
  In very simple terms, this property can be checked at the level of the elementary example provided by Eq.~(\ref{G(1,1,1,1,1)-exp4d}). 
For this, let's assume that the transcendentality level of $\zeta_s$ is $s$ and the one of $\veps$ is $-1$. 
Then, we see that all displayed terms of the $\veps$-expansion have transcendentality $3$. Such an observation strongly constrains the coefficients of the series and, when judiciously used, sometimes allows
to reconstruct a whole series from the knowledge of the first few terms, see examples in \cite{Fleischer:1998nb}. 

It turns out that, unfortunately, none these powerful theorems and beautiful observations hold 
in the case of odd-dimensional field theories and related expansions in the vicinity of non-integer indices that we will be concerned with in the following.  
 From the few existing studies, one may anticipate that odd-dimensional theories are ``transcendentally'' more complex that even dimensional ones 
\cite{Broadhurst:1996yc}. As a simple example, let's reconsider Eq.~(\ref{G(1,1,1,1,1)}) in $D=3-2\veps$. In this case:
\begin{flalign}
G(3-2\veps,1,1,1,1,1) = -2\,\frac{\Gamma(1+2\veps)}{1+2\veps}\,\left[  2\veps\,\frac{\Gamma^4(1/2-\veps)\Gamma^2(1/2+\veps)}{\Gamma^2(1-2\veps)\Gamma(1+2\veps)} 
+ \frac{1+6\veps}{\veps\,(1+2\veps)}\, \frac{\Gamma^3(1/2-\veps)}{\Gamma(1/2-3\veps)} \right ]\, ,
\label{G(1,1,1,1,1)-G3d}
\end{flalign}
where the first argument in $G$ emphasizes that we work near $3$ dimensions. From Eq.~(\ref{G(1,1,1,1,1)-G3d}), we first see the appearance of a $1/\veps$-pole. The latter is of IR nature
and arises from the last $G$-function in Eq.~(\ref{G(1,1,1,1,1)}). Moreover, as $G$-functions in $D$-dimensions are expressed in terms of $\Gamma$-functions with arguments depending on $D/2$, 
we see the appearance of half-integer indices in Eq.~(\ref{G(1,1,1,1,1)-G3d}) around which the expansion has to be made. With the help of Eq.~(\ref{gamma-expansion2}), this leads to:
\begin{flalign}
&G(3-2\veps,1,1,1,1,1) = -2\pi\,\frac{e^{-2\gamma_E \veps + \zeta_2 \veps^2}}{1+2\veps}\,
\left[ \frac{1}{\veps} + 4 + \big(2\pi^2 - 9\zeta_2 -8 \big) \veps 
\right .
\nonum \\
&\left . \qquad \qquad \qquad \qquad \qquad \qquad  
+ \frac{4}{3}\,\big( 12 - 44 \zeta_3 -27\zeta_2 + 6\pi^2\,\log 2 \big) \, \veps^2 + \Ord(\veps^3) \right]\, ,
\label{G(1,1,1,1,1)-exp3d}
\end{flalign}
where we have used the fact that $\psi(1/2)=-\gamma_E - 2 \log 2$, $\psi'(1/2) = 3 \zeta_2$ and $\psi''(1/2)=-14\zeta_3$.\footnote{We have: $\psi^{(n)}(1/2)=(-1)^{n+1}\,n!\,(2^{n+1}-1)\,\zeta_{n+1}$ for $n \in \N^*$.}
As anticipated, Eq.~(\ref{G(1,1,1,1,1)-exp3d}) has no property of ``maximal transcendentality''. It also features numbers such as $\pi$ and $\log 2$
which arise from derivatives of the $\Gamma$-function at half-integer argument. In principle, 
according to \cite{Broadhurst:1996yc}, even more complicated  numbers may appear at a higher number of loops, the first of which is:
\be
U(3,1) = \sum_{m>n>0} \frac{(-1)^{m+n}}{m^3 n} = \frac{1}{2}\,\zeta(4) -2 \left \{ \text{Li}_4(1/2) +\frac{1}{24}\,\log^2 2 \, \left( \log^2 2 - \pi^2 \right) \right \} \, ,
\ee
with a non-trivial, \eg, beyond MZV, $\text{Li}_4(1/2)$.

\begin{figure}
  \begin{center}
  \includegraphics{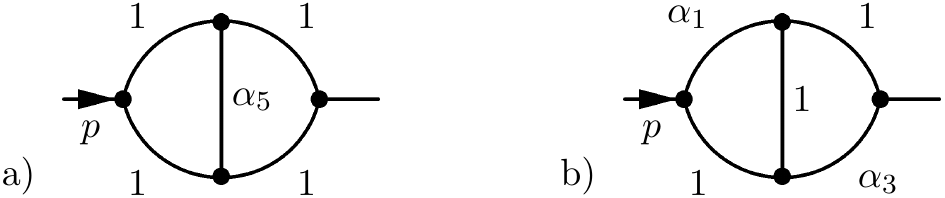}
  \caption{\label{fig:beyondtriangle}
   Examples of two-loop massless p-type diagrams which are beyond IBP identities and uniqueness.}
  \end{center}
\end{figure}

Pursuing with our historical overview, there are some specific cases where an exact evaluation of the diagram can be found. 
One of the simplest non-trivial, {\it i.e.}, beyond the rules of triangle see Fig.~\ref{fig:beyondtriangle},
diagram which may be considered is the one with an arbitrary index on the central line:
\begin{flalign}
G(D,1,1,1,1,\al_5) \quad = \quad
\parbox{16mm}{
    \begin{fmfgraph*}(16,14)
      \fmfleft{ve}
      \fmfright{vo}
      \fmftop{vn}
      \fmftop{vs}
      \fmffreeze
      \fmfforce{(0w,0.5h)}{ve}
      \fmfforce{(1.0w,0.5h)}{vo}
      \fmfforce{(.5w,0.95h)}{vn}
      \fmfforce{(.5w,0.05h)}{vs}
      \fmffreeze
      \fmf{plain,left=0.8}{ve,vo}
	    \fmf{phantom,left=0.7,label=\small{$1$},l.d=-0.01w}{ve,vn}
	    \fmf{phantom,right=0.7,label=\small{$1$},l.d=-0.01w}{vo,vn}
      \fmf{plain,left=0.8}{vo,ve}
	    \fmf{phantom,left=0.7,label=\small{$1$},l.d=-0.01w}{vo,vs}
	    \fmf{phantom,right=0.7,label=\small{$1$},l.d=-0.01w}{ve,vs}
	    \fmf{plain,label=\small{$\al_5$},l.d=0.05w}{vs,vn}
      \fmffreeze
      \fmfdot{ve,vn,vo,vs}
    \end{fmfgraph*}
}\quad \, .
\label{def:I(al)}
\end{flalign}
\vskip 2mm

\ni Early calculations by Vasil'ev, Pis'mak and Khonkonen~\cite{Vasiliev:1981dg} focused on the case where the index $\al_5$ is related to the dimensionality of the system as follows:
$\al_5 = \lambda = D/2 - 1$. Using the method of uniqueness in real space, they have shown that~\cite{Vasiliev:1981dg} (see also discussions in Refs.~\cite{Vasiliev:1992wr,Kivel:1993wq}):
\be
G(D=2\lambda+2,1,1,1,1,\lambda) = 3\,\frac{\Gamma(\lambda)\Gamma(1-\lambda)}{\Gamma(2\lambda)} \,\Big[ \psi'(\lambda) - \psi'(1) \Big] \, ,
\label{res:I(lambda)}
\ee
where $\psi'(x)$ is the trigamma function. Notice that this result has been recently recovered, hopefully in a simpler way, in Ref.~\cite{Kotikov:2013kcl} using the method of uniqueness
in momentum space. For an arbitrary index $\al_5$, the diagram is beyond IBP as well as uniqueness. A one-fold series representation of Eq.~(\ref{def:I(al)}) has first been given by Kazakov in Ref.~\cite{Kazakov:1983pk}.
His expression reads:
%
%
%
\begin{flalign}
&G(2\lambda+2,1,1,1,1,\al_5) =
-2\, \frac{\Gamma^2(\lambda)\Gamma(1-\lambda)\Gamma(\lambda-\al_5) \Gamma(1-2\lambda+\al_5)}{\Gamma(2\lambda)} \Biggl[
\frac{1}{\Gamma(\al_5)\Gamma(3\lambda-\al_5-1)}
\nonum\\
&\qquad \times
\sum_{n=1}^{\infty}\,(-1)^n \frac{\Gamma(n+2\lambda-1)}{\Gamma(n+1-\lambda)}\,\left(\frac{1}{n-\lambda+\al_5} + \frac{1}{n-1+2\lambda-\al_5}\right)
-\cos [\pi \lambda] \Biggr]\, ,
\label{res:I(al):Kazakov}
\end{flalign}
where the one-fold series can be represented as a combination of two ${}_3F_2$-hypergeometric functions of argument $-1$ (see Sec.~\ref{sec:functional} 
for a proof via functional equations).~\footnote{According to \cite{Kalmykov:2008gq}, Regge proposed that any Feynman diagram can be understood
in terms of some hypergeometric functions, see \cite{Kalmykov:2008gq} and references therein for more on the hypergeometric function approach to Feynman 
diagrams.}

\begin{figure}
  \begin{center}
\includegraphics{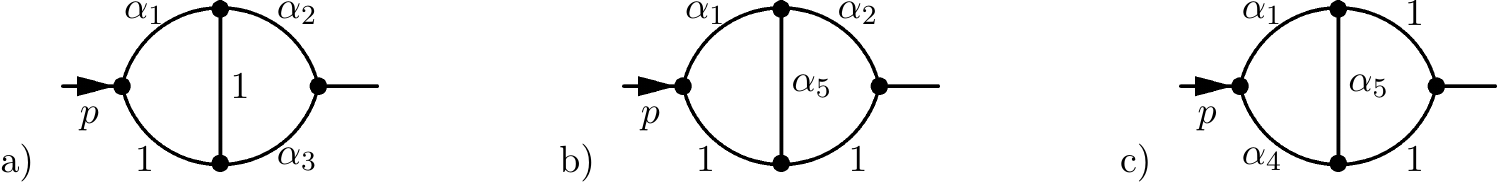}
  \caption{\label{fig:beyondtriangle+}
  Examples of more complicated two-loop massless p-type diagrams.}
  \end{center}
\end{figure}

Later, a whole class of complicated diagrams where two adjacent indices are integers and the three others are arbitrary, see Fig.~\ref{fig:beyondtriangle+},
could be computed exactly by Kotikov~\cite{Kotikov:1995cw} on the basis of a new development of the Gegenbauer polynomial technique.
For this class of diagrams, similar results have been obtained in Ref.~\cite{Broadhurst:1996ur} using an Ansatz to solve the
recurrence relations arising from IBP for the 2-loop diagram.
All these results are expressed in terms of (combinations of) generalized hypergeometric functions, ${}_3F_2$ with argument $1$.
As a matter of fact, from \cite{Kotikov:1995cw}, the diagram of Eq.~(\ref{def:I(al)}) could be expressed as:
\begin{flalign}
&G(2\lambda+2,1,1,1,1,\al_5) =
-2\, \Gamma(\lambda)\Gamma(\lambda-\al_5) \Gamma(1-2\lambda+\al_5) \times
\label{res:I(al):Kotikov}\\
&\qquad \times \left [ \frac{\Gamma(\lambda)}{\Gamma(2\lambda)\Gamma(3\lambda-\al_5-1)}\,
\sum_{n=0}^{\infty}\,\frac{\Gamma(n+2\lambda)\Gamma(n+1)}{n!\,\Gamma(n+1+\al_5)}\,\frac{1}{n+1-\lambda+\al_5}
+\frac{\pi \cot \pi (2\lambda-\al_5)}{\Gamma(2\lambda)} \right ]\, ,
\nonum
\end{flalign}
see Sec.~\ref{sec:meth:Gegenbauer} for more details. 
Notice that the equality of the two representations (\ref{res:I(al):Kazakov}) and (\ref{res:I(al):Kotikov}) was proven only recently \cite{Kotikov:2016rgs}, see also 
this reference for other representations of this diagram. This proof provides the following relation, conjectured in Ref.~\cite{Kotikov:1995cw},  
between two ${}_3F_2$-hypergeometric functions of argument $-1$ and a single  ${}_3F_2$-hypergeometric function of argument $1$:
%
\begin{flalign}
&{}_3F_2(2A,B,1;B+1,2-A;-1)+ \frac{B}{1+A-B} \, {}_3F_2(2A,1+A-B,1;2+A-B,2-A;-1) \nonumber \\
&= B \cdot \frac{\Gamma(2-A)\Gamma(B+A-1)\Gamma(B-A)\Gamma(1+A-B)}{\Gamma(2A)\Gamma(1+B-2A)} - \frac{1-A}{B+A-1} \, {}_3F_2(2A,B,1;B+1,A+B;1) \, ,
\label{id:F32}
\end{flalign}
%
where $A$, $B$ and $C$ are arbitrary. As far as we know, such a relation does not appear in standard textbooks.

On the basis of Ref.~\cite{Kotikov:1995cw}, the case of:
\be
G(D,\al_1,1,\al_3,1,1) \quad = \quad
\parbox{16mm}{
    \begin{fmfgraph*}(16,14)
      \fmfleft{ve}
      \fmfright{vo}
      \fmftop{vn}
      \fmftop{vs}
      \fmffreeze
      \fmfforce{(0w,0.5h)}{ve}
      \fmfforce{(1.0w,0.5h)}{vo}
      \fmfforce{(.5w,0.95h)}{vn}
      \fmfforce{(.5w,0.05h)}{vs}
      \fmffreeze
      \fmf{plain,left=0.8}{ve,vo}
	    \fmf{phantom,left=0.7,label=\small{$\al_1$},l.d=-0.01w}{ve,vn}
	    \fmf{phantom,right=0.7,label=\small{$1$},l.d=-0.01w}{vo,vn}
      \fmf{plain,left=0.8}{vo,ve}
	    \fmf{phantom,left=0.7,label=\small{$\al_3$},l.d=-0.01w}{vo,vs}
	    \fmf{phantom,right=0.7,label=\small{$1$},l.d=-0.01w}{ve,vs}
	    \fmf{plain,label=\small{$1$},l.d=0.05w}{vs,vn}
      \fmffreeze
      \fmfdot{ve,vn,vo,vs}
    \end{fmfgraph*}
}\quad \, ,
\label{def:I(al,beta)}
\ee
\vskip 2mm

\ni could also be computed explicitly \cite{Kotikov:2013eha}. It was shown to be expressed in terms of two generalized hypergeometric functions, ${}_3F_2$ with argument $1$~\cite{Kotikov:2013eha}. 
The result reads \cite{Kotikov:2013eha}:
\be
G(D,\al_1,1,\al_3,1,1) =
\frac{1}{\tilde{\al}_1-1} \frac{1}{1-\tilde{\al}_3} \,
\frac{\Gamma(\tilde{\al}_1)\Gamma(\tilde{\al}_3)
\Gamma(3-\tilde{\al}_1-\tilde{\al}_3)}{\Gamma(\al_1)
\Gamma(\lambda-2+\tilde{\al}_1+\tilde{\al}_3)}
\frac{\Gamma(\lambda)}{\Gamma(2\lambda)} \, I(\tilde{\al}_1,\tilde{\al}_3) \, ,
\label{res:Gab}
\ee
where $\tilde{\al} = D/2-\al$, $\lambda=D/2-1$, $D=4-2\veps$ and the function $I(\tilde{\al}_1,\tilde{\al}_3)$ can be written in four different forms; for example:
\bea
&&I(\tilde{\al}_1,\tilde{\al}_3) =
\frac{\Gamma(1+\lambda-\tilde{\al}_1)}{
\Gamma(3-\tilde{\al}_1-\tilde{\al}_3)}
\frac{\pi \sin[\pi(\tilde{\al}_3- \tilde{\al}_1+\lambda)]}{\sin[\pi (\lambda-1+\tilde{\al}_3)]\sin[\pi \tilde{\al}_1]}
+ \sum_{n=0}^{\infty} \frac{\Gamma(n+2\lambda)}{n!(n+\lambda+\tilde{\al}_1-1)}
\label{res:Gab:I1} \\
&& \times \biggl( \frac{\Gamma(n+1)}{\Gamma(n+2+\lambda-\tilde{\al}_3)}
- \frac{\Gamma(n-2+\lambda +\tilde{\al}_1+\tilde{\al}_3)\Gamma(2-\tilde{\al}_3) \Gamma(\lambda)
}{\Gamma(n-1+2\lambda+\tilde{\al}_1)\Gamma(3-\tilde{\al}_1-\tilde{\al}_3)\Gamma(\lambda+\tilde{\al}_1-1)}
\frac{\sin[\pi(\tilde{\al}_3+\lambda-1)]}{\sin[\pi \tilde{\al}_1]}
\biggl)
\, ,
\nonum
\eea
\bea
&&I(\tilde{\al}_1,\tilde{\al}_3) =
\frac{\Gamma(1+\lambda - \tilde{\al}_1)}{
\Gamma(3-\tilde{\al}_1-\tilde{\al}_3)}
\frac{\pi \sin[\pi \tilde{\al}_1]}{\sin[\pi (\lambda-1+\tilde{\al}_3)] \sin[\pi (\tilde{\al}_1+\tilde{\al}_3+\lambda-1) ]}
\label{res:Gab:I2} \\&&
+ \sum_{n=0}^{\infty} \frac{\Gamma(n+2\lambda)}{n!
}
\biggl(\frac{1}{n+\lambda+\tilde{\al}_1-1}
\frac{\Gamma(n+1)}{\Gamma(n+2+\lambda-\tilde{\al}_3)}
\nonumber \\&&
+ \frac{1}{n+\lambda+1-\tilde{\al}_1}
\frac{\Gamma(n+2-\tilde{\al}_1)\Gamma(2-\tilde{\al}_3)\Gamma(\lambda)
}{\Gamma(n+3+\lambda-\tilde{\al}_1-\tilde{\al}_3)\Gamma(3-\tilde{\al}_1-\tilde{\al}_3)\Gamma(\lambda+\tilde{\al}_1-1)}
\frac{\sin[\pi(\tilde{\al}_3+\lambda-1)]}{\sin[\pi (\tilde{\al}_1+\tilde{\al}_3+\lambda-1)]}
\biggl)
\, ,
\nonum
\eea
see \cite{Kotikov:2013eha} for other representations in terms of $\psi$-functions. Of course, Eq.~(\ref{res:Gab}) allows to recover all previously known cases
for integer indices. 

As an application, we may consider the peculiar case where $\al_1=\veps$ and $\al_3=2\veps$ which appears in the computation of the renormalization group function of the $\Phi^4$-model at 
6 loops \cite{Kompaniets:2016hct,Kompaniets:2017yct}. Using Eqs.~(\ref{res:Gab}) and (\ref{res:Gab:I2}), the first terms of the expansion are easily found and read:
%
%
%
\begin{flalign}
&G(4-2\veps,\veps,1,2\veps,1,1) = G^2(\veps) \, \bigg [  - \frac{7}{30 \veps} - \frac{9}{5} - \frac{367}{30}\,\veps + \left( \frac{239}{15} \zeta_3 - \frac{1187}{15} \right)\,\veps^2 \bigg .
\nonum \\
&\qquad \qquad \qquad \qquad \qquad \qquad \qquad \bigg .  + \left( \frac{239}{10} \zeta_4 + \frac{576}{5}\zeta_3 - \frac{15031}{30} \right)\, \veps^3 + \Ord(\veps^4) \bigg ] \, ,
\label{exp:G1e2e}
\end{flalign}
where the so-called $G$-scheme \cite{Chetyrkin:1980pr} has been used where:
\be
G(\veps) = \veps\,G(1,1) = \frac{\Gamma^2(1-\veps)\Gamma(1+\veps)}{\Gamma(2-2\veps)}\, .
\label{G-scheme}
\ee
In numerical form, the terms in the brackets of Eq.~(\ref{exp:G1e2e}) read:
\begin{flalign}
&G(4-2\veps,\veps,1,2\veps,1,1) = G^2(\veps) \,\bigg [- \frac{0.23333333333333333333}{\veps} - 1.8 - 12.2333333333333333336\,\veps \bigg .
\nonum \\
&\qquad \bigg . -59.98056000965713105\,\veps^2 -336.68885280365186888\, \veps^3 + \Ord(\veps^4)  \bigg ]\, ,
\label{exp:G1e2e:num}
\end{flalign}
and are in good agreement with estimates from the sector decomposition technique
:~\footnote{Unpublished result from M.~Kompaniets reproduced with his kind permission.}
%
\begin{flalign}
&G(4-2\veps,\veps,1,2\veps,1,1) = G^2(\veps) \, \bigg [-\frac{0.23333324(12)}{\veps}  -1.7999970(31) -12.233338(24) \,\veps  \bigg .
\nonum \\ 
&\qquad \bigg . - 59.98056(12) \,\veps^2 - 336.6893(7) \, \veps^3 + \Ord(\veps^4)  \bigg ]\, . 
\end{flalign}

\begin{figure}
  \begin{center}
  \includegraphics{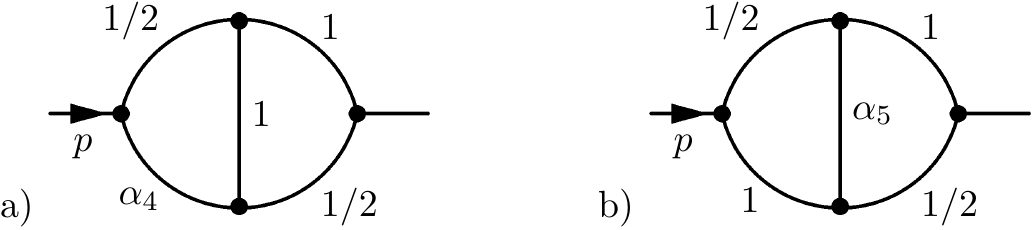}
  \caption{\label{fig:morecomplicateddiags} Examples of complicated diagrams appearing in Refs.~\cite{Kotikov:2016wrb,Kotikov:2016prf}.}
  \end{center}
\end{figure}

The more complicated two-loop massless p-type diagrams of Fig.~\ref{fig:beyondtriangle+} are also in principle computable with the technique of Ref.~\cite{Kotikov:1995cw}.\footnote{We were
informed by M.~Kompaniets that, for even space dimensions, it is also possible to compute these integrals with the help of HyperInt \cite{Panzer:2014caa} using the technique developed in \cite{Kompaniets:2016hct}.}
The explicit computation has not been carried out yet but it is expected that they will also be expressed in terms of a linear combination of generalized hypergeometric functions, ${}_3F_2$ with argument $1$.
In some cases, one may expect that simpler expressions may be obtained, \eg, in the case of the diagrams of Fig.~\ref{fig:morecomplicateddiags} which appear in the study of three-dimensional QED \cite{Kotikov:2016wrb,Kotikov:2016prf}. Though this task has not been carried out explicitly yet, it has been shown in \cite{Kotikov:2016wrb} that the Gegenbauer polynomial 
technique, see Sec.~\ref{sec:meth:Gegenbauer}, provides a series representation for these diagrams which is a convenient starting point to compute them numerically for some specific values of $\al_4$ and $\al_5$.

\end{fmffile}

\begin{fmffile}{fmf-review2}

\section{Methods of calculations}
\label{sec:meth}

In this section we provide an overview of some useful methods to compute Feynman diagrams. Among these methods, some of them (parametric integration, Gegenbauer polynomial technique) involve explicit integrations. 
Other methods are {\it algebraic} and involve identities between different diagrams which are conveniently expressed in graphical form. 
These identities transform a diagram into another one (with different indices) and sometimes allow its exact computation without performing any explicit integration 
(assuming we know the one-loop $G$-functions). We shall refer to them as the {\it standard rules of perturbation theory for massless Feynman diagrams} \cite{Kazakov:1984km}.

\subsection{Parametric integration}
\label{sec:meth:param-int}

This is probably one of the oldest techniques known in Feynman diagram calculation, see, \eg, the textbooks \cite{grozin2007lectures,smirnov2013analytic}. 
 It is based on so-called {\it Schwinger-trick} (see below) and  amounts to represent a diagram, originally expressed in position or momentum space, in the space of {\it Feynman parameters} (or parametric space).
The method is very useful in the case of massive Feynman diagrams. 
Many recent developments, even for massless multi-loop diagrams, are based on this technique, see, \eg, 
	\cite{Brown:2008um,Brown:2009ta}, \cite{Binoth:2000ps}-\cite{Golz:2015rea},\cite{Kompaniets:2016hct,Kompaniets:2017yct}.

The Schwinger-trick is based on the integral representation of the $\Gamma$-function:
\be
\frac{1}{A_j^{\al_j}} = \frac{1}{\Gamma(\al_j)}\,\int_0^{+\infty}\D t\, e^{-A_j t}\,t^{\al_j-1}\, ,
\label{al-parametriz}
\ee
where $A_j=k_j^2+m_j^2$ and $\al_j$ is the index of the propagator. It immediately allows to compute the massive tadpole diagram, Eq.~(\ref{massive-one-loop-tadpole}).
It's generalization to the product of an arbitrary number of propagators with arbitrary exponents can be written as:
\be
\frac{1}{A_1^{\al_1} \cdots A_N^{\al_N}} = \frac{\Gamma(\al)}{\Gamma(\al_1) \cdots \Gamma(\al_N)} \, 
\int_{0}^1 \D u_1 \cdots \int_{0}^1 \D u_N \frac{\delta(1- \sum_{j=1}^N u_j) u_1^{\al_1-1} \cdots u_N^{\al_N-1}}{(u_1 A_1 + \cdots + u_N A_N)^{\al}}\, ,
\label{feyn-param}
\ee
where $\al=\sum_{j=1}^N \al_j$ and the $u_i$ are the so-called Feynman parameters.  The one-loop p-type integral can be computed straightforwardly using Eq.~(\ref{feyn-param}) yielding Eq.~(\ref{one-loop-G}).

Let's consider an $L$-loop diagram with external momenta collected in the vector $\vec p$ and $N$ internal propagators whose indices are collected in the vector $\vec \al$. In momentum space, this diagram reads:
\be
J_L(D,\vec p,\vec \al\,) = \int \frac{[\D^D k_1] \cdots [\D^D k_L]}{A_1^{\al_1} \cdots A_N^{\al_N}}\, .
\label{JL}
\ee
Using Eq.~(\ref{feyn-param}), Eq.~(\ref{JL}) can then be represented in parametric space under the general form:
\be
J_L(D,\vec p, \vec \al\,) = (4\pi)^{-LD}\,\frac{\Gamma(\al)}{\Gamma(\al_1) \cdots \Gamma(\al_N)}\,\int_0^1 \D^N u \, \delta \left(1- \sum_{j=1}^N u_j \right)\,\Pi_{j=1}^N u_j^{\al_j-1} \, 
\frac{\mathcal{U}^{\al - (L+1)D/2}}{\mathcal{F}(\vec{p})^{\al - LD/2}}\, ,
\label{expr:parametric}
\ee
where $\mathcal{U}$ and $\mathcal{F}$ are polynomials in the Feynman parameters and $\mathcal{F}$ also depends on the external momenta and masses.
For an arbitrary Feynman diagram, the Symansik polynomials $\mathcal{U}$ and $\mathcal{F}$ are not so easy to compute starting from the momentum representation.
Some more efficient derivations are based on the topology of the diagram under consideration, see, \eg, \cite{Bogner:2007cr}. Let's mention that in the case of the massless
2-loop p-type integral, these polynomials read:
\begin{subequations}
\label{symanzikU+F}
\bea
&&\mathcal{U} = (u_1 + u_2 + u_3 + u_4) u_5 + (u_1+u_4)(u_2+u_3)\, ,
\label{symanzikU}\\
&&\mathcal{F} = \bigg( (u_1+u_4)u_2u_3 +(u_2+u_3)u_1u_4 + (u_1+u_2)(u_3+u_4)u_5 \bigg) p^2\, . 
\label{symanzikF}
\eea
\end{subequations}

\subsection{Integration by parts}
\label{sec:meth:IBP}

Thanks to it's simplicity and efficiency, integration by parts (IBP) is one of the most widely used methods in multi-loop calculations. It has been introduced by Vasil'ev, Pis'mak and Khonkonen~\cite{Vasiliev:1981dg},
Tkachov \cite{Tkachov:1981wb} and Chetyrkin and Tkachov~\cite{Chetyrkin:1981qh}. It allows to reduce a complicated Feynman diagram in terms of a limited number of master integrals;
such reduction is now automated via it's implementation in computer programs with the help of various algorithms, see, \eg,
\cite{Laporta:2001dd}-\cite{Lee:2013mka}.
As reviewed in Sec.~\ref{Sec:hist}, in some simple cases, the master integrals themselves can be computed from IBP alone. In general, however, other methods have to be used often in combination with IBP (see next items).

\subsubsection{Presentation of the method}

IBP recurrence relations in momentum-space are essentially based on the translational invariance of dimensionally regularized integrals:
\be
\int\,\D^D k\,f(k) = \int\,\D^D k\,f(k+q) \quad \Ra \quad 0 = \int\,\D^D k\,\frac{\partial}{\partial k^\mu}\,f(k) \, .
\ee
In the following, we shall mainly be concerned with the application of the IBP procedure to the 2-loop massless p-type diagram of Eq.~(\ref{def:two-loop-p-int}).~\footnote{The original references 
\cite{Tkachov:1981wb,Chetyrkin:1981qh} were actually focusing on IBP relations for 3-loop massless p-type diagrams.}
The IBP relations for this diagram then follow from:
\be
0 = (\partial_{\mathcal{C}} \cdot P)\,J(D,p,\al_1,\al_2,\al_3,\al_4,\al_5)\, ,
\label{sec:meth:IBP:general}
\ee
where $\mathcal{C}$ is a closed oriented contour and $P$ some momentum. In the above identity it is understood that differentiation goes before integration.
Let's define $p_i$ as the momentum carried by the line of index $\al_i$:
\be
p_1=k-p,\quad p_2=q-p,\quad p_3=q,\quad p_4=k,\quad p_5=k-q\, ,
\label{sec:meth:IBP:parametrization}
\ee
where $p$ is the external momentum. Different IBP relations come from different choices for the contour $\mathcal{C}$ and the momentum $P$ (in the following, 
the line carrying momentum $P$ will be referred to as the {\it distinguished line}). According to Chetyrkin and  Tkachov \cite{Chetyrkin:1981qh}, 
for an $L$-loop p-integral one can write down $(L+1)^2$ independent IBP identities. This comes from the fact that there are $L+1$ possible choices for $\mathcal{C}$ ($L$ internal and $1$ external)
and a similar number for $P$ ($L$ loop momenta and one external momentum). So, for the 2-loop massless p-type diagram there are {\it a priori} $9$ IBP relations.

Let's first consider the case corresponding to $P=p_5$ and $\mathcal{C}=\{+p_1,\,+p_4,\,+p_5\}$. Along this contour the derivative reads:
\be
\partial_{\mathcal{C}} = + \frac{\partial}{\partial p_1} + \frac{\partial}{\partial p_4} + \frac{\partial}{\partial p_5}\, ,
\ee
where the sign before $p_i$ or $\partial / \partial p_i$ is plus if $p_i$ flows in the direction of $\mathcal{C}$ and minus otherwise.
With the help of Eq.~(\ref{sec:meth:IBP:general}), this leads to:
\be
\int \int \D^D k \, \D^D q\, \frac{\partial}{\partial k^\mu} \left \{ \frac{1}{(k-p)^{2\al_1}\,(q-p)^{2\al_2}\,q^{2\al_3}\,k^{2\al_4}}
\,\frac{(k-q)^\mu}{(k-q)^{2\al_5}} \right \} = 0 \, .
\label{sec:IBP:example}
\ee
Using equalities of the type:
\bea
\frac{\partial k^\nu}{\partial k^\mu} = \delta_\mu^\nu \quad (\delta_\mu^\mu = D) \, ,
\qquad
\frac{\partial (k-p)^{-2\al}}{\partial k^\mu} = - 2\al \,\frac{(k-p)_\mu}{(k-p)^{2\al+2}} \, ,
\label{sec:meth:IBP:equalities}
\eea
and canceling squared combinations of momentum in the numerator and denominator yields, in graphical form:
%
\begin{flalign}
(D-\al_1-\al_4-2\al_5) \quad
\parbox{16mm}{
    \begin{fmfgraph*}(16,14)
      \fmfleft{i}
      \fmfright{o}
      \fmfleft{ve}
      \fmfright{vo}
      \fmftop{vn}
      \fmftop{vs}
      \fmffreeze
      \fmfforce{(-0.1w,0.5h)}{i}
      \fmfforce{(1.1w,0.5h)}{o}
      \fmfforce{(0w,0.5h)}{ve}
      \fmfforce{(1.0w,0.5h)}{vo}
      \fmfforce{(.5w,0.95h)}{vn}
      \fmfforce{(.5w,0.05h)}{vs}
      \fmffreeze
      \fmf{plain}{i,ve}
      \fmf{plain,left=0.8}{ve,vo}
	    \fmf{phantom,left=0.5,label=\footnotesize{$\al_1$},l.d=-0.01w}{ve,vn}
	    \fmf{phantom,right=0.5,label=\footnotesize{$\al_2$},l.d=-0.01w}{vo,vn}
      \fmf{plain,left=0.8}{vo,ve}
	    \fmf{phantom,left=0.5,label=\footnotesize{$\al_3$},l.d=-0.01w}{vo,vs}
	    \fmf{phantom,right=0.5,label=\footnotesize{$\al_4$},l.d=-0.01w}{ve,vs}
	    \fmf{plain,label=\footnotesize{$\al_5$},l.d=0.05w}{vs,vn}
      \fmf{plain}{vo,o}
      \fmffreeze
      \fmfdot{ve,vn,vo,vs}
    \end{fmfgraph*}
}
\quad & =  \,\, \al_1 \left[\quad
\parbox{16mm}{
    \begin{fmfgraph*}(16,14)
      \fmfleft{ve}
      \fmfright{vo}
      \fmftop{vn}
      \fmftop{vs}
      \fmffreeze
      \fmfforce{(0w,0.5h)}{ve}
      \fmfforce{(1.0w,0.5h)}{vo}
      \fmfforce{(.5w,0.95h)}{vn}
      \fmfforce{(.5w,0.05h)}{vs}
      \fmffreeze
      \fmf{plain,left=0.8}{ve,vo}
	    \fmf{phantom,left=0.5,label=\footnotesize{$+$},l.d=-0.01w}{ve,vn}
      \fmf{phantom,right=0.4}{vo,vn}
      \fmf{plain,left=0.8}{vo,ve}
      \fmf{phantom,left=0.5}{vo,vs}
      \fmf{phantom,right=0.5}{ve,vs}
	    \fmf{plain,label=\footnotesize{$-$},l.d=0.05w}{vs,vn}
      \fmffreeze
      \fmfdot{ve,vn,vo,vs}
    \end{fmfgraph*}
} \qquad - \qquad
\parbox{16mm}{
    \begin{fmfgraph*}(16,14)
      \fmfleft{ve}
      \fmfright{vo}
      \fmftop{vn}
      \fmftop{vs}
      \fmffreeze
      \fmfforce{(0w,0.5h)}{ve}
      \fmfforce{(1.0w,0.5h)}{vo}
      \fmfforce{(.5w,0.95h)}{vn}
      \fmfforce{(.5w,0.05h)}{vs}
      \fmffreeze
      \fmf{plain,left=0.8}{ve,vo}
	    \fmf{phantom,left=0.4,label=\footnotesize{$+$},l.d=-0.01w}{ve,vn}
	    \fmf{phantom,right=0.4,label=\footnotesize{$-$},l.d=-0.01w}{vo,vn}
      \fmf{plain,left=0.8}{vo,ve}
      \fmf{phantom,left=0.5}{vo,vs}
      \fmf{phantom,right=0.5}{ve,vs}
      \fmf{plain}{vs,vn}
      \fmffreeze
      \fmfdot{ve,vn,vo,vs}
    \end{fmfgraph*}
} \quad
\right] \quad +
\nonum
\\ &\qquad \nonum
\\ & +  \,\,
\al_4 \left[\quad
\parbox{16mm}{
    \begin{fmfgraph*}(16,14)
      \fmfleft{ve}
      \fmfright{vo}
      \fmftop{vn}
      \fmftop{vs}
      \fmffreeze
      \fmfforce{(0w,0.5h)}{ve}
      \fmfforce{(1.0w,0.5h)}{vo}
      \fmfforce{(.5w,0.95h)}{vn}
      \fmfforce{(.5w,0.05h)}{vs}
      \fmffreeze
      \fmf{plain,left=0.8}{ve,vo}
      \fmf{phantom,left=0.5}{ve,vn}
      \fmf{phantom,right=0.5}{vo,vn}
      \fmf{plain,left=0.8}{vo,ve}
      \fmf{phantom,left=0.4}{vo,vs}
	    \fmf{phantom,right=0.5,label=\footnotesize{$+$},l.d=-0.01w}{ve,vs}
	    \fmf{plain,label=\footnotesize{$-$},l.d=0.05w}{vs,vn}
      \fmffreeze
      \fmfdot{ve,vn,vo,vs}
    \end{fmfgraph*}
} \qquad - \qquad
\parbox{16mm}{
    \begin{fmfgraph*}(16,14)
      \fmfleft{ve}
      \fmfright{vo}
      \fmftop{vn}
      \fmftop{vs}
      \fmffreeze
      \fmfforce{(0w,0.5h)}{ve}
      \fmfforce{(1.0w,0.5h)}{vo}
      \fmfforce{(.5w,0.95h)}{vn}
      \fmfforce{(.5w,0.05h)}{vs}
      \fmffreeze
      \fmf{plain,left=0.8}{ve,vo}
      \fmf{phantom,left=0.5}{ve,vn}
      \fmf{phantom,right=0.5}{vo,vn}
      \fmf{plain,left=0.8}{vo,ve}
	    \fmf{phantom,left=0.4,label=\footnotesize{$-$},l.d=-0.01w}{vo,vs}
	    \fmf{phantom,right=0.4,label=\footnotesize{$+$},l.d=-0.01w}{ve,vs}
      \fmf{plain}{vs,vn}
      \fmf{plain}{vs,vn}
      \fmffreeze
      \fmfdot{ve,vn,vo,vs}
    \end{fmfgraph*}
} \quad
\right]\, ,
\label{def:IBP-5-left}
\end{flalign}
where $\pm$ on the right-hand side of the equation denotes the increase or decrease of a line index by $1$ with respect to its value on the left-hand side.
Similarly, for $P=p_5$ but a contour running along the right triangle, the following identity is obtained:
\begin{flalign}
(D-\al_2-\al_3-2\al_5) \quad
\parbox{16mm}{
    \begin{fmfgraph*}(16,14)
      \fmfleft{i}
      \fmfright{o}
      \fmfleft{ve}
      \fmfright{vo}
      \fmftop{vn}
      \fmftop{vs}
      \fmffreeze
      \fmfforce{(-0.1w,0.5h)}{i}
      \fmfforce{(1.1w,0.5h)}{o}
      \fmfforce{(0w,0.5h)}{ve}
      \fmfforce{(1.0w,0.5h)}{vo}
      \fmfforce{(.5w,0.95h)}{vn}
      \fmfforce{(.5w,0.05h)}{vs}
      \fmffreeze
      \fmf{plain}{i,ve}
      \fmf{plain,left=0.8}{ve,vo}
	    \fmf{phantom,left=0.5,label=\footnotesize{$\al_1$},l.d=-0.01w}{ve,vn}
	    \fmf{phantom,right=0.5,label=\footnotesize{$\al_2$},l.d=-0.01w}{vo,vn}
      \fmf{plain,left=0.8}{vo,ve}
	    \fmf{phantom,left=0.5,label=\footnotesize{$\al_3$},l.d=-0.01w}{vo,vs}
	    \fmf{phantom,right=0.5,label=\footnotesize{$\al_4$},l.d=-0.01w}{ve,vs}
	    \fmf{plain,label=\footnotesize{$\al_5$},l.d=0.05w}{vs,vn}
      \fmf{plain}{vo,o}
      \fmffreeze
      \fmfdot{ve,vn,vo,vs}
    \end{fmfgraph*}
}
\quad & =  \,\, \al_2 \left[\quad
\parbox{16mm}{
    \begin{fmfgraph*}(16,14)
      \fmfleft{ve}
      \fmfright{vo}
      \fmftop{vn}
      \fmftop{vs}
      \fmffreeze
      \fmfforce{(0w,0.5h)}{ve}
      \fmfforce{(1.0w,0.5h)}{vo}
      \fmfforce{(.5w,0.95h)}{vn}
      \fmfforce{(.5w,0.05h)}{vs}
      \fmffreeze
      \fmf{plain,left=0.8}{ve,vo}
      \fmf{phantom,left=0.5}{ve,vn}
	    \fmf{phantom,right=0.4,label=\footnotesize{$+$},l.d=-0.01w}{vo,vn}
      \fmf{plain,left=0.8}{vo,ve}
      \fmf{phantom,left=0.5}{vo,vs}
      \fmf{phantom,right=0.5}{ve,vs}
	    \fmf{plain,label=\footnotesize{$-$},l.d=0.05w}{vs,vn}
      \fmffreeze
      \fmfdot{ve,vn,vo,vs}
    \end{fmfgraph*}
} \qquad - \qquad
\parbox{16mm}{
    \begin{fmfgraph*}(16,14)
      \fmfleft{ve}
      \fmfright{vo}
      \fmftop{vn}
      \fmftop{vs}
      \fmffreeze
      \fmfforce{(0w,0.5h)}{ve}
      \fmfforce{(1.0w,0.5h)}{vo}
      \fmfforce{(.5w,0.95h)}{vn}
      \fmfforce{(.5w,0.05h)}{vs}
      \fmffreeze
      \fmf{plain,left=0.8}{ve,vo}
	    \fmf{phantom,left=0.4,label=\footnotesize{$-$},l.d=-0.01w}{ve,vn}
	    \fmf{phantom,right=0.4,label=\footnotesize{$+$},l.d=-0.01w}{vo,vn}
      \fmf{plain,left=0.8}{vo,ve}
      \fmf{phantom,left=0.5}{vo,vs}
      \fmf{phantom,right=0.5}{ve,vs}
      \fmf{plain}{vs,vn}
      \fmffreeze
      \fmfdot{ve,vn,vo,vs}
    \end{fmfgraph*}
} \quad
\right] \quad +
\nonum
\\ &\qquad \nonum
\\ & +  \,\,
\al_3 \left[\quad
\parbox{16mm}{
    \begin{fmfgraph*}(16,14)
      \fmfleft{ve}
      \fmfright{vo}
      \fmftop{vn}
      \fmftop{vs}
      \fmffreeze
      \fmfforce{(0w,0.5h)}{ve}
      \fmfforce{(1.0w,0.5h)}{vo}
      \fmfforce{(.5w,0.95h)}{vn}
      \fmfforce{(.5w,0.05h)}{vs}
      \fmffreeze
      \fmf{plain,left=0.8}{ve,vo}
      \fmf{phantom,left=0.5}{ve,vn}
      \fmf{phantom,right=0.5}{vo,vn}
      \fmf{plain,left=0.8}{vo,ve}
	    \fmf{phantom,left=0.4,label=\footnotesize{$+$},l.d=-0.01w}{vo,vs}
      \fmf{phantom,right=0.5}{ve,vs}
	    \fmf{plain,label=\footnotesize{$-$},l.d=0.05w}{vs,vn}
      \fmffreeze
      \fmfdot{ve,vn,vo,vs}
    \end{fmfgraph*}
} \qquad - \qquad
\parbox{16mm}{
    \begin{fmfgraph*}(16,14)
      \fmfleft{ve}
      \fmfright{vo}
      \fmftop{vn}
      \fmftop{vs}
      \fmffreeze
      \fmfforce{(0w,0.5h)}{ve}
      \fmfforce{(1.0w,0.5h)}{vo}
      \fmfforce{(.5w,0.95h)}{vn}
      \fmfforce{(.5w,0.05h)}{vs}
      \fmffreeze
      \fmf{plain,left=0.8}{ve,vo}
      \fmf{phantom,left=0.5}{ve,vn}
      \fmf{phantom,right=0.5}{vo,vn}
      \fmf{plain,left=0.8}{vo,ve}
	    \fmf{phantom,left=0.4,label=\footnotesize{$+$},l.d=-0.01w}{vo,vs}
	    \fmf{phantom,right=0.4,label=\footnotesize{$-$},l.d=-0.01w}{ve,vs}
      \fmf{plain}{vs,vn}
      \fmf{plain}{vs,vn}
      \fmffreeze
      \fmfdot{ve,vn,vo,vs}
    \end{fmfgraph*}
} \quad
\right]\, .
\label{def:IBP-5-right}
\end{flalign}
Actually, Eqs.~(\ref{def:IBP-5-right}) and Eq.~(\ref{def:IBP-5-left}) are related to each other by using the symmetries of the diagram.
As can be seen from Eqs.~(\ref{def:IBP-5-left}) and (\ref{def:IBP-5-right}), the distinguished line in the above examples is the vertical one.

As another example, we may take $P=p_4$ (\ie, the distinguished line now becomes the one with index $\al_4$)
and a contour running along the left triangle; this yields:
\begin{flalign}
(D-\al_1-\al_5-2\al_4) \quad
\parbox{16mm}{
    \begin{fmfgraph*}(16,14)
      \fmfleft{ve}
      \fmfright{vo}
      \fmftop{vn}
      \fmftop{vs}
      \fmffreeze
      \fmfforce{(0w,0.5h)}{ve}
      \fmfforce{(1.0w,0.5h)}{vo}
      \fmfforce{(.5w,0.95h)}{vn}
      \fmfforce{(.5w,0.05h)}{vs}
      \fmffreeze
      \fmf{plain,left=0.8}{ve,vo}
	    \fmf{phantom,left=0.5,label=\footnotesize{$\al_1$},l.d=-0.01w}{ve,vn}
	    \fmf{phantom,right=0.5,label=\footnotesize{$\al_2$},l.d=-0.01w}{vo,vn}
      \fmf{plain,left=0.8}{vo,ve}
	    \fmf{phantom,left=0.5,label=\footnotesize{$\al_3$},l.d=-0.01w}{vo,vs}
	    \fmf{phantom,right=0.5,label=\footnotesize{$\al_4$},l.d=-0.01w}{ve,vs}
	    \fmf{plain,label=\footnotesize{$\al_5$},l.d=0.05w}{vs,vn}
      \fmffreeze
      \fmfdot{ve,vn,vo,vs}
    \end{fmfgraph*}
}
\quad & =  \,\, \al_1 \left[\quad
\parbox{16mm}{
    \begin{fmfgraph*}(16,14)
      \fmfleft{ve}
      \fmfright{vo}
      \fmftop{vn}
      \fmftop{vs}
      \fmffreeze
      \fmfforce{(0w,0.5h)}{ve}
      \fmfforce{(1.0w,0.5h)}{vo}
      \fmfforce{(.5w,0.95h)}{vn}
      \fmfforce{(.5w,0.05h)}{vs}
      \fmffreeze
      \fmf{plain,left=0.8}{ve,vo}
	    \fmf{phantom,left=0.5,label=\footnotesize{$+$},l.d=-0.01w}{ve,vn}
      \fmf{phantom,right=0.4}{vo,vn}
      \fmf{plain,left=0.8}{vo,ve}
      \fmf{phantom,left=0.5}{vo,vs}
	    \fmf{phantom,right=0.5,label=\footnotesize{$-$},l.d=-0.01w}{ve,vs}
      \fmf{plain}{vs,vn}
      \fmffreeze
      \fmfdot{ve,vn,vo,vs}
    \end{fmfgraph*}
} \qquad - \quad (p^2)\times ~
\parbox{16mm}{
    \begin{fmfgraph*}(16,14)
      \fmfleft{ve}
      \fmfright{vo}
      \fmftop{vn}
      \fmftop{vs}
      \fmffreeze
      \fmfforce{(0w,0.5h)}{ve}
      \fmfforce{(1.0w,0.5h)}{vo}
      \fmfforce{(.5w,0.95h)}{vn}
      \fmfforce{(.5w,0.05h)}{vs}
      \fmffreeze
      \fmf{plain,left=0.8}{ve,vo}
	    \fmf{phantom,left=0.4,label=\footnotesize{$+$},l.d=-0.01w}{ve,vn}
      \fmf{phantom,right=0.4}{vo,vn}
      \fmf{plain,left=0.8}{vo,ve}
      \fmf{phantom,left=0.5}{vo,vs}
      \fmf{phantom,right=0.5}{ve,vs}
      \fmf{plain}{vs,vn}
      \fmffreeze
      \fmfdot{ve,vn,vo,vs}
    \end{fmfgraph*}
} \quad
\right] \quad +
\nonum
\\ &\qquad \nonum
\\ & +  \,\,
\al_5 \left[\quad
\parbox{16mm}{
    \begin{fmfgraph*}(16,14)
      \fmfleft{ve}
      \fmfright{vo}
      \fmftop{vn}
      \fmftop{vs}
      \fmffreeze
      \fmfforce{(0w,0.5h)}{ve}
      \fmfforce{(1.0w,0.5h)}{vo}
      \fmfforce{(.5w,0.95h)}{vn}
      \fmfforce{(.5w,0.05h)}{vs}
      \fmffreeze
      \fmf{plain,left=0.8}{ve,vo}
      \fmf{phantom,left=0.5}{ve,vn}
      \fmf{phantom,right=0.5}{vo,vn}
      \fmf{plain,left=0.8}{vo,ve}
      \fmf{phantom,left=0.4}{vo,vs}
	    \fmf{phantom,right=0.5,label=\footnotesize{$-$},l.d=-0.01w}{ve,vs}
	    \fmf{plain,label=\footnotesize{$+$},l.d=0.05w}{vs,vn}
      \fmffreeze
      \fmfdot{ve,vn,vo,vs}
    \end{fmfgraph*}
} \qquad - \qquad
\parbox{16mm}{
    \begin{fmfgraph*}(16,14)
      \fmfleft{ve}
      \fmfright{vo}
      \fmftop{vn}
      \fmftop{vs}
      \fmffreeze
      \fmfforce{(0w,0.5h)}{ve}
      \fmfforce{(1.0w,0.5h)}{vo}
      \fmfforce{(.5w,0.95h)}{vn}
      \fmfforce{(.5w,0.05h)}{vs}
      \fmffreeze
      \fmf{plain,left=0.8}{ve,vo}
      \fmf{phantom,left=0.5}{ve,vn}
      \fmf{phantom,right=0.5}{vo,vn}
      \fmf{plain,left=0.8}{vo,ve}
	    \fmf{phantom,left=0.4,label=\footnotesize{$-$},l.d=-0.01w}{vo,vs}
      \fmf{phantom,right=0.4}{ve,vs}
	    \fmf{plain,label=\footnotesize{$+$},l.d=0.05w}{vs,vn}
      \fmf{plain}{vs,vn}
      \fmffreeze
      \fmfdot{ve,vn,vo,vs}
    \end{fmfgraph*}
} \quad
\right]\, .
\label{def:IBP-4-left}
\end{flalign}
Other identities can be obtained from Eq.~(\ref{def:IBP-4-left}) by using the symmetries of the diagrams and will not be displayed.


Another set of useful identities, 
the so-called ``homogeneity'' relations \cite{Chetyrkin:1981qh}, follow from the fact that the dimensionality $d_F$ of $J$ (in units of momentum) is known in terms of the $\al_i$ and $D$ from simple dimensional analysis:
\be
d_F = 2\left( D - \sum_{j=1}^5 \al_j \right)\, .
\ee
These relations read \cite{Chetyrkin:1981qh}:
\begin{subequations}
\label{sec:IBP:2loop-homog}
\bea
d_F J &=& \left( p \cdot \frac{d}{d p} \right)\,J = - p \cdot \left( \frac{\partial}{\partial p_1} + \frac{\partial}{\partial p_2} \right)\,J \, ,
\label{sec:IBP:2loop-homoga} \\
(D+d_F)\,\frac{p_1 \cdot p}{p^2}\, J &=& \left( \frac{d}{d p} \cdot p_1 \right)\,J = -\left( \frac{\partial}{\partial p_1} + \frac{\partial}{\partial p_2} \right) \cdot p_1 \,J \, ,
\label{sec:IBP:2loop-homogb} \\
(D+d_F)\,\frac{p_2 \cdot p}{p^2}\, J &=& \left( \frac{d}{d p} \cdot p_2 \right)\,J = -\left( \frac{\partial}{\partial p_1} + \frac{\partial}{\partial p_2} \right) \cdot p_2 \,J \, ,
\label{sec:IBP:2loop-homogc} 
\eea
\end{subequations}
where the chosen circuit is $\mathcal{C}=\{p,\,p_1,\,p_2\}$ and successive relations correspond to $P=p$, $P=p_1$ and $P=p_2$, respectively.
In graphical notations, Eq.~(\ref{sec:IBP:2loop-homogc}) reads:
\begin{flalign}
&\left( \frac{D}{2}+\al_1-\al_3-\al_4-\al_5 \right) \quad
\parbox{16mm}{
    \begin{fmfgraph*}(16,14)
      \fmfleft{ve}
      \fmfright{vo}
      \fmftop{vn}
      \fmftop{vs}
      \fmffreeze
      \fmfforce{(0w,0.5h)}{ve}
      \fmfforce{(1.0w,0.5h)}{vo}
      \fmfforce{(.5w,0.95h)}{vn}
      \fmfforce{(.5w,0.05h)}{vs}
      \fmffreeze
      \fmf{plain,left=0.8}{ve,vo}
	    \fmf{phantom,left=0.5,label=\footnotesize{$\al_1$},l.d=-0.01w}{ve,vn}
	    \fmf{phantom,right=0.5,label=\footnotesize{$\al_2$},l.d=-0.01w}{vo,vn}
      \fmf{plain,left=0.8}{vo,ve}
	    \fmf{phantom,left=0.5,label=\footnotesize{$\al_3$},l.d=-0.01w}{vo,vs}
	    \fmf{phantom,right=0.5,label=\footnotesize{$\al_4$},l.d=-0.01w}{ve,vs}
	    \fmf{plain,label=\footnotesize{$\al_5$},l.d=0.05w}{vs,vn}
      \fmffreeze
      \fmfdot{ve,vn,vo,vs}
    \end{fmfgraph*}
} 
 \quad  = \quad
\al_2 \left[ \quad
\parbox{16mm}{
    \begin{fmfgraph*}(16,14)
      \fmfleft{ve}
      \fmfright{vo}
      \fmftop{vn}
      \fmftop{vs}
      \fmffreeze
      \fmfforce{(0w,0.5h)}{ve}
      \fmfforce{(1.0w,0.5h)}{vo}
      \fmfforce{(.5w,0.95h)}{vn}
      \fmfforce{(.5w,0.05h)}{vs}
      \fmffreeze
      \fmf{plain,left=0.8}{ve,vo}
      \fmf{phantom,left=0.5}{ve,vn}
	    \fmf{phantom,right=0.4,label=\footnotesize{$+$},l.d=-0.01w}{vo,vn}
      \fmf{plain,left=0.8}{vo,ve}
      \fmf{phantom,left=0.5}{vo,vs}
      \fmf{phantom,right=0.5}{ve,vs}
	    \fmf{plain,label=\footnotesize{$-$},l.d=0.05w}{vs,vn}
      \fmffreeze
      \fmfdot{ve,vn,vo,vs}
    \end{fmfgraph*}
} \quad - \quad
\parbox{16mm}{
    \begin{fmfgraph*}(16,14)
      \fmfleft{ve}
      \fmfright{vo}
      \fmftop{vn}
      \fmftop{vs}
      \fmffreeze
      \fmfforce{(0w,0.5h)}{ve}
      \fmfforce{(1.0w,0.5h)}{vo}
      \fmfforce{(.5w,0.95h)}{vn}
      \fmfforce{(.5w,0.05h)}{vs}
      \fmffreeze
      \fmf{plain,left=0.8}{ve,vo}
	    \fmf{phantom,left=0.4,label=\footnotesize{$-$},l.d=-0.01w}{ve,vn}
	    \fmf{phantom,right=0.4,label=\footnotesize{$+$},l.d=-0.01w}{vo,vn}
      \fmf{plain,left=0.8}{vo,ve}
      \fmf{phantom,left=0.5}{vo,vs}
      \fmf{phantom,right=0.5}{ve,vs}
      \fmf{plain}{vs,vn}
      \fmffreeze
      \fmfdot{ve,vn,vo,vs}
    \end{fmfgraph*}
}
\quad \right] \quad +
\label{sec:IBP:2loop-homogc-graph}
\\
& \quad
\nonum \\
& \qquad \qquad + \left(\frac{3D}{2} - \sum_{i=1}^5\al_i \right) \,(k^2)^{-1}\, \left[ \quad
\parbox{16mm}{
    \begin{fmfgraph*}(16,14)
      \fmfleft{ve}
      \fmfright{vo}
      \fmftop{vn}
      \fmftop{vs}
      \fmffreeze
      \fmfforce{(0w,0.5h)}{ve}
      \fmfforce{(1.0w,0.5h)}{vo}
      \fmfforce{(.5w,0.95h)}{vn}
      \fmfforce{(.5w,0.05h)}{vs}
      \fmffreeze
      \fmf{plain,left=0.8}{ve,vo}
      \fmf{phantom,left=0.5}{ve,vn}
      \fmf{phantom,right=0.4}{vo,vn}
      \fmf{plain,left=0.8}{vo,ve}
      \fmf{phantom,left=0.5}{vo,vs}
	    \fmf{phantom,right=0.5,label=\footnotesize{$-$},l.d=-0.01w}{ve,vs}
      \fmf{plain}{vs,vn}
      \fmffreeze
      \fmfdot{ve,vn,vo,vs}
    \end{fmfgraph*}
} \quad - \quad
\parbox{16mm}{
    \begin{fmfgraph*}(16,14)
      \fmfleft{ve}
      \fmfright{vo}
      \fmftop{vn}
      \fmftop{vs}
      \fmffreeze
      \fmfforce{(0w,0.5h)}{ve}
      \fmfforce{(1.0w,0.5h)}{vo}
      \fmfforce{(.5w,0.95h)}{vn}
      \fmfforce{(.5w,0.05h)}{vs}
      \fmffreeze
      \fmf{plain,left=0.8}{ve,vo}
	    \fmf{phantom,left=0.4,label=\footnotesize{$-$},l.d=-0.01w}{ve,vn}
      \fmf{phantom,right=0.4}{vo,vn}
      \fmf{plain,left=0.8}{vo,ve}
      \fmf{phantom,left=0.5}{vo,vs}
      \fmf{phantom,right=0.5}{ve,vs}
      \fmf{plain}{vs,vn}
      \fmffreeze
      \fmfdot{ve,vn,vo,vs}
    \end{fmfgraph*}
} \quad \right] \, .
\nonum 
\end{flalign}
This identity is particularly useful in order to express a diagram as a function of another diagram with one index decreased.

Other relations follow from double differentiation with respect to the external momentum:
\be
d_F(d_F+D-2)\,J = p^2 \frac{d^2 \,J}{dp_\mu\,dp^\mu} = p^2 \, \left( \frac{\partial}{\partial p_1} + \frac{\partial}{\partial p_2} \right)^2 \,J\, .
\label{sec:IBP:2loop-dblext}
\ee
In graphical form, this reads:
\bea
& & d_F(d_F+D-2) \quad
\parbox{16mm}{
    \begin{fmfgraph*}(16,14)
      \fmfleft{ve}
      \fmfright{vo}
      \fmftop{vn}
      \fmftop{vs}
      \fmffreeze
      \fmfforce{(0w,0.5h)}{ve}
      \fmfforce{(1.0w,0.5h)}{vo}
      \fmfforce{(.5w,0.95h)}{vn}
      \fmfforce{(.5w,0.05h)}{vs}
      \fmffreeze
      \fmf{plain,left=0.8}{ve,vo}
	    \fmf{phantom,left=0.5,label=\footnotesize{$\al_1$},l.d=-0.01w}{ve,vn}
	    \fmf{phantom,right=0.5,label=\footnotesize{$\al_2$},l.d=-0.01w}{vo,vn}
      \fmf{plain,left=0.8}{vo,ve}
	    \fmf{phantom,left=0.5,label=\footnotesize{$\al_3$},l.d=-0.01w}{vo,vs}
	    \fmf{phantom,right=0.5,label=\footnotesize{$\al_4$},l.d=-0.01w}{ve,vs}
	    \fmf{plain,label=\footnotesize{$\al_5$},l.d=0.05w}{vs,vn}
      \fmffreeze
      \fmfdot{ve,vn,vo,vs}
    \end{fmfgraph*}
} 
\quad =  \,\, -  \quad
4 \al_1 \al_2 \, \quad
\parbox{16mm}{
    \begin{fmfgraph*}(16,14)
      \fmfleft{ve}
      \fmfright{vo}
      \fmftop{vn}
      \fmftop{vs}
      \fmffreeze
      \fmfforce{(0w,0.5h)}{ve}
      \fmfforce{(1.0w,0.5h)}{vo}
      \fmfforce{(.5w,0.95h)}{vn}
      \fmfforce{(.5w,0.05h)}{vs}
      \fmffreeze
      \fmf{plain,left=0.8}{ve,vo}
	    \fmf{phantom,left=0.5,label=\footnotesize{$+$},l.d=-0.01w}{ve,vn}
	    \fmf{phantom,right=0.5,label=\footnotesize{$+$},l.d=-0.01w}{vo,vn}
      \fmf{plain,left=0.8}{vo,ve}
      \fmf{phantom,left=0.4}{vo,vs}
      \fmf{phantom,right=0.5}{ve,vs}
	    \fmf{plain,label=\footnotesize{$-$},l.d=0.05w}{vs,vn}
      \fmffreeze
      \fmfdot{ve,vn,vo,vs}
    \end{fmfgraph*}
} \quad (k^2) \qquad + 
\label{sec:IBP:dblext-graph} \\
& & \quad
\nonum \\
& & \,\, +  \quad
2(2\al_1+2\al_2+2-D)\, \left[\quad
\al_1 \quad
\parbox{16mm}{
    \begin{fmfgraph*}(16,14)
      \fmfleft{ve}
      \fmfright{vo}
      \fmftop{vn}
      \fmftop{vs}
      \fmffreeze
      \fmfforce{(0w,0.5h)}{ve}
      \fmfforce{(1.0w,0.5h)}{vo}
      \fmfforce{(.5w,0.95h)}{vn}
      \fmfforce{(.5w,0.05h)}{vs}
      \fmffreeze
      \fmf{plain,left=0.8}{ve,vo}
	    \fmf{phantom,left=0.5,label=\footnotesize{$+$},l.d=-0.01w}{ve,vn}
      \fmf{phantom,right=0.4}{vo,vn}
      \fmf{plain,left=0.8}{vo,ve}
      \fmf{phantom,left=0.5}{vo,vs}
      \fmf{phantom,right=0.5}{ve,vs}
      \fmf{plain}{vs,vn}
      \fmffreeze
      \fmfdot{ve,vn,vo,vs}
    \end{fmfgraph*}
} \quad + \quad \al_2\,\,\,\,\,\,
\parbox{16mm}{
    \begin{fmfgraph*}(16,14)
      \fmfleft{ve}
      \fmfright{vo}
      \fmftop{vn}
      \fmftop{vs}
      \fmffreeze
      \fmfforce{(0w,0.5h)}{ve}
      \fmfforce{(1.0w,0.5h)}{vo}
      \fmfforce{(.5w,0.95h)}{vn}
      \fmfforce{(.5w,0.05h)}{vs}
      \fmffreeze
      \fmf{plain,left=0.8}{ve,vo}
      \fmf{phantom,left=0.4}{ve,vn}
	    \fmf{phantom,right=0.4,label=\footnotesize{$+$},l.d=-0.01w}{vo,vn}
      \fmf{plain,left=0.8}{vo,ve}
      \fmf{phantom,left=0.5}{vo,vs}
      \fmf{phantom,right=0.5}{ve,vs}
      \fmf{plain}{vs,vn}
      \fmffreeze
      \fmfdot{ve,vn,vo,vs}
    \end{fmfgraph*}
} \quad
\right] \quad (k^2)\, .
\nonum
\eea

Finally, let us mention the IBP relations also apply to diagrams with numerator \cite{Kazakov:1986mu} and/or with mass \cite{Kotikov:1990kg}.

\subsubsection{Example of an important application}
\label{sec:meth:uniquenessApp}

In order to provide a concrete example of IBP relations and their application, let's reconsider the diagram displayed in Fig.~\ref{fig:triangle}a. The latter is defined as:
\bea
I(\al_1, \al_4) = 
J(D,p,\al_1,1,1,\al_4,1) =  \qquad \parbox{16mm}{ 
    \begin{fmfgraph*}(16,14)
      \fmfleft{i}
      \fmfright{o}
      \fmfleft{ve}
      \fmfright{vo}
      \fmftop{vn}
      \fmftop{vs}
      \fmffreeze
      \fmfforce{(-0.3w,0.5h)}{i}
      \fmfforce{(1.3w,0.5h)}{o}
      \fmfforce{(0w,0.5h)}{ve}
      \fmfforce{(1.0w,0.5h)}{vo}
      \fmfforce{(.5w,0.95h)}{vn}
      \fmfforce{(.5w,0.05h)}{vs}
      \fmffreeze
	    \fmf{fermion,label=\footnotesize{$p$}}{i,ve}
      \fmf{plain,left=0.8}{ve,vo}
	    \fmf{phantom,left=0.7,label=\footnotesize{$\al_1$},l.d=-0.01h}{ve,vn}
      \fmf{phantom,right=0.7}{vo,vn}
      \fmf{plain,left=0.8}{vo,ve}
      \fmf{phantom,left=0.7}{vo,vs}
	    \fmf{phantom,right=0.7,label=\footnotesize{$\al_4$},l.d=-0.01h}{ve,vs}
      \fmf{plain}{vs,vn}
      \fmf{plain}{vo,o}
      \fmffreeze
      \fmfdot{ve,vn,vo,vs}
    \end{fmfgraph*}
}
\qquad \, ,
\label{IBPex}
\eea
where one triangle (the one on the right in the above equation) has all lines with indices equal to $1$ (anticipating the next section, we shall refer to it as an {\it ordinary triangle}). Applying the IBP procedure with the vertical line being the distinguished one and the left triangle as the contour, we have:
\begin{flalign}
(D-2 -\al_1 - \al_4) \, I(\al_1, \al_4) &= \al_1 \left( 
\quad  
\parbox{16mm}{
\begin{fmfgraph*}(24,14)
    \fmfleft{i}
    \fmfright{o}
    \fmfleft{ve}
    \fmfright{vo}
    \fmftop{v}
    \fmffreeze
    \fmfforce{(-0.1w,0.5h)}{i}
    \fmfforce{(1.1w,0.5h)}{o}
    \fmfforce{(0w,0.5h)}{ve}
    \fmfforce{(1.0w,0.5h)}{vo}
    \fmfforce{(.5w,0.5h)}{v}
    \fmffreeze
    \fmf{plain}{i,ve}
	\fmf{plain,left=0.8,label=\footnotesize{$\al_1+1$},l.d=0.1h}{ve,v}
	\fmf{plain,left=0.8,label=\footnotesize{$\al_4$},l.d=0.1h}{v,ve}
    \fmf{plain,left=0.8}{v,vo}
    \fmf{plain,left=0.8}{vo,v}
    \fmf{plain}{vo,o}
    \fmffreeze
    \fmfdot{ve,v,vo}
  \end{fmfgraph*}
}
  \qquad \qquad
-
\qquad
\parbox{16mm}{
\begin{fmfgraph*}(16,14)
    \fmfleft{i}
    \fmfright{o}
    \fmfleft{ve}
    \fmfright{vo}
    \fmftop{v}
    \fmffreeze
    \fmfforce{(-0.1w,0.5h)}{i}
    \fmfforce{(1.1w,0.5h)}{o}
    \fmfforce{(0w,0.5h)}{ve}
    \fmfforce{(1.0w,0.5h)}{vo}
    \fmfforce{(.5w,0.05h)}{v}
    \fmffreeze
    \fmf{plain}{i,ve}
	\fmf{plain,left=0.8,label=\footnotesize{$\al_1+1$},l.d=0.1h}{ve,vo}
    \fmf{plain,left=0.8}{vo,ve}
    \fmf{plain,left=0.5}{v,vo}
	\fmf{phantom,left=0.1,label=\footnotesize{$\al_4$},l.d=0.1h}{v,ve}
    \fmf{plain}{vo,o}
    \fmffreeze
    \fmfdot{ve,v,vo}
  \end{fmfgraph*}
 } \quad
\right) \qquad +
\nonumber 
\\ &\qquad 
\nonum
\\
&+  \al_4 \left( 
\quad
\parbox{16mm}{
\begin{fmfgraph*}(24,14)
    \fmfleft{i}
    \fmfright{o}
    \fmfleft{ve}
    \fmfright{vo}
    \fmftop{v}
    \fmffreeze
    \fmfforce{(-0.1w,0.5h)}{i}
    \fmfforce{(1.1w,0.5h)}{o}
    \fmfforce{(0w,0.5h)}{ve}
    \fmfforce{(1.0w,0.5h)}{vo}
    \fmfforce{(.5w,0.5h)}{v}
    \fmffreeze
    \fmf{plain}{i,ve}
	\fmf{plain,left=0.8,label=\footnotesize{$\al_1$},l.d=0.1h}{ve,v}
	\fmf{plain,left=0.8,label=\footnotesize{$\al_4+1$},l.d=0.1h}{v,ve}
    \fmf{plain,left=0.8}{v,vo}
    \fmf{plain,left=0.8}{vo,v}
    \fmf{plain}{vo,o}
    \fmffreeze
    \fmfdot{ve,v,vo}
  \end{fmfgraph*}
}
  \qquad \qquad
-
  \qquad
\parbox{16mm}{
\begin{fmfgraph*}(16,14)
    \fmfleft{i}
    \fmfright{o}
    \fmfleft{ve}
    \fmfright{vo}
    \fmftop{v}
    \fmffreeze
    \fmfforce{(-0.1w,0.5h)}{i}
    \fmfforce{(1.1w,0.5h)}{o}
    \fmfforce{(0w,0.5h)}{ve}
    \fmfforce{(1.0w,0.5h)}{vo}
    \fmfforce{(.5w,0.05h)}{v}
    \fmffreeze
    \fmf{plain}{i,ve}
	\fmf{plain,left=0.8,label=\footnotesize{$\al_1$},l.d=0.1h}{ve,vo}
    \fmf{plain,left=0.8}{vo,ve}
    \fmf{plain,left=0.5}{v,vo}
	\fmf{phantom,left=0.1,label=\footnotesize{$\al_4+1$},l.d=0.1h}{v,ve}
    \fmf{plain}{vo,o}
    \fmffreeze
    \fmfdot{ve,v,vo}
  \end{fmfgraph*}
 }
\quad \right) \, .
\label{IBPex1}
\end{flalign}
Evaluating the diagrams appearing on the rhs of the above equation yields the advertised result 
of Eq.~(\ref{sec:IBP:application1}) that we reproduce here for clarity ($D=4-2\ep$):
\begin{flalign}
  C_D[I(\al_1,\al_4)]  &=  \frac{G(D,1,1)}{D - 2 -\al_1 - \al_4}\,\biggl[ \al_1 \,G(D,1+\al_1,\al_4) - \al_1 \,G(D,1+\al_1,\al_4 + \ep) 
\nonum \\
&\quad + \al_4 \,G(D,1+\al_4,\al_1) - \al_4 \,G(D,1+\al_4,\al_1+\ep) \biggr]\, . 
\label{IBPex2}
\end{flalign}
A particularly important application of $I(\al_1,\al_4)$ is related to the master integral: $I(1+a_1\ep,1+a_4\ep)$, the $\ep$-expansion of which can be calculated using Eq.~(\ref{IBPex2}).  
After a little algebra, this leads to:
\begin{flalign}
C_D[I(1+a_1\ep,1+a_4\ep)] &= \frac{\hat{K}_2}{1-2\ep} \, \biggl[A_0 \,\zeta_2 + A_1 \,\zeta_4 \ep+ A_2 \,\zeta_5 \ep^2 +
\Bigl[A_3 \,\zeta_6 - A_4 \,\zeta_3^2 \Bigr]\, \ep^3 \nonumber \\
&+ \Bigl[A_5\, \zeta_7 - A_6 \, \zeta_3 \zeta_4  \Bigr]\, \ep^4 + O(\ep^5)\biggr]\, ,
\label{IBPex3}
\end{flalign}
with  
\begin{flalign}
&A_0=6\, , 
\nonum \\ 
&A_1=9\, , 
\nonum \\
&A_2= 42 + 30 \,\overline{a} + 10 \,\overline{a}^2 -10\, a_1\,a_4\, , 
\nonum \\
&A_3= \frac{5}{2} \Bigl(A_2 - 6\Bigr)\, , \label{IBPex4} \\
	&A_4= 46 + 42 \,\overline{a} + 14 \,\overline{a}^2 -14\,a_1\,a_4\, ,
\nonum \\ 
	&A_5= 294 + 402\, \overline{a} + 260\, \overline{a}^2 -134\, a_1\,a_4 + 84\, \overline{a}^3 - 84\, \overline{a} \,a_1 \,a_4 +14\,
	\overline{a}^4 - 28\, \overline{a}^2\,a_1\,a_4 + 14\, a_1^2\,a_4^2 \, ,
	\nonum \\
&A_6= 3 \Bigl(A_4 - 1\Bigr)\, ,
\nonum
\end{flalign}
where
%
\fleq{
\overline{a}=a_1+a_4,\qquad  ~~ \hat{K}_n = \exp\left[-n\left(\gamma_E \ep + \frac{\zeta_2}{2} \ep^2\right)\right]\, .
\label{IBPex5}
}
Notice that, for $a_1=a_4=0$, we recover the result in (\ref{G(1,1,1,1,1)-exp4d}).
%

\subsection{The method of uniqueness}
\label{sec:meth:uniqueness}

The method of uniqueness is a powerful (but not very well known) technique devoted to the computation of {\it massless} multi-loop Feynman diagrams.
This method owes its name to the so-called {\it uniqueness relation}, otherwise known as the {\it star-triangle} or {\it Yang-Baxter
relation}, which is used in theories with conformal symmetry. Historically,
such relation was probably first used to compute three-dimensional integrals by D'Eramo, Peleti and Parisi \cite{D'Eramo:1971zz}.
Within the framework of multi-loop calculations, the method has first been introduced by Vasil'ev, Pis'mak and Khonkonen \cite{Vasiliev:1981dg}.
It allows, in principle, the computation of complicated Feynman diagrams using sequences of
simple transformations (including integration by parts) without performing any explicit integration.
A diagram is straightforwardly integrated once the appropriate sequence is found. In a sense, the method greatly simplifies
multi-loop calculations \cite{Vasiliev:1981dg}-\cite{Kazakov:1983pk}.
As a matter of fact, the first analytical expression for the
five-loop $\beta$-function of the $\Phi^4$-model was derived by Kazakov using this technique \cite{Kazakov:1984km,Kazakov:1983pk}.
For a given diagram, the task of finding the sequence of transformations is, however, highly non-trivial. 
In the following, we will briefly present the method in momentum space in very close analogy with the beautiful lectures of Kazakov \cite{Kazakov:1984bw}
where the method was presented in coordinate space, see also \cite{Isaev:2003tk} for a recent short review. 

In momentum space, a triangle made of scalar propagators with three arbitrary indices is defined as:
\vspace{3mm}
\be
\parbox{16mm}{
  \begin{fmfgraph*}(16,16)
    \fmfleft{l}
    \fmfleft{vl}
    \fmfright{r}
    \fmfright{vr}
    \fmftop{t}
    \fmftop{vt}
    \fmffreeze
    \fmfforce{(0.0w,0.0h)}{l}
    \fmfforce{(0.25w,0.25h)}{vl}
    \fmfforce{(1.0w,0.0h)}{r}
    \fmfforce{(0.75w,0.25h)}{vr}
    \fmfforce{(.5w,1.0h)}{t}
    \fmfforce{(.5w,0.75h)}{vt}
    \fmffreeze
	  \fmf{plain,label=\footnotesize{$\al_2$},label.side=left}{vr,vl}
	  \fmf{plain,label=\footnotesize{$\al_3$},label.side=left}{vl,vt}
	  \fmf{plain,label=\footnotesize{$\al_1$},label.side=left}{vt,vr}
    \fmf{plain}{vl,l}
    \fmf{plain}{vr,r}
    \fmf{plain}{vt,t}
    \fmffreeze
    \fmfdot{vl,vt,vr}
    \fmflabel{$p_1$}{l}
    \fmflabel{$p_2$}{t}
    \fmflabel{$p_3=-p_1-p_2$}{r}
  \end{fmfgraph*}
} \,\,\, = \,\,\, \int\, \frac{[\D^D k]}{k^{2\al_2} (k-p_1)^{2\al_3} (k-p_1-p_2)^{2\al_1}} \, .
\label{def:triangle}
\ee
\vskip 4mm

\ni On the other hand, a vertex made of scalar propagators with three arbitrary indices is defined as:
\vspace{3mm}
\be
\parbox{16mm}{
  \begin{fmfgraph*}(16,16)
    \fmfleft{l}
    \fmfright{r}
    \fmftop{t}
    \fmfleft{v}
    \fmffreeze
    \fmfforce{(0.0w,0.0h)}{l}
    \fmfforce{(1.0w,0.0h)}{r}
    \fmfforce{(.5w,1.0h)}{t}
    \fmfforce{(.5w,0.5h)}{v}
    \fmffreeze
	  \fmf{plain,label=\footnotesize{$\beta_2$},label.side=right,l.d=0.05w}{v,t}
	  \fmf{plain,label=\footnotesize{$\beta_3$},label.side=right,l.d=0.05w}{v,r}
	  \fmf{plain,label=\footnotesize{$\beta_1$},label.side=right,l.d=0.05w}{v,l}
    \fmffreeze
    \fmfdot{v}
  \end{fmfgraph*}
} \qquad = \qquad \frac{1}{p_1^{2\beta_1} p_2^{2\beta_2} (p_1+p_2)^{2\beta_3}}\, .
\label{def:star}
\ee
Both triangle and vertex diagrams have indices $\sum_{j=1}^3 \al_j$, corresponding to the sum of the indices of their constituent lines.
In momentum space, {\it ordinary} triangles and vertices, that is triangle and vertices made of ordinary lines, have index $3$.
Of great importance in the following will be the notion of a {\it unique} triangle and a {\it unique} vertex.
In momentum space, these diagrams are said to be {\it unique} if their indices are equal
to $D$ and $D/2$, respectively; see Tab.~\ref{meth:tab:indices} for a summary.

\begin{center}
\renewcommand{\tabcolsep}{1.2cm}
\renewcommand{\arraystretch}{1.5}
\begin{table}
    \begin{tabular}{  l || c | c | c }
      \hline
        ~~      &       {\bf Line}                      &       {\bf Triangle}          &       {\bf Vertex} \\
      \hline \hline
      Arbitrary &       $\al$                           & $\sum_{j=1}^3 \al_j$          &       $\sum_{j=1}^3 \al_j$    \\
      \hline
      Ordinary  &       $1$                             & $3$                           &       $3$     \\
      \hline
      Unique    &       $D/2$                           & $D$                           &       $D/2$  \\
      \hline
    \end{tabular}
    \caption{Indices of lines, triangles and vertices in $p$-space.}
    \label{meth:tab:indices}
\end{table}
\end{center}

The uniqueness (or star-triangle) relation connects a unique triangle to a unique vertex and reads:
\be
\parbox{16mm}{
  \begin{fmfgraph*}(16,16)
    \fmfleft{l}
    \fmfright{r}
    \fmftop{t}
    \fmfleft{v}
    \fmffreeze
    \fmfforce{(-0.1w,0.142265h)}{l}
    \fmfforce{(1.1w,0.142265h)}{r}
    \fmfforce{(.5w,1.1h)}{t}
    \fmfforce{(.5w,0.488675h)}{v}
    \fmffreeze
	  \fmf{plain,label=\footnotesize{${\al}_2$},label.side=left,l.d=0.05w}{v,t}
	  \fmf{plain,label=\footnotesize{${\al}_3$},label.side=right,l.d=0.05w}{r,v}
	  \fmf{plain,label=\footnotesize{${\al}_1$},label.side=left,l.d=0.05w}{v,l}
    \fmffreeze
    \fmfdot{v}
  \end{fmfgraph*}
}\, \qquad \underset{\underset{j}{\sum} \al_j = D/2}{=} \qquad \frac{(4\pi)^{D/2}}{G(\tilde{\al}_1,\tilde{\al}_2)}\, \quad
\parbox{16mm}{
  \begin{fmfgraph*}(16,16)
    \fmfleft{l}
    \fmfleft{vl}
    \fmfright{r}
    \fmfright{vr}
    \fmftop{t}
    \fmftop{vt}
    \fmffreeze
    \fmfforce{(-0.1w,0.042265h)}{l}
    \fmfforce{(0.1w,0.2h)}{vl}
    \fmfforce{(1.1w,0.042265h)}{r}
    \fmfforce{(0.9w,0.2h)}{vr}
    \fmfforce{(.5w,1.1h)}{t}
    \fmfforce{(.5w,0.966025h)}{vt}
    \fmffreeze
	  \fmf{plain,label=\footnotesize{$\tilde{\al}_2$},label.side=left,l.d=0.05w}{vr,vl}
	  \fmf{plain,label=\footnotesize{$\tilde{\al}_3$},label.side=left,l.d=0.05w}{vl,vt}
	  \fmf{plain,label=\footnotesize{$\tilde{\al}_1$},label.side=left,l.d=0.05w}{vt,vr}
    \fmf{plain}{l,vl}
    \fmf{plain}{r,vr}
    \fmf{plain}{t,vt}
    \fmffreeze
    \fmfdot{vl,vt,vr}
  \end{fmfgraph*}
}\, \qquad ,
\label{def:star-triangle}
\ee
where $\tilde{\al}_i=D/2-\al_i$ is the index dual to $\al_i$ and the condition $\sum_{j=1}^3 \al_j= D/2$ implies that the vertex is unique.
This relation can be proved by performing an inversion of all integration variables in the triangle: $k_\mu \ra k_\mu/(k)^2$,
keeping the external momenta fixed. Upon using the fact that the triangle is unique,  $\sum_j \tilde{\al}_j =D$, the integral simplifies and reduces to a simple vertex.

\subsection{Transformation of indices}
\label{sec:transformation}

In order to illustrate the power of the method of uniqueness, we now proceed on giving several useful transformations of indices. 
Following Kazakov (see also Sec.~\ref{sec:symmetries}) and considering the general massless two-loop p-type diagram with arbitrary indices, \eg, Eq.~(\ref{def:two-loop-p-int}), 
we shall denote 
the index of the left (right) triangle as: $t_1=\al_1+\al_4+\al_5$ ($t_2=\al_2+\al_3+\al_5$) and the index of the upper (lower) vertex as: 
$v_1=\al_1+\al_2+\al_5$ ($v_2=\al_3+\al_4+\al_5$). The following notations will also be useful: $\overline{t}_i=t_i -D/2$, $\overline{v}_i=v_i -D/2$ $(i=1,2)$ and,
as previously defined, $\tilde{\al}_j = D/2-\al_j$ $(j=1,...,5)$.

\subsubsection{Splitting a line into loop}
\label{sec:line}

This transformation allows to derive Eq.~(\ref{transf4}) which was first obtained by Gorishny and Isaev \cite{Gorishnii:1984te}.
From the two-loop p-type diagram with arbitrary indices, we start by replacing the central line by a loop~\footnote{In coordinate space, it corresponds to the insertion of a
point into this line (see the table of such transformations in Ref.~\cite{Vasiliev:1981dg} and also Ref.~\cite{Kazakov:1984bw} for a review).} in such
a way that the right triangle is unique. This yields:
\bea
\parbox{16mm}{
    \begin{fmfgraph*}(16,14)
      \fmfleft{i}
      \fmfright{o}
      \fmfleft{ve}
      \fmfright{vo}
      \fmftop{vn}
      \fmftop{vs}
      \fmffreeze
      \fmfforce{(-0.3w,0.5h)}{i}
      \fmfforce{(1.3w,0.5h)}{o}
      \fmfforce{(0w,0.5h)}{ve}
      \fmfforce{(1.0w,0.5h)}{vo}
      \fmfforce{(.5w,0.95h)}{vn}
      \fmfforce{(.5w,0.05h)}{vs}
      \fmffreeze
      \fmf{plain}{i,ve}
      \fmf{plain,left=0.8}{ve,vo}
	    \fmf{phantom,left=0.5,label=\footnotesize{$\al_1$},l.d=-0.01w}{ve,vn}
	    \fmf{phantom,right=0.5,label=\footnotesize{$\al_2$},l.d=-0.01w}{vo,vn}
      \fmf{plain,left=0.8}{vo,ve}
	    \fmf{phantom,left=0.5,label=\footnotesize{$\al_3$},l.d=-0.01w}{vo,vs}
	    \fmf{phantom,right=0.5,label=\footnotesize{$\al_4$},l.d=-0.01w}{ve,vs}
	    \fmf{plain,label=\footnotesize{$\al_5$},l.d=0.05w}{vs,vn}
      \fmf{plain}{vo,o}
      \fmffreeze
      \fmfdot{ve,vn,vo,vs}
    \end{fmfgraph*}
}  \qquad = \quad \frac{(4\pi)^{D/2}}{G(D,\beta,\gamma)} \qquad ~
\parbox{18mm}{
    \begin{fmfgraph*}(18,16)
     \fmfleft{i}
      \fmfright{o}
      \fmfleft{ve}
      \fmfright{vo}
      \fmftop{vn}
      \fmftop{vs}
      \fmffreeze
      \fmfforce{(-0.3w,0.5h)}{i}
      \fmfforce{(1.3w,0.5h)}{o}
      \fmfforce{(0w,0.5h)}{ve}
      \fmfforce{(1.0w,0.5h)}{vo}
      \fmfforce{(.5w,0.95h)}{vn}
      \fmfforce{(.5w,0.05h)}{vs}
      \fmffreeze
      \fmf{plain}{i,ve}
      \fmf{plain,left=0.8}{ve,vo}
	    \fmf{phantom,left=0.7,label=\footnotesize{$\al_1$},l.d=-0.0w}{ve,vn}
	    \fmf{phantom,right=0.7,label=\footnotesize{$\al_2$},l.d=-0.0w}{vo,vn}
      \fmf{plain,left=0.8}{vo,ve}
	    \fmf{phantom,left=0.7,label=\footnotesize{$\al_3$},l.d=-0.0w}{vo,vs}
	    \fmf{phantom,right=0.7,label=\footnotesize{$\al_4$},l.d=-0.0w}{ve,vs}
	    \fmf{plain,left=0.3,label=\footnotesize{$\beta$},l.d=0.05w}{vs,vn}
	    \fmf{plain,left=0.3,label=\footnotesize{$\gamma$},l.d=0.05w}{vn,vs}
      \fmf{plain}{vo,o}
      \fmffreeze
      \fmfdot{ve,vn,vo,vs}
    \end{fmfgraph*}
} \qquad \, ,
\label{GI1}
\eea
%
where $\beta = t_2 -D/2 \equiv \overline{t}_2$ and $\gamma=D-\al_2-\al_3  \equiv \tilde{\al}_2+\tilde{\al}_3$.
The right triangle being unique, we can use Eq.~(\ref{def:star-triangle}) to simplify the diagram:
\bea
\parbox{16mm}{
    \begin{fmfgraph*}(16,14)
      \fmfleft{i}
      \fmfright{o}
      \fmfleft{ve}
      \fmfright{vo}
      \fmftop{vn}
      \fmftop{vs}
      \fmffreeze
      \fmfforce{(-0.3w,0.5h)}{i}
      \fmfforce{(1.3w,0.5h)}{o}
      \fmfforce{(0w,0.5h)}{ve}
      \fmfforce{(1.0w,0.5h)}{vo}
      \fmfforce{(.5w,0.95h)}{vn}
      \fmfforce{(.5w,0.05h)}{vs}
      \fmffreeze
      \fmf{plain}{i,ve}
      \fmf{plain,left=0.8}{ve,vo}
	    \fmf{phantom,left=0.5,label=\footnotesize{$\al_1$},l.d=-0.01w}{ve,vn}
	    \fmf{phantom,right=0.5,label=\footnotesize{$\al_2$},l.d=-0.01w}{vo,vn}
      \fmf{plain,left=0.8}{vo,ve}
	    \fmf{phantom,left=0.5,label=\footnotesize{$\al_3$},l.d=-0.01w}{vo,vs}
	    \fmf{phantom,right=0.5,label=\footnotesize{$\al_4$},l.d=-0.01w}{ve,vs}
	    \fmf{plain,label=\footnotesize{$\al_5$},l.d=0.05w}{vs,vn}
      \fmf{plain}{vo,o}
      \fmffreeze
      \fmfdot{ve,vn,vo,vs}
    \end{fmfgraph*}
}\qquad = \quad \frac{G(D,\al_2,\gamma)}{G(D,\overline{t}_2,\gamma)} \,\,(p^2)^{D/2-\al_2-\al_3}\qquad 
\parbox{16mm}{
    \begin{fmfgraph*}(16,14)
      \fmfleft{i}
      \fmfright{o}
      \fmfleft{ve}
      \fmfright{vo}
      \fmftop{vn}
      \fmftop{vs}
      \fmffreeze
      \fmfforce{(-0.3w,0.5h)}{i}
      \fmfforce{(1.3w,0.5h)}{o}
      \fmfforce{(0w,0.5h)}{ve}
      \fmfforce{(1.0w,0.5h)}{vo}
      \fmfforce{(.5w,0.95h)}{vn}
      \fmfforce{(.5w,0.05h)}{vs}
      \fmffreeze
      \fmf{plain}{i,ve}
      \fmf{plain,left=0.8}{ve,vo}
	    \fmf{phantom,left=0.5,label=\footnotesize{$\al_1$},l.d=-0.01w}{ve,vn}
	    \fmf{phantom,right=0.5,label=\footnotesize{$\tilde{\al}_3$},l.d=-0.01w}{vo,vn}
      \fmf{plain,left=0.8}{vo,ve}
	    \fmf{phantom,left=0.5,label=\footnotesize{$\tilde{\al}_2$},l.d=-0.01w}{vo,vs}
	    \fmf{phantom,right=0.5,label=\footnotesize{$\al_4$},l.d=-0.01w}{ve,vs}
	    \fmf{plain,label=\footnotesize{$\overline{t}_2$},l.d=0.05w}{vs,vn}
      \fmf{plain}{vo,o}
      \fmffreeze
      \fmfdot{ve,vn,vo,vs}
    \end{fmfgraph*}
} \qquad \qquad .
\label{GI2}
\eea
Focusing for simplicity on the coefficient functions, all the dependence on the external momentum disappears.
Together with the simplification of the $G$-functions this yields the advertised Eq.~(\ref{transf4}) that we reproduce for clarity: 
\fleq{
C_D[J(D,p,\al_1,\al_2,\al_3,\al_4,\al_5)] = a(\al_2)a(\al_3)a(\al_5)a(D-t_2)\,C_D[J(D,p,\al_1,\tilde{\al}_3,\tilde{\al}_2,\al_4,\overline{t}_2)] \, .
\nonum
}
%
Notice that, for $\al_1=\al_4=\overline{t}_2=1$, \ie, $\al_1=\al_4=1$ and $t_2=1+D/2$ (that is, the sum of indices $\al_2$, $\al_3$ and $\al_5$ is equal to $D/2+1$), 
the left triangle on the rhs of Eq.~(\ref{GI2}) corresponds to the ordinary one, see Tab.~\ref{meth:tab:indices}. Moreover, for $\tilde{\al}_2=\tilde{\al}_3=\overline{t}_2=1$, \ie, $\al_2=\al_3=D/2-1=1-\ep$
and $\al_5=3-D/2=1+\ep$, it is the right triangle on the rhs of Eq.~(\ref{GI2}) that corresponds to the ordinary one. In these cases, the diagrams can be evaluated with the help of
the IBP relations and expressed in terms of $\Gamma$-functions in agreement with subsection \ref{sec:meth:uniquenessApp}.

We can now repeat the above manipulations with a lateral line (we use the one having the index $\al_1$), \ie,
\bea
\parbox{16mm}{
    \begin{fmfgraph*}(16,14)
      \fmfleft{i}
      \fmfright{o}
      \fmfleft{ve}
      \fmfright{vo}
      \fmftop{vn}
      \fmftop{vs}
      \fmffreeze
      \fmfforce{(-0.3w,0.5h)}{i}
      \fmfforce{(1.3w,0.5h)}{o}
      \fmfforce{(0w,0.5h)}{ve}
      \fmfforce{(1.0w,0.5h)}{vo}
      \fmfforce{(.5w,0.95h)}{vn}
      \fmfforce{(.5w,0.05h)}{vs}
      \fmffreeze
      \fmf{plain}{i,ve}
      \fmf{plain,left=0.8}{ve,vo}
	    \fmf{phantom,left=0.5,label=\footnotesize{$\al_1$},l.d=-0.01w}{ve,vn}
	    \fmf{phantom,right=0.5,label=\footnotesize{$\al_2$},l.d=-0.01w}{vo,vn}
      \fmf{plain,left=0.8}{vo,ve}
	    \fmf{phantom,left=0.5,label=\footnotesize{$\al_3$},l.d=-0.01w}{vo,vs}
	    \fmf{phantom,right=0.5,label=\footnotesize{$\al_4$},l.d=-0.01w}{ve,vs}
	    \fmf{plain,label=\footnotesize{$\al_5$},l.d=0.05w}{vs,vn}
      \fmf{plain}{vo,o}
      \fmffreeze
      \fmfdot{ve,vn,vo,vs}
    \end{fmfgraph*}
}  \qquad = \quad  \frac{(4\pi)^{D/2}}{G(D,\beta_1,\gamma_1)} \qquad
\parbox{16mm}{
    \begin{fmfgraph*}(16,14)
      \fmfleft{i}
      \fmfright{o}
      \fmfleft{ve}
      \fmfright{vo}
      \fmftop{vn}
      \fmftop{vs}
      \fmffreeze
      \fmfforce{(-0.3w,0.5h)}{i}
      \fmfforce{(1.3w,0.5h)}{o}
      \fmfforce{(0w,0.5h)}{ve}
      \fmfforce{(1.0w,0.5h)}{vo}
      \fmfforce{(.5w,0.95h)}{vn}
      \fmfforce{(.5w,0.05h)}{vs}
      \fmffreeze
      \fmf{plain}{i,ve}
      \fmf{plain,left=0.8}{ve,vo}
	    \fmf{phantom,left=0.5,label=\footnotesize{$\beta_1$},l.d=-0.01w}{ve,vn}	    
	    \fmf{phantom,right=0.5,label=\footnotesize{$\al_2$},l.d=-0.01w}{vo,vn}
      \fmf{plain,left=0.8}{vo,ve}
	    \fmf{phantom,left=0.5,label=\footnotesize{$\al_3$},l.d=-0.01w}{vo,vs}
	    \fmf{phantom,right=0.5,label=\footnotesize{$\al_4$},l.d=-0.01w}{ve,vs}
	    \fmf{plain,label=\footnotesize{$\al_5$},l.d=0.05w}{vs,vn}
      \fmf{plain}{vo,o}
      \fmf{plain,right=0.5}{ve,vn}
	    \fmf{phantom,right=0.5,label=\footnotesize{$\gamma_1$},l.d=0.2h,l.s=left}{ve,vs}
      \fmffreeze
      \fmfdot{ve,vn,vo,vs}
    \end{fmfgraph*}
}  \qquad \, ,
\eea
where $\beta_1 = t_1-D/2=\overline{t}_1$ 
and $\gamma_1=D-\al_4-\al_5 =\tilde{\al}_4 +\tilde{\al}_5 $. The triangle having the lines with the indices  $\gamma_1$, $\al_4$ and $\al_5$,
is unique so we can use Eq.~(\ref{def:star-triangle}) to simplify the diagram:
\bea
\parbox{16mm}{
    \begin{fmfgraph*}(16,14)
      \fmfleft{i}
      \fmfright{o}
      \fmfleft{ve}
      \fmfright{vo}
      \fmftop{vn}
      \fmftop{vs}
      \fmffreeze
      \fmfforce{(-0.3w,0.5h)}{i}
      \fmfforce{(1.3w,0.5h)}{o}
      \fmfforce{(0w,0.5h)}{ve}
      \fmfforce{(1.0w,0.5h)}{vo}
      \fmfforce{(.5w,0.95h)}{vn}
      \fmfforce{(.5w,0.05h)}{vs}
      \fmffreeze
      \fmf{plain}{i,ve}
      \fmf{plain,left=0.8}{ve,vo}
	    \fmf{phantom,left=0.5,label=\footnotesize{$\al_1$},l.d=-0.01w}{ve,vn}
	    \fmf{phantom,right=0.5,label=\footnotesize{$\al_2$},l.d=-0.01w}{vo,vn}
      \fmf{plain,left=0.8}{vo,ve}
	    \fmf{phantom,left=0.5,label=\footnotesize{$\al_3$},l.d=-0.01w}{vo,vs}
	    \fmf{phantom,right=0.5,label=\footnotesize{$\al_4$},l.d=-0.01w}{ve,vs}
	    \fmf{plain,label=\footnotesize{$\al_5$},l.d=0.05w}{vs,vn}
      \fmf{plain}{vo,o}
      \fmffreeze
      \fmfdot{ve,vn,vo,vs}
    \end{fmfgraph*}
}  \qquad = \quad \frac{G(D,\al_4,\gamma_1)}{G(D,\overline{t}_1,\gamma_1)} 
\qquad
\parbox{16mm}{
    \begin{fmfgraph*}(16,14)
      \fmfleft{i}
      \fmfright{o}
      \fmfleft{ve}
      \fmfright{vo}
      \fmftop{vn}
      \fmftop{vs}
      \fmffreeze
      \fmfforce{(-0.3w,0.5h)}{i}
      \fmfforce{(1.3w,0.5h)}{o}
      \fmfforce{(0w,0.5h)}{ve}
      \fmfforce{(1.0w,0.5h)}{vo}
      \fmfforce{(.5w,0.95h)}{vn}
      \fmfforce{(.5w,0.05h)}{vs}
      \fmffreeze
      \fmf{plain}{i,ve}
      \fmf{plain,left=0.8}{ve,vo}
	    \fmf{phantom,left=0.5,label=\footnotesize{$\overline{t}_1$},l.d=-0.01w}{ve,vn}
	    \fmf{phantom,right=0.5,label=\footnotesize{$\al_2$},l.d=-0.01w}{vo,vn}
      \fmf{plain,left=0.8}{vo,ve}
	    \fmf{phantom,left=0.5,label=\footnotesize{$\overline{\al}_3$},l.d=-0.01w}{vo,vs}
	    \fmf{phantom,right=0.5,label=\footnotesize{$\tilde{\al}_5$},l.d=-0.01w}{ve,vs}
	    \fmf{plain,label=\footnotesize{$\tilde{\al}_4$},l.d=0.05w}{vs,vn}
      \fmf{plain}{vo,o}
      \fmffreeze
      \fmfdot{ve,vn,vo,vs}
    \end{fmfgraph*}
}  \qquad \, ,
\label{Split}
\eea
where $\overline{\al}_3=\al_3+D/2-\gamma_1=\al_3+\al_4+\al_5-D/2=v_2-D/2 \equiv \overline{v}_2$.
Eq.~(\ref{Split}) gives for the corresponding coefficient function:
%
\fleq{
C_D[J(D,p,\al_1,\al_2,\al_3,\al_4,\al_5)] = a(\al_1)a(\al_4)a(\al_5)a(D-t_1)\,C_D[J(D,p,\overline{t}_1,\al_2,\overline{v}_2,
\tilde{\al}_5,\tilde{\al}_4)]\, .
\nonum
}
In this example, we have now an ordinary left triangle when: $\tilde{\al}_4=\tilde{\al}_5=\overline{t}_1=1$, 
\ie, $\al_2=\al_5=D/2-1=1-\ep$ and $\al_1=3-D/2=1+\ep$. On the other hand, the right triangle becomes ordinary when:
$\al_2=\tilde{\al}_3=\tilde{\al}_4=1$, \ie, 
$\tilde{\al}_4= \al_2=1$ and $v_2=\al_3 + \al_4 + \al_5=D/2+1$, and, finally,
$\al_2=1$, $\al_4=D/2-1=1-\ep$ and $\al_3+\al_5=2$. In all these cases, the diagrams may again be evaluated with the help of
the IBP relations and expressed in terms of $\Gamma$-functions in agreement with subsection \ref{sec:meth:uniquenessApp}.

\subsubsection{Adding a new propagator and a new loop}
\label{sec:additional}

Another transformation allowing to change the indices of the two-loop p-type diagram (see also Eq.~(\ref{def:two-loop-G-func})): 
\bea
\parbox{16mm}{
    \begin{fmfgraph*}(16,14)
      \fmfleft{i}
      \fmfright{o}
      \fmfleft{ve}
      \fmfright{vo}
      \fmftop{vn}
      \fmftop{vs}
      \fmffreeze
      \fmfforce{(-0.3w,0.5h)}{i}
      \fmfforce{(1.3w,0.5h)}{o}
      \fmfforce{(0w,0.5h)}{ve}
      \fmfforce{(1.0w,0.5h)}{vo}
      \fmfforce{(.5w,0.95h)}{vn}
      \fmfforce{(.5w,0.05h)}{vs}
      \fmffreeze
      \fmf{plain}{i,ve}
      \fmf{plain,left=0.8}{ve,vo}
	    \fmf{phantom,left=0.5,label=\footnotesize{$\al_1$},l.d=-0.01w}{ve,vn}
	    \fmf{phantom,right=0.5,label=\footnotesize{$\al_2$},l.d=-0.01w}{vo,vn}
      \fmf{plain,left=0.8}{vo,ve}
	    \fmf{phantom,left=0.5,label=\footnotesize{$\al_3$},l.d=-0.01w}{vo,vs}
	    \fmf{phantom,right=0.5,label=\footnotesize{$\al_4$},l.d=-0.01w}{ve,vs}
	    \fmf{plain,label=\footnotesize{$\al_5$},l.d=0.05w}{vs,vn}
      \fmf{plain}{vo,o}
      \fmffreeze
      \fmfdot{ve,vn,vo,vs}
    \end{fmfgraph*}
} \qquad  =  \quad \frac{p^{-2\delta}}{(4\pi)^{D}} \, C_D[J(D,p,\al_1,\al_2,\al_3,\al_4,\al_5)] \, ,
\label{Dia}
\eea
where $\delta = \sum_{i=1}^{5} \al_i -D$, consists in adding an additional propagator with index $\tilde{\al}_2-\al_3$
to both its lhs and rhs. This leads to:
\bea
 \frac{p^{-2\delta}}{(4\pi)^{D}} \, C_D[J(D,p,\al_1,\al_2,\al_3,\al_4,\al_5)]~\frac{1}{p^{2(\tilde{\al}_2- \al_3)}} \quad = \qquad
\parbox{16mm}{
    \begin{fmfgraph*}(16,14)
      \fmfleft{i}
      \fmfright{o}
      \fmfleft{ve}
      \fmfright{vo}
      \fmftop{vn}
      \fmftop{vs}
      \fmffreeze
      \fmfforce{(-0.3w,0.5h)}{i}
      \fmfforce{(2w,0.5h)}{o1}
      \fmfforce{(2.3w,0.5h)}{o}	    
      \fmfforce{(0w,0.5h)}{ve}
      \fmfforce{(1.0w,0.5h)}{vo}
      \fmfforce{(.5w,0.95h)}{vn}
      \fmfforce{(.5w,0.05h)}{vs}
      \fmffreeze
      \fmf{plain}{i,ve}
      \fmf{plain,left=0.8}{ve,vo}
	    \fmf{phantom,left=0.5,label=\footnotesize{$\al_1$},l.d=-0.01w}{ve,vn}
	    \fmf{phantom,right=0.5,label=\footnotesize{$\al_2$},l.d=-0.01w}{vo,vn}
      \fmf{plain,left=0.8}{vo,ve}
	    \fmf{phantom,left=0.5,label=\footnotesize{$\al_3$},l.d=-0.01w}{vo,vs}
	    \fmf{phantom,right=0.5,label=\footnotesize{$\al_4$},l.d=-0.01w}{ve,vs}
	    \fmf{plain,label=\footnotesize{$\al_5$},l.d=0.05w}{vs,vn}
	    \fmf{plain,label=\footnotesize{$\tilde{\al}_2-\al_3$},l.d=0.1h}{vo,o1}
      \fmf{plain}{o1,o}
      \fmffreeze
      \fmfdot{ve,vn,vo,vs,o1}
    \end{fmfgraph*}
} \qquad \qquad \qquad \, ,
\label{addProp}
\eea
with the vertex of indices $\al_2$, $\al_3$ and $\tilde{\al}_2- \al_3$ being unique. We can therefore use
Eq.~(\ref{def:star-triangle}) from right to left in order to obtain: 
\bea
 \mbox{ rhs of Eq.~(\ref{addProp})} \quad  = \quad \frac{(4\pi)^{D/2}}{G(D,\tilde{\al}_2,\tilde{\gamma})} \qquad
 \parbox{18mm}{
    \begin{fmfgraph*}(18,16)
     \fmfleft{i}
      \fmfright{o}
      \fmfleft{ve}
      \fmfright{vo}
      \fmftop{vn}
      \fmftop{vs}
      \fmffreeze
      \fmfforce{(-0.3w,0.5h)}{i}
      \fmfforce{(1.3w,0.5h)}{o}
      \fmfforce{(0w,0.5h)}{ve}
      \fmfforce{(1.0w,0.5h)}{vo}
      \fmfforce{(.5w,0.95h)}{vn}
      \fmfforce{(.5w,0.05h)}{vs}
      \fmffreeze
      \fmf{plain}{i,ve}
      \fmf{plain,left=0.8}{ve,vo}
	    \fmf{phantom,left=0.7,label=\footnotesize{$\al_1$},l.d=-0.0w}{ve,vn}
	    \fmf{phantom,right=0.7,label=\footnotesize{$\tilde{\al}_3$},l.d=-0.0w}{vo,vn}
      \fmf{plain,left=0.8}{vo,ve}
	    \fmf{phantom,left=0.7,label=\footnotesize{$\tilde{\al}_2$},l.d=-0.0w}{vo,vs}
	    \fmf{phantom,right=0.7,label=\footnotesize{$\al_4$},l.d=-0.0w}{ve,vs}
	    \fmf{plain,left=0.3,label=\footnotesize{$\al_5$},l.d=0.05w}{vs,vn}
	    \fmf{plain,left=0.3,label=\footnotesize{$\tilde{\gamma}$},l.d=0.05w}{vn,vs}
      \fmf{plain}{vo,o}
      \fmffreeze
      \fmfdot{ve,vn,vo,vs}
    \end{fmfgraph*}
}\qquad \, ,
\eea
where $\tilde{\gamma} = D/2 - \tilde{\al}_2+ \al_3=\al_2+ \al_3 $.
Finally, calculating the internal loop, yields:
%
\fleq{
 \frac{p^{-2\delta}}{(4\pi)^{D}} \, C_D[J(D,p,\al_1,\al_2,\al_3,\al_4,\al_5)]~\frac{1}{p^{2(\tilde{\al}_2- \al_3)}} ~ = 
~   \frac{G(D,\al_5,\al_2+\al_3)}{G(D,\tilde{\al}_2,\al_2+\al_3)}
\qquad
\parbox{16mm}{
    \begin{fmfgraph*}(16,14)
      \fmfleft{i}
      \fmfright{o}
      \fmfleft{ve}
      \fmfright{vo}
      \fmftop{vn}
      \fmftop{vs}
      \fmffreeze
      \fmfforce{(-0.3w,0.5h)}{i}
      \fmfforce{(1.3w,0.5h)}{o}
      \fmfforce{(0w,0.5h)}{ve}
      \fmfforce{(1.0w,0.5h)}{vo}
      \fmfforce{(.5w,0.95h)}{vn}
      \fmfforce{(.5w,0.05h)}{vs}
      \fmffreeze
      \fmf{plain}{i,ve}
      \fmf{plain,left=0.8}{ve,vo}
	    \fmf{phantom,left=0.5,label=\footnotesize{$\al_1$},l.d=-0.01w}{ve,vn}
	    \fmf{phantom,right=0.5,label=\footnotesize{$\tilde{\al}_3$},l.d=-0.01w}{vo,vn}
      \fmf{plain,left=0.8}{vo,ve}
	    \fmf{phantom,left=0.5,label=\footnotesize{$\tilde{\al}_2$},l.d=-0.01w}{vo,vs}
	    \fmf{phantom,right=0.5,label=\footnotesize{$\al_4$},l.d=-0.01w}{ve,vs}
	    \fmf{plain,label=\footnotesize{$\beta$},l.d=0.05w}{vs,vn}
      \fmf{plain}{vo,o}
      \fmffreeze
      \fmfdot{ve,vn,vo,vs}
    \end{fmfgraph*}
} \qquad \, ,
\label{addProp1}
}
with $\beta = t_2-D/2 \equiv \overline{t}_2$ (see the previous subsection). Hence, for the corresponding coefficient function, we have:
\fleq{
C_D[J(D,p,\al_1,\al_2,\al_3,\al_4,\al_5)] = a(\al_2)a(\al_3)a(\al_5)a(D-t_2)\,C_D[J(D,p,\al_1,\tilde{\al}_3,\tilde{\al}_2,\al_4,\overline{t}_2)]\, .
\nonum
}

In a sense, the procedure of adding an additional propagator 
can be considered as the inverse of the transform given in the previous subsection, see Eq.~(\ref{GI1}). 
From Eq.~(\ref{addProp1}), we can then see that the left triangle becomes the ordinary one when $\al_1=\al_4=\overline{t}_2=1$, \ie, $\al_1=\al_4=1$ and $t_2=1+D/2$ 
(so the sum of the indices $\al_2$, $\al_3$ and $\al_5$ is equal to $D/2+1$).
Moreover, the right triangle becomes the ordinary one when  $\tilde{\al}_2=\tilde{\al}_3=\overline{t}_2=1$, \ie, $\al_2=\al_3=D/2-1=1-\ep$
and $\al_5=3-D/2=1+\ep$.
In all these cases, the diagrams may again be evaluated with the help of
the IBP relations and expressed in terms of $\Gamma$-functions in agreement with subsection \ref{sec:meth:uniquenessApp}.

Now let's add an additional line of index $\tilde{\al}_1 + \tilde{\al}_2$ to both sides of Eq.~(\ref{Dia}). This yields: 
%
\fleq{
&\qquad \nonum \\
&\frac{p^{-2\,\overline{v}_2}}{(4\pi)^{3D/2}}\,C_D[J(D,p,\al_1,\al_2,\al_3,\al_4,\al_5)]\,G(D,\delta,\tilde{\al}_1 + \tilde{\al}_2) 
\quad = \qquad
\parbox{16mm}{
    \begin{fmfgraph*}(16,14)
      \fmfleft{i}
      \fmfright{o}
      \fmfleft{ve}
      \fmfright{vo}
      \fmftop{vn}
      \fmftop{vs}
      \fmffreeze
      \fmfforce{(-0.3w,0.5h)}{i}
      \fmfforce{(1.3w,0.5h)}{o}
      \fmfforce{(0w,0.5h)}{ve}
      \fmfforce{(1.0w,0.5h)}{vo}
      \fmfforce{(.5w,0.95h)}{vn}
      \fmfforce{(.5w,0.05h)}{vs}
      \fmffreeze
      \fmf{plain}{i,ve}
      \fmf{plain,left=0.8}{ve,vo}
	    \fmf{phantom,left=0.5,label=\footnotesize{$\al_1$},l.d=-0.01w}{ve,vn}
	    \fmf{phantom,right=0.5,label=\footnotesize{$\al_2$},l.d=-0.01w}{vo,vn}
      \fmf{plain,left=0.8}{vo,ve}
	    \fmf{phantom,left=0.5,label=\footnotesize{$\al_3$},l.d=-0.01w}{vo,vs}
	    \fmf{phantom,right=0.5,label=\footnotesize{$\al_4$},l.d=-0.01w}{ve,vs}
	    \fmf{plain,label=\footnotesize{$\al_5$},l.d=0.05w}{vs,vn}
      \fmf{plain}{vo,o}
	    \fmf{phantom,left=0.8,label=\footnotesize{$\tilde{\al}_1+\tilde{\al}_2$}}{ve,vo}    
      \fmffreeze
      \fmfdot{ve,vn,vo,vs}
      \fmfposition
\fmfi{plain}{vloc(__vo) ..controls vloc(__o) and (xpart(vloc(__o)),1.5h) ..(xpart(vloc(__vn)),1.5h)}
\fmfi{plain}{vloc(__ve) ..controls vloc(__i) and (xpart(vloc(__i)),1.5h) ..(xpart(vloc(__vn)),1.5h)}
    \end{fmfgraph*}
} \qquad \, .
}
%
The upper triangle is unique which leads to the replacement of the rhs by the diagram:
\bea
\frac{1}{(4\pi)^{D/2}} \, G(D,\al_1,\al_2) \qquad
\parbox{16mm}{
    \begin{fmfgraph*}(16,14)
      \fmfleft{i}
      \fmfright{o}
      \fmfleft{ve}
      \fmfright{vo}
      \fmftop{vn}
      \fmftop{vs}
      \fmffreeze
      \fmfforce{(-0.3w,0.5h)}{i}
      \fmfforce{(1.3w,0.5h)}{o}
      \fmfforce{(0w,0.5h)}{ve}
      \fmfforce{(1.0w,0.5h)}{vo}
      \fmfforce{(.5w,0.95h)}{vn}
      \fmfforce{(.5w,0.05h)}{vs}
      \fmffreeze
      \fmf{plain}{i,ve}
      \fmf{plain,left=0.8}{ve,vo}
	    \fmf{phantom,left=0.5,label=\footnotesize{$\tilde{\al}_2$},l.d=-0.01w}{ve,vn}
	    \fmf{phantom,right=0.5,label=\footnotesize{$\tilde{\al}_1$},l.d=-0.01w}{vo,vn}
      \fmf{plain,left=0.8}{vo,ve}
	    \fmf{phantom,left=0.5,label=\footnotesize{$\al_3$},l.d=-0.01w}{vo,vs}
	    \fmf{phantom,right=0.5,label=\footnotesize{$\al_4$},l.d=-0.01w}{ve,vs}
	    \fmf{plain,label=\footnotesize{$\overline{v}_1$},l.d=0.05w}{vs,vn}
      \fmf{plain}{vo,o}
      \fmffreeze
      \fmfdot{ve,vn,vo,vs}
    \end{fmfgraph*}
} \qquad \, .
\eea
This finally leads to the identity:
%
\fleq{
C_D[J(D,p,\al_1,\al_2,\al_3,\al_4,\al_5)] = \frac{ G(\al_1,\al_2)}{G(\delta,\tilde{\al}_1+ \tilde{\al}_2)} \,
C_D[J(D,p,\tilde{\al}_2,\tilde{\al}_1,\al_3,\al_4,\overline{v}_1)] \, ,
\nonum
}
and
\fleq{
  C_D[J(D,p,\al_1,\al_2,\al_3,\al_4,\al_5)] = a(\al_1)a(\al_2)a(\overline{v}_2)a(3D/2-\sum_{i=1}^{5}\al_i)\,C_D[J(D,p,\tilde{\al}_2,\tilde{\al}_1,
    \al_3,\al_4, \overline{v}_1)] \, .
\nonum
}
From these results, we observe that the left triangle becomes an ordinary one for
$\tilde{\al}_2=\al_4=\overline{v}_1=1$, \ie, $\al_4=1$, $\al_2=D/2-1=1-\ep$ and $\al_1+\al_5=2$. 
Moreover, the right triangle becomes an ordinary one for  $\tilde{\al}_1=\al_3=\overline{v}_1=1$, \ie, $\al_3=1$, $\al_1=D/2-1=1-\ep$
and $\al_2+\al_5=2$.
 Once again, in all these cases, the final results may be evaluated with the help of
the IBP relations and expressed in terms of $\Gamma$-functions in agreement with subsection \ref{sec:meth:uniquenessApp}.

Let's add that similar transformations but for diagrams containing lines with arrows (see definition of such line in (\ref{def:lines})
can be found in \cite{Kotikov:1987mw}.\footnote{The journal version of \cite{Kotikov:1987mw} contains mostly the formulas without graphics. The corresponding graphical
    representations can be found in the preprint version of \cite{Kotikov:1987mw} (see, for example, the corresponding KEK scanned document).}

\subsubsection{Conformal transform of the inversion}
\label{sec:meth:conformal}

In order to implement this transformation, we perform the inversion:
\bea
k_i^{\mu} \to \frac{k_i^{\mu}}{k_i^{2}}\, ,
\label{inv}
\eea
of the integration variables and the external momenta. So, we have:
\bea
 k_i^{2} \to \frac{1}{k_i^{2}}, \qquad (k_i-k_j)^{2} \to \frac{(k_i-k_j)^{2}}{k_i^{2}k_j^2}, \qquad \D^D k_i \to \frac{\D^D k_i}{k_i^{2D}}\, ,
\label{inv1}
\eea
and, thus,
%
\fleq{	
{\textstyle \int  \frac{[\D^D k_1] [\D^D k_2]}{k_1^{2\al_1}k_2^{2\al_2}(k_1-k_2)^{2\al_5}(p-k_1)^{2\al_4}(p-k_2)^{2\al_3}}  \quad \to  \quad
\int [\D^D k_1] [\D^D k_2]  \frac{k_1^{2\al_1}\,k_2^{2\al_2}\, (k_1^2k_2^2)^{2\al_5}\,(p^2k_1^2)^{2\al_4}\,(p^2k_2^2)^{2\al_3}}{k_1^{2D}k_2^{2D}(k_1-k_2)^{2\al_5}(p-k_1)^{2\al_4}(p-k_2)^{2\al_3}} \, ,}
\label{inv12}
}
%
or, graphically:
%
\bea
\parbox{16mm}{
    \begin{fmfgraph*}(16,14)
      \fmfleft{i}
      \fmfright{o}
      \fmfleft{ve}
      \fmfright{vo}
      \fmftop{vn}
      \fmftop{vs}
      \fmffreeze
      \fmfforce{(-0.3w,0.5h)}{i}
      \fmfforce{(1.3w,0.5h)}{o}
      \fmfforce{(0w,0.5h)}{ve}
      \fmfforce{(1.0w,0.5h)}{vo}
      \fmfforce{(.5w,0.95h)}{vn}
      \fmfforce{(.5w,0.05h)}{vs}
      \fmffreeze
      \fmf{plain}{i,ve}
      \fmf{plain,left=0.8}{ve,vo}
	    \fmf{phantom,left=0.5,label=\footnotesize{$\al_1$},l.d=-0.01w}{ve,vn}
	    \fmf{phantom,right=0.5,label=\footnotesize{$\al_2$},l.d=-0.01w}{vo,vn}
      \fmf{plain,left=0.8}{vo,ve}
	    \fmf{phantom,left=0.5,label=\footnotesize{$\al_3$},l.d=-0.01w}{vo,vs}
	    \fmf{phantom,right=0.5,label=\footnotesize{$\al_4$},l.d=-0.01w}{ve,vs}
	    \fmf{plain,label=\footnotesize{$\al_5$},l.d=0.05w}{vs,vn}
      \fmf{plain}{vo,o}
      \fmffreeze
      \fmfdot{ve,vn,vo,vs}
    \end{fmfgraph*}
} \qquad \qquad \to \qquad \qquad
\parbox{16mm}{
    \begin{fmfgraph*}(16,14)
      \fmfleft{i}
      \fmfright{o}
      \fmfleft{ve}
      \fmfright{vo}
      \fmftop{vn}
      \fmftop{vs}
      \fmffreeze
      \fmfforce{(-0.3w,0.5h)}{i}
      \fmfforce{(1.3w,0.5h)}{o}
      \fmfforce{(0w,0.5h)}{ve}
      \fmfforce{(1.0w,0.5h)}{vo}
      \fmfforce{(.5w,0.95h)}{vn}
      \fmfforce{(.5w,0.05h)}{vs}
      \fmffreeze
      \fmf{plain}{i,ve}
      \fmf{plain,left=0.8}{ve,vo}
	    \fmf{phantom,left=0.5,label=\footnotesize{$D-t_1$},l.d=-0.01w}{ve,vn}
	    \fmf{phantom,right=0.5,label=\footnotesize{$D-t_2$},l.d=-0.01w}{vo,vn}
      \fmf{plain,left=0.8}{vo,ve}
	    \fmf{phantom,left=0.5,label=\footnotesize{$\al_3$},l.d=-0.01w}{vo,vs}
	    \fmf{phantom,right=0.5,label=\footnotesize{$\al_4$},l.d=-0.01w}{ve,vs}
	    \fmf{plain,label=\footnotesize{$\al_5$},l.d=0.05w}{vs,vn}
      \fmf{plain}{vo,o}
      \fmffreeze
      \fmfdot{ve,vn,vo,vs}
    \end{fmfgraph*}
} \qquad  \, .
\eea
The last result can be rewritten as:
\bea
C_D[J(D,p,\al_1,\al_2,\al_3,\al_4,\al_5)] \underset{(\text{In})}{=}
C_D[J(D,p,D-t_1,
D-t_2,
\al_3,\al_4,\al_5)] \, .
\nonum
\eea
This equation shows that the left triangle becomes an ordinary one for
$\al_4=\al_5=D-t_1=1$, \ie, $\al_4=\al_5=1$ and $\al_1=D-3=1-2\ep$. 
 Moreover, the right triangle becomes an ordinary one for  $\al_3=\al_5=D-t_2=1$, \ie, $\al_3=\al_5=1$ and $\al_2=D-3=1-2\ep$.
 In these cases, the final results may be evaluated with the help of
the IBP relations and expressed in terms of $\Gamma$-functions in agreement with subsection \ref{sec:meth:uniquenessApp}.

\subsection{Fourier transform and duality}
\label{sec:meth:FT+Du}

Up to now, all diagrams were expressed in momentum space. Useful identities can be obtained by relating $p$-space and $x$-space diagrams.
Following \cite{Kazakov:1984bw}, we briefly present them in this paragraph.

Recall, from Eq.~(\ref{def:two-loop-p-int}), that the 2-loop massless propagator-type diagram
in $p$-space was defined as:
\be
J(D,p,\{\al_i\}) = \int \frac{[\D^Dk][\D^Dq]}{[(k-p)^2]^{\al_1}[(q-p)^2]^{\al_2}[q^2]^{\al_3}[k^2]^{\al_4}[(k-q)^2]^{\al_5}}
\quad = \qquad
\parbox{16mm}{
    \begin{fmfgraph*}(16,14)
      \fmfleft{i}
      \fmfright{o}
      \fmfleft{ve}
      \fmfright{vo}
      \fmftop{vn}
      \fmftop{vs}
      \fmffreeze
      \fmfforce{(-0.3w,0.5h)}{i}
      \fmfforce{(1.3w,0.5h)}{o}
      \fmfforce{(0w,0.5h)}{ve}
      \fmfforce{(1.0w,0.5h)}{vo}
      \fmfforce{(.5w,0.95h)}{vn}
      \fmfforce{(.5w,0.05h)}{vs}
      \fmffreeze
	    \fmf{fermion,label=\footnotesize{$p$}}{i,ve}
      \fmf{plain,left=0.8}{ve,vo}
	    \fmf{phantom,left=0.7,label=\footnotesize{$\al_1$},l.d=-0.1w}{ve,vn}
	    \fmf{phantom,right=0.7,label=\footnotesize{$\al_2$},l.d=-0.1w}{vo,vn}
      \fmf{plain,left=0.8}{vo,ve}
	    \fmf{phantom,left=0.7,label=\footnotesize{$\al_3$},l.d=-0.1w}{vo,vs}
	    \fmf{phantom,right=0.7,label=\footnotesize{$\al_4$},l.d=-0.1w}{ve,vs}
	    \fmf{plain,label=\footnotesize{$\al_5$},l.d=0.05w}{vs,vn}
      \fmf{plain}{vo,o}
      \fmffreeze
      \fmfdot{ve,vn,vo,vs}
    \end{fmfgraph*}
} \qquad \,.
\label{def:two-loop-p-int2}
\ee
Equivalently, all calculations may be done in position space. In $x$-space, the 2-loop massless propagator-type diagram
is defined as:
\be
J(D,z,\{\overline{\al}_i\}) = \int \frac{[\D^D x][\D^D y]}{[y^2]^{\overline{\al}_1}[(y-z)^2]^{\overline{\al}_2}[(z-x)^2]^{\overline{\al}_3}[x^2]^{\overline{\al}_4}[(x-y)^2]^{\overline{\al}_5}}
\quad = \qquad
\parbox{16mm}{
    \begin{fmfgraph*}(16,14)
      \fmfleft{i}
      \fmfright{o}
      \fmfleft{ve}
      \fmfright{vo}
      \fmftop{vn}
      \fmftop{vs}
      \fmftop{vt}
      \fmfbottom{vb}		
      \fmffreeze
      \fmfforce{(-0.3w,0.5h)}{i}
      \fmfforce{(1.3w,0.5h)}{o}
      \fmfforce{(0w,0.5h)}{ve}
      \fmfforce{(1.0w,0.5h)}{vo}
      \fmfforce{(.5w,0.95h)}{vn}
      \fmfforce{(.5w,0.05h)}{vs}
      \fmfforce{(.5w,1.25h)}{vt}
      \fmfforce{(.5w,-0.3h)}{vb}
      \fmffreeze
	    \fmf{plain,label=\footnotesize{$0$},l.s=left}{i,ve}
      \fmf{plain,left=0.8}{ve,vo}
	    \fmf{phantom,left=0.7,label=\footnotesize{$\overline{\al}_1$},l.d=-0.1w}{ve,vn}
	    \fmf{phantom,right=0.7,label=\footnotesize{$\overline{\al}_2$},l.d=-0.1w}{vo,vn}
      \fmf{plain,left=0.8}{vo,ve}
	    \fmf{phantom,left=0.7,label=\footnotesize{$\overline{\al}_3$},l.d=-0.1w}{vo,vs}
	    \fmf{phantom,right=0.7,label=\footnotesize{$\overline{\al}_4$},l.d=-0.1w}{ve,vs}
	    \fmf{plain,label=\footnotesize{$\overline{\al}_5$},l.d=0.05w}{vs,vn}
	    \fmf{plain,label=\footnotesize{$z$},l.s=left}{vo,o}
	    \fmf{phantom,label=\footnotesize{$y$},l.d=-0.1w}{vn,vt}
	    \fmf{phantom,label=\footnotesize{$x$},l.d=-0.1w}{vs,vb}
      \fmffreeze
      \fmfdot{ve,vn,vo,vs}
    \end{fmfgraph*}
} \qquad \, ,
\label{def:two-loop-p-int-x}
\ee
where $0$ denotes the so-called ``root vertex'' and the $\overline{\al}_i$ are arbitrary indices. 
It is actually straightforward to show that the  $p$-space and $x$-space diagrams are related provided that $\overline{\al}_i=\tilde{\al}_i$ where $\tilde{\al} = D/2 -\al$ 
is the index which is dual (in the sense of Fourier transform) to $\al$. This follows from the Fourier transform, Eq.~(\ref{FT-inv-line}), that we reproduce here for clarity:
\be
\frac{1}{[k^2]^{\al}} = \frac{a(\al)}{(2\pi)^{D/2}}\,\int \D^D x \,\frac{e^{\I k x}}{[x^2]^{D/2-\al}}, \qquad a(\al) = \frac{\Gamma(D/2-\al)}{\Gamma(\al)}\, .
\ee
Hence, for a given diagram, the Fourier transform allows to relate its $p$-space and $x$-space coefficient functions and the relation reads:  
\be
\text{C}_D[\,J(D,p,\al_1,\al_2,\al_3,\al_4,\al_5)\,] \underset{(\text{FT})}{=}
\frac{\prod_{j=1}^5\,a(\al_j)}{a\left(\sum_{j=1}^5\al_j-D\right)}\,\text{C}_D[\,J(D,z,\tilde{\al}_1,\tilde{\al}_2,\tilde{\al}_3,\tilde{\al}_4,\tilde{\al}_5)\,].
\label{FT-inv-line2}
\ee
Graphically, this can be represented as:
\bea
\parbox{16mm}{
    \begin{fmfgraph*}(16,14)
      \fmfleft{ve}
      \fmfright{vo}
      \fmftop{vn}
      \fmftop{vs}
      \fmffreeze
      \fmfforce{(0w,0.5h)}{ve}
      \fmfforce{(1.0w,0.5h)}{vo}
      \fmfforce{(.5w,0.95h)}{vn}
      \fmfforce{(.5w,0.05h)}{vs}
      \fmffreeze
      \fmf{plain,left=0.8}{ve,vo}
	    \fmf{phantom,left=0.7,label=\footnotesize{$\al_1$},l.d=-0.1w}{ve,vn}
	    \fmf{phantom,right=0.7,label=\footnotesize{$\al_2$},l.d=-0.1w}{vo,vn}
      \fmf{plain,left=0.8}{vo,ve}
	    \fmf{phantom,left=0.7,label=\footnotesize{$\al_3$},l.d=-0.1w}{vo,vs}
	    \fmf{phantom,right=0.7,label=\footnotesize{$\al_4$},l.d=-0.1w}{ve,vs}
	    \fmf{plain,label=\footnotesize{$\al_5$},l.d=0.05w}{vs,vn}
      \fmffreeze
      \fmfdot{ve,vn,vo,vs}
    \end{fmfgraph*}
} \quad = \quad \frac{\prod_{j=1}^5\,a(\al_j)}{a\left(\sum_{j=1}^5\al_j-D\right)}\,\qquad
\parbox{16mm}{
    \begin{fmfgraph*}(16,14)
      \fmfleft{i}
      \fmfright{o}
      \fmfleft{ve}
      \fmfright{vo}
      \fmftop{vn}
      \fmftop{vs}
      \fmffreeze
      \fmfforce{(-0.3w,0.5h)}{i}
      \fmfforce{(1.3w,0.5h)}{o}
      \fmfforce{(0w,0.5h)}{ve}
      \fmfforce{(1.0w,0.5h)}{vo}
      \fmfforce{(.5w,0.95h)}{vn}
      \fmfforce{(.5w,0.05h)}{vs}
      \fmffreeze
	    \fmf{phantom,label=\footnotesize{$0$},l.d=-0.1w,l.s=left}{i,ve}
      \fmf{plain,left=0.8}{ve,vo}
	    \fmf{phantom,left=0.7,label=\footnotesize{$\tilde{\al}_1$},l.d=-0.1w}{ve,vn}
	    \fmf{phantom,right=0.7,label=\footnotesize{$\tilde{\al}_2$},l.d=-0.1w}{vo,vn}
      \fmf{plain,left=0.8}{vo,ve}
	    \fmf{phantom,left=0.7,label=\footnotesize{$\tilde{\al}_3$},l.d=-0.1w}{vo,vs}
	    \fmf{phantom,right=0.7,label=\footnotesize{$\tilde{\al}_4$},l.d=-0.1w}{ve,vs}
	    \fmf{plain,label=\footnotesize{$\tilde{\al}_5$},l.d=0.05w}{vs,vn}
	    \fmf{phantom,label=\footnotesize{$z$},l.d=-0.1w,l.s=left}{vo,o}
      \fmffreeze
      \fmfdot{ve,vn,vo,vs}
    \end{fmfgraph*}
} \qquad \, ,
\eea
where all external legs were amputated as the diagrams correspond to coefficient functions
but we have explicitly indicated the location of the external vertices in the $x$-space function to distinguish it from its $p$-space counterpart.

Another useful transformation is the so-called duality transformation. It is based on the fact that the loop momenta are dummy integration variables.
They can therefore be replaced by dummy coordinate integration variables. Such an innocent looking change of
variables yields a dual diagram with some indices exchanged with respect to the original diagram. At the level of coefficient functions, the relation is given by:
\bea
\text{C}_D[\,J(D,p,\al_1,\al_2,\al_3,\al_4,\al_5)\,] & \underset{(\text{Du})}{=} & \text{C}_D[\,J(D,z,\al_1,\al_4,\al_3,\al_2,\al_5)\,]
\label{duality-def} \\
& \underset{(\text{Du})}{=} & \text{C}_D[\,J(D,z,\al_3,\al_2,\al_1,\al_4,\al_5)\,]
\nonum \\
& \underset{(\text{Du})}{=} & \text{C}_D[\,J(D,z,\al_2,\al_3,\al_4,\al_1,\al_5)\,].
\nonum
\eea
Notice that in the first line, the duality transformation exchanges indices 2 and 4. The two other equalities follow from the symmetries of the diagram
($1 \leftrightarrow 2$, $3 \leftrightarrow 4$ and $1 \leftrightarrow 4$, $2 \leftrightarrow 3$).

For later purposes, let's add that, for an arbitrary planar propagator-type diagram, the dual one can be constructed by 
first putting a point in every loop of the diagram and two points outside of it. Then, one has to connect all the points by lines such that every line of the initial diagram is crossed once. 
The new lines produce the dual diagram with indices equal~\footnote{The duality transformation
  defined here follows from Kotikov~\cite{Kotikov:1987mw,Kotikov:1995cw} and  Kazakov and Kotikov~(\cite{Kazakov:1986mu}) and differs from
  the duality transformation
considered by Kazakov~\cite{Kazakov:1984bw} which corresponds to duality plus Fourier transform, see Eq.~(\ref{Fourier+duality-def})} to that of the crossed
old lines. This is demonstrated by considering the following cases:
\begin{subequations}
\bea
&& \qquad \nonum \\
&& \parbox{16mm}{
    \begin{fmfgraph*}(16,14)
      \fmfleft{i}
      \fmfright{o}
      \fmfleft{ve}
      \fmfright{vo}
      \fmftop{vn}
      \fmftop{vs}
      \fmffreeze
      \fmfforce{(-0.2w,0.5h)}{i}
      \fmfforce{(1.2w,0.5h)}{o}
      \fmfforce{(0w,0.5h)}{ve}
      \fmfforce{(1.0w,0.5h)}{vo}
      \fmfforce{(.3w,0.9h)}{v1}
      \fmfforce{(.7w,0.9h)}{v2}
      \fmfforce{(.5w,0.05h)}{vs}
      \fmffreeze
      \fmf{plain}{i,ve}
      \fmf{plain}{vo,o}
      \fmf{plain,left=0.8}{ve,vo}
	    \fmf{phantom,left=0.5,label=\footnotesize{$\al_i$},l.d=-0.01w}{ve,v1}
      \fmf{plain,left=0.8}{vo,ve}
      \fmf{plain}{vs,v1}
      \fmf{plain}{vs,v2}
      \fmffreeze
      \fmfdot{ve,v1,v2,vo,vs}
    \end{fmfgraph*}
}
\qquad \Rightarrow \qquad
\parbox{16mm}{
    \begin{fmfgraph*}(16,14)
      \fmfleft{i}
      \fmfright{o}
      \fmfleft{ve}
      \fmfright{vo}
      \fmftop{vn}
      \fmftop{vs}
      \fmffreeze
      \fmfforce{(-0.2w,0.5h)}{i}
      \fmfforce{(1.2w,0.5h)}{o}
      \fmfforce{(0w,0.5h)}{ve}
      \fmfforce{(1.0w,0.5h)}{vo}
      \fmfforce{(.2w,0.5h)}{vde}
      \fmfforce{(.8w,0.5h)}{vdo}
      \fmfforce{(.5w,0.7h)}{vdc}
      \fmfforce{(.5w,1.3h)}{vdn}
      \fmfforce{(.5w,-0.3h)}{vds}
      \fmfforce{(.3w,0.9h)}{v1}
      \fmfforce{(.7w,0.9h)}{v2}
      \fmfforce{(.5w,0.05h)}{vs}
      \fmffreeze
      \fmf{plain}{i,ve}
      \fmf{plain}{vo,o}
      \fmf{plain,left=0.8}{ve,vo}
      \fmf{plain,left=0.8}{vo,ve}
      \fmf{plain}{vs,v1}
      \fmf{plain}{vs,v2}
      \fmf{dashes}{vde,vdc}
      \fmf{dashes}{vdo,vdc}
      \fmf{dashes}{vdn,vdc}
      \fmf{dashes}{vds,vde}
      \fmf{dashes}{vds,vdo}
      \fmf{dashes,left=0.8}{vde,vdn}
      \fmf{dashes,right=0.8}{vdo,vdn}
      \fmffreeze
      \fmfdot{ve,v1,v2,vo,vs}
      \fmfdot{vde,vdo,vdc,vdn,vds}
    \end{fmfgraph*}
}
\qquad \Rightarrow \qquad
\parbox{16mm}{
    \begin{fmfgraph*}(16,14)
      \fmftop{dt}
      \fmftop{db}
      \fmfleft{vde}
      \fmfright{vdo}
      \fmftop{vdn}
      \fmftop{vds}
      \fmffreeze
      \fmfforce{(.5w,1.2h)}{dt}
      \fmfforce{(.5w,-0.2h)}{db}
      \fmfforce{(.0w,0.5h)}{vde}
      \fmfforce{(1.0w,0.5h)}{vdo}
      \fmfforce{(.5w,0.5h)}{vdc}
      \fmfforce{(.5w,1.0h)}{vdn}
      \fmfforce{(.5w,0.0h)}{vds}
      \fmffreeze
      \fmf{plain}{vdn,dt}
      \fmf{plain}{db,vds}
      \fmf{plain}{vde,vdc}
      \fmf{plain}{vdo,vdc}
      \fmf{plain}{vdn,vdc}
      \fmf{plain,left=0.9}{vde,vdo}
      \fmf{plain,right=0.9}{vde,vdo}
	    \fmf{phantom,left=0.5,label=\footnotesize{$\al_i$},l.d=-0.01w}{vde,vdn}
      \fmffreeze
      \fmfdot{vde,vdo,vdc,vdn,vds}
    \end{fmfgraph*}
} \qquad \, ,
\\
&& \qquad \nonum \\
&& \qquad \nonum \\
&& \qquad \nonum \\
&&\parbox{16mm}{
    \begin{fmfgraph*}(16,14)
      \fmfleft{i}
      \fmfright{o}
      \fmfleft{ve}
      \fmfright{vo}
      \fmffreeze
      \fmfforce{(-0.2w,0.5h)}{i}
      \fmfforce{(1.2w,0.5h)}{o}
      \fmfforce{(0w,0.5h)}{ve}
      \fmfforce{(1.0w,0.5h)}{vo}
      \fmfforce{(.3w,0.9h)}{vn1}
      \fmfforce{(.7w,0.9h)}{vn2}
      \fmfforce{(.3w,0.1h)}{vs1}
      \fmfforce{(.7w,0.1h)}{vs2}
      \fmffreeze
      \fmf{plain}{i,ve}
      \fmf{plain}{vo,o}
      \fmf{plain,left=0.8}{ve,vo}
	    \fmf{phantom,left=0.5,label=\footnotesize{$\al_i$},l.d=-0.01w}{ve,vn1}
      \fmf{plain,left=0.8}{vo,ve}
      \fmf{plain}{vs1,vn1}
      \fmf{plain}{vs2,vn2}
      \fmffreeze
      \fmfdot{ve,vs1,vs2,vn1,vn2,vo}
    \end{fmfgraph*}
}
\qquad \Rightarrow \qquad
\parbox{16mm}{
    \begin{fmfgraph*}(16,14)
      \fmfleft{i}
      \fmfright{o}
      \fmfleft{ve}
      \fmfright{vo}
      \fmftop{vdn}
      \fmftop{vds}
      \fmftop{vde}
      \fmftop{vdo}
      \fmftop{vdc}
      \fmffreeze
      \fmfforce{(-0.2w,0.5h)}{i}
      \fmfforce{(1.2w,0.5h)}{o}
      \fmfforce{(0w,0.5h)}{ve}
      \fmfforce{(1.0w,0.5h)}{vo}
      \fmfforce{(.3w,0.9h)}{vn1}
      \fmfforce{(.7w,0.9h)}{vn2}
      \fmfforce{(.3w,0.1h)}{vs1}
      \fmfforce{(.7w,0.1h)}{vs2}
      \fmfforce{(.2w,0.5h)}{vde}
      \fmfforce{(.5w,0.5h)}{vdc}
      \fmfforce{(.8w,0.5h)}{vdo}
      \fmfforce{(.5w,1.2h)}{vdn}
      \fmfforce{(.5w,-0.2h)}{vds}
      \fmffreeze
      \fmf{plain}{i,ve}
      \fmf{plain}{vo,o}
      \fmf{plain,left=0.8}{ve,vo}
      \fmf{plain,left=0.8}{vo,ve}
      \fmf{plain}{vs1,vn1}
      \fmf{plain}{vs2,vn2}
      \fmf{dashes}{vde,vdo}
      \fmf{dashes}{vds,vdn}
      \fmf{dashes,left}{vde,vdn,vdo}
      \fmf{dashes,right}{vde,vds,vdo}
      \fmffreeze
      \fmfdot{ve,vs1,vs2,vn1,vn2,vo}
      \fmfdot{vde,vds,vdo,vdn,vdc}
    \end{fmfgraph*}
}
\qquad \Rightarrow \qquad
\parbox{16mm}{
    \begin{fmfgraph*}(16,14)
      \fmftop{dt}
      \fmfright{db}
      \fmftop{vdn}
      \fmftop{vds}
      \fmftop{vde}
      \fmftop{vdo}
      \fmftop{vdc}
      \fmffreeze
      \fmfforce{(0.5w,1.2h)}{dt}
      \fmfforce{(.5w,-0.2h)}{db}
      \fmfforce{(.0w,0.5h)}{vde}
      \fmfforce{(.5w,0.5h)}{vdc}
      \fmfforce{(1.0w,0.5h)}{vdo}
      \fmfforce{(.5w,1.0h)}{vdn}
      \fmfforce{(.5w,0.0h)}{vds}
      \fmffreeze
      \fmf{plain}{vdn,dt}
      \fmf{plain}{vds,db}
      \fmf{plain,left=0.9}{vde,vdo}
      \fmf{plain,left=0.9}{vdo,vde}
      \fmf{plain}{vde,vdo}
      \fmf{plain}{vds,vdn}
	    \fmf{phantom,left=0.5,label=\footnotesize{$\al_i$},l.d=-0.01w}{vde,vdn}
      \fmffreeze
      \fmfdot{vde,vds,vdo,vdn,vdc}
    \end{fmfgraph*}
} \qquad .
 \\
&& \qquad \nonum
\eea
\end{subequations}
These examples show that, in the general case, not only indices are changed but also the topology of the diagram.
Hence, the fact that the 2-loop p-type diagram preserves its topology under the duality transformation is a rather special case.

Both Fourier transform and duality transform relate diagrams which are in different spaces with different integration rules. By combining them,
it is possible to relate the coefficient functions of two $p$-space diagrams with changed indices:
\be
\text{C}_D[J(D,p,\al_1,\al_2,\al_3,\al_4,\al_5)] \underset{(\text{FT}+\text{Du})}{=} \frac{\Pi_i a(\al_i)}{a(\sum_i \al_i - D)}\,\text{C}_D[J(D,p,\tilde{\al}_2,\tilde{\al}_3,\tilde{\al}_4,\tilde{\al}_1,\tilde{\al}_5)].
\label{Fourier+duality-def}
\ee
Other similar transformations can be obtained from the symmetries of the diagram.

To conclude this section, it is in practice often convenient to perform calculations in dual $x$-space (see
  \cite{Kazakov:1986mu,Kazakov:1987jk,Kotikov:1987mw}), because
  the dual transform does not change the indices of the considered diagrams and does not introduce any additional coefficient
  in front of the diagram. Such type of calculations have been carried out \cite{Kazakov:1986mu,Kazakov:1987jk,Kotikov:1987mw} in order to evaluate
  two-loop four-point diagrams, in a special kinematics, which contribute to the cross-sections of the deep-inelastic scattering at parton level
  (see \cite{Kazakov:1987jk,Kazakov:1986vz}).
  Nowadays, the dual transform is very popular for the evaluation of the so-called conformal integrals (see \cite{Heslop:2018zut}
  and references therein).

  We would also like to note that it is convenient \cite{Kotikov:1990zs,Kotikov:1990zk}
  to use the dual transform even for diagrams having massive propagators, where the mass has formally inverse-mass dimensionality.
  The recent application of the dual transform to such type of diagrams can be found in the review \cite{Henn:2014yza}. 

\subsection{Functional equations}
\label{sec:functional}

Up to now, all calculations could be performed exactly due to the
fact that after some transformations, \eg, integration by parts,
uniqueness relation and transformation of indices, the diagrams on the
rhs could be reduced to sequences of chains and simple loops.
Using the results of Eqs.~(\ref{IBPex3}) and (\ref{IBPex4}) we are for example in a position 
to compute the expansion for the rather general integral corresponding to $\al_i=1+a_i \ep$ $(i=1,2,3,4)$ and $\al_5=1$.
However, sometimes, this cannot be achieved, especially when the number of
loops is large enough ($\geq 5$). 
 Actually, as discussed in previous sections, even the 2-loop massless propagator type diagram is beyond IBP and
 uniqueness for arbitrary values of indices. This is already the case for:
 $\al_i=1$ $(i=1,2,3,4)$ and $\al_5=1+a_5 \ep$ that we shall now proceed on investigating 
 with the help of the so-called functional equations \cite{Kazakov:1983pk,Kazakov:1984bw}.

Let's then consider the simplest non-trivial case previously discussed of a two-loop p-type 
diagram with an arbitrary index on the central line $I(\al_5) = J(D,p,1,1,1,1,\al_5)$ (see Fig.~\ref{fig:beyondtriangle}a):
\bea
\parbox{16mm}{
    \begin{fmfgraph*}(16,14)
      \fmfleft{i}
      \fmfright{o}
      \fmfleft{ve}
      \fmfright{vo}
      \fmftop{vn}
      \fmftop{vs}
      \fmffreeze
      \fmfforce{(-0.3w,0.5h)}{i}
      \fmfforce{(1.3w,0.5h)}{o}
      \fmfforce{(-0.2w,0.5h)}{ve}
      \fmfforce{(1.2w,0.5h)}{vo}
      \fmfforce{(.5w,0.95h)}{vn}
      \fmfforce{(.5w,0.05h)}{vs}
      \fmffreeze
      \fmf{plain}{i,ve}
      \fmf{plain,left=0.57}{ve,vo}
      \fmf{plain,left=0.57}{vo,ve}
            \fmf{plain,label=\footnotesize{$1+a$},l.d=0.05w}{vs,vn}
      \fmf{plain}{vo,o}
      \fmffreeze
      \fmfdot{ve,vn,vo,vs}
    \end{fmfgraph*}
} 
\qquad = \quad I(1+a)\, ,
\nonumber
\eea
where we have chosen $\al_5=1+a$.
We start by implementing the following sequence of transformations (here In denotes inversion transformation and AL the addition of a new loop): 
%
\fleq{
&\qquad \nonum \\
&\parbox{16mm}{
    \begin{fmfgraph*}(16,14)
      \fmfleft{i}
      \fmfright{o}
      \fmfleft{ve}
      \fmfright{vo}
      \fmftop{vn}
      \fmftop{vs}
      \fmffreeze
      \fmfforce{(-0.3w,0.5h)}{i}
      \fmfforce{(1.3w,0.5h)}{o}
      \fmfforce{(-0.2w,0.5h)}{ve}
      \fmfforce{(1.2w,0.5h)}{vo}
      \fmfforce{(.5w,0.95h)}{vn}
      \fmfforce{(.5w,0.05h)}{vs}
      \fmffreeze
      \fmf{plain}{i,ve}
      \fmf{plain,left=0.57}{ve,vo}
      \fmf{plain,left=0.57}{vo,ve}
            \fmf{plain,label=\footnotesize{$1+a$},l.d=0.05w}{vs,vn}
      \fmf{plain}{vo,o}
      \fmffreeze
      \fmfdot{ve,vn,vo,vs}
    \end{fmfgraph*}
}
\quad \quad \underset{(\text{In})}{=} \quad \quad
\parbox{16mm}{
    \begin{fmfgraph*}(16,14)
      \fmfleft{i}
      \fmfright{o}
      \fmfleft{ve}
      \fmfright{vo}
      \fmftop{vn}
      \fmftop{vs}
      \fmffreeze
      \fmfforce{(-0.3w,0.5h)}{i}
      \fmfforce{(1.3w,0.5h)}{o}
      \fmfforce{(-0.2w,0.5h)}{ve}
      \fmfforce{(1.2w,0.5h)}{vo}
      \fmfforce{(.5w,0.95h)}{vn}
      \fmfforce{(.5w,0.05h)}{vs}
      \fmffreeze
      \fmf{plain}{i,ve}
      \fmf{plain,left=0.57}{ve,vo}
            \fmf{phantom,left=0.5,label=\footnotesize{$\overline{\al}$},l.d=-0.01w}{ve,vn}
      \fmf{phantom,right=0.5,label=\footnotesize{$\overline{\al}$},l.d=-0.01w}{vo,vn}
      \fmf{plain,left=0.57}{vo,ve}
            \fmf{plain,label=\footnotesize{$1+a$},l.d=0.05w}{vs,vn}
      \fmf{plain}{vo,o}
      \fmffreeze
      \fmfdot{ve,vn,vo,vs}
    \end{fmfgraph*}
}  \quad \quad \underset{(\text{AL})}{=} \quad \quad
\parbox{16mm}{
    \begin{fmfgraph*}(16,14)
      \fmfleft{i}
      \fmfright{o}
      \fmfleft{ve}
      \fmfright{vo}
      \fmftop{vn}
      \fmftop{vs}
      \fmffreeze
      \fmfforce{(-0.1w,0.5h)}{i}
      \fmfforce{(1.1w,0.5h)}{o}
      \fmfforce{(0w,0.5h)}{ve}
      \fmfforce{(1.0w,0.5h)}{vo}
      \fmfforce{(.5w,0.95h)}{vn}
      \fmfforce{(.5w,0.05h)}{vs}
      \fmffreeze
      \fmf{plain}{i,ve}
      \fmf{plain,left=0.8}{ve,vo}
            \fmf{phantom,left=0.5,label=\footnotesize{$D/2-\overline{\al}$},l.d=-0.01w}{ve,vn}
            \fmf{phantom,right=0.5,label=\footnotesize{$D/2-\overline{\al}$},l.d=-0.01w}{vo,vn}
      \fmf{plain,left=0.8}{vo,ve}
            \fmf{plain,label=\footnotesize{$\overline{v}_1$},l.d=0.05w}{vs,vn}
      \fmf{plain}{vo,o}
      \fmffreeze
      \fmfdot{ve,vn,vo,vs}
    \end{fmfgraph*}
} \quad \quad \, ,
\nonumber \\
&\qquad \nonum
}
where $\overline{\al}=D-t_1=D-3-a$, $v_1=1+a+2\,\overline{\al}$ and, thus, $\overline{v}_1=v_1-D/2=1-3\ep-a$.
This yields the first equation for $I(1+a)$:
\bea
I(1+a) = I(1-a-3\ep)\, .
\label{Fa}
\eea
In order to get another equation, we start by applying integration by parts to the right triangle with the vertical distinguished line. This yields:
\bea
&&\qquad \nonum \\
&&I(1+a) \, (D-4 -2a) = 2 \left[ \quad
\parbox{16mm}{
    \begin{fmfgraph*}(16,14)
      \fmfleft{i}
      \fmfright{o}
      \fmfleft{ve}
      \fmfright{vo}
      \fmftop{vn}
      \fmftop{vs}
      \fmffreeze
      \fmfforce{(-0.1w,0.5h)}{i}
      \fmfforce{(1.1w,0.5h)}{o}
      \fmfforce{(0w,0.5h)}{ve}
      \fmfforce{(1.0w,0.5h)}{vo}
      \fmfforce{(.5w,0.95h)}{vn}
      \fmfforce{(.5w,0.05h)}{vs}
      \fmffreeze
      \fmf{plain}{i,ve}
      \fmf{plain,left=0.8}{ve,vo}
	    \fmf{phantom,left=0.5,label=\footnotesize{$2$},l.d=-0.01w}{ve,vn}
      \fmf{plain,left=0.8}{vo,ve}
	    \fmf{plain,label=\footnotesize{$a$},l.d=0.05w}{vs,vn}
      \fmf{plain}{vo,o}
      \fmffreeze
      \fmfdot{ve,vn,vo,vs}
    \end{fmfgraph*}
} \quad - \quad
\parbox{16mm}{
\begin{fmfgraph*}(16,14)
    \fmfleft{i}
    \fmfright{o}
    \fmfleft{ve}
    \fmfright{vo}
    \fmftop{v}
    \fmffreeze
    \fmfforce{(-0.1w,0.5h)}{i}
    \fmfforce{(1.1w,0.5h)}{o}
    \fmfforce{(0w,0.5h)}{ve}
    \fmfforce{(1.0w,0.5h)}{vo}
    \fmfforce{(.5w,0h)}{v}
    \fmffreeze
    \fmf{plain}{i,ve}
	\fmf{plain,left=0.8,label=\footnotesize{$2$},l.d=0.1h}{ve,vo}
    \fmf{plain,left=0.9}{vo,ve}
    \fmf{plain,left=0.5}{v,vo}
    \fmf{phantom,left=0.3,label=\scriptsize{$1+a$},l.d=0.1h}{v,vo}
    \fmf{plain}{vo,o}
    \fmffreeze
    \fmfdot{ve,v,vo}
  \end{fmfgraph*}
 } \quad \right]\, .
\label{IBP1}\\
&&\qquad \nonum
\eea
Similarly, using the integration by parts to the right triangle with the lateral distinguished line for the diagram $I(a)$, we obtain:
%
\fleq{
&I(a) \, (D-3 -a) = \quad 
\parbox{16mm}{
\begin{fmfgraph*}(16,14)
    \fmfleft{i}
    \fmfright{o}
    \fmfleft{ve}
    \fmfright{vo}
    \fmftop{v}
    \fmffreeze
    \fmfforce{(-0.1w,0.5h)}{i}
    \fmfforce{(1.1w,0.5h)}{o}
    \fmfforce{(0w,0.5h)}{ve}
    \fmfforce{(1.0w,0.5h)}{vo}
    \fmfforce{(.5w,0h)}{v}
    \fmffreeze
    \fmf{plain}{i,ve}
    \fmf{plain,left=0.8}{ve,vo}
    \fmf{plain,left=0.9}{vo,ve}
    \fmf{plain,right=0.5}{v,ve}
	\fmf{phantom,right=0.3,label=\footnotesize{$a$},l.d=0.1h}{v,ve}
	\fmf{phantom,right=0.3,label=\footnotesize{$2$},l.d=0.1h}{ve,v}
    \fmf{plain}{vo,o}
    \fmffreeze
    \fmfdot{ve,v,vo}
  \end{fmfgraph*}
 } \quad
- ~p^2 \quad 
\parbox{16mm}{
    \begin{fmfgraph*}(16,14)
      \fmfleft{i}
      \fmfright{o}
      \fmfleft{ve}
      \fmfright{vo}
      \fmftop{vn}
      \fmftop{vs}
      \fmffreeze
      \fmfforce{(-0.1w,0.5h)}{i}
      \fmfforce{(1.1w,0.5h)}{o}
      \fmfforce{(0w,0.5h)}{ve}
      \fmfforce{(1.0w,0.5h)}{vo}
      \fmfforce{(.5w,0.95h)}{vn}
      \fmfforce{(.5w,0.05h)}{vs}
      \fmffreeze
      \fmf{plain}{i,ve}
      \fmf{plain,left=0.8}{ve,vo}
      \fmf{plain,left=0.8}{vo,ve}
	    \fmf{phantom,right=0.5,label=\footnotesize{$2$},l.d=-0.01w}{ve,vs}
	    \fmf{plain,label=\footnotesize{$a$},l.d=0.05w}{vs,vn}
      \fmf{plain}{vo,o}
      \fmffreeze
      \fmfdot{ve,vn,vo,vs}
    \end{fmfgraph*}
}
\quad + a  \left[ \quad
\parbox{16mm}{
\begin{fmfgraph*}(16,14)
    \fmfleft{i}
    \fmfright{o}
    \fmfleft{ve}
    \fmfright{vo}
    \fmftop{v}
    \fmffreeze
    \fmfforce{(-0.1w,0.5h)}{i}
    \fmfforce{(1.1w,0.5h)}{o}
    \fmfforce{(0w,0.5h)}{ve}
    \fmfforce{(1.0w,0.5h)}{vo}
    \fmfforce{(.5w,0h)}{v}
    \fmffreeze
    \fmf{plain}{i,ve}
    \fmf{plain,left=0.8}{ve,vo}
    \fmf{plain,left=0.9}{vo,ve}
    \fmf{plain,right=0.5}{v,ve}
	\fmf{phantom,right=0.3,label=\scriptsize{$1+a$},l.d=0.1h}{v,ve}
    \fmf{plain}{vo,o}
    \fmffreeze
    \fmfdot{ve,v,vo}
  \end{fmfgraph*}
 } \quad
- \quad
\parbox{16mm}{
\begin{fmfgraph*}(16,14)
    \fmfleft{i}
    \fmfright{o}
    \fmfleft{ve}
    \fmfright{vo}
    \fmftop{v}
    \fmffreeze
    \fmfforce{(-0.1w,0.5h)}{i}
    \fmfforce{(1.1w,0.5h)}{o}
    \fmfforce{(0w,0.5h)}{ve}
    \fmfforce{(1.0w,0.5h)}{vo}
    \fmfforce{(.5w,0h)}{v}
    \fmffreeze
    \fmf{plain}{i,ve}
    \fmf{plain,left=0.8}{ve,vo}
    \fmf{plain,left=0.9}{vo,ve}
    \fmf{plain,left=0.5}{v,vo}
	\fmf{phantom,left=0.3,label=\scriptsize{$1+a$},l.d=0.1h}{v,vo}
    \fmf{plain}{vo,o}
    \fmffreeze
    \fmfdot{ve,v,vo}
  \end{fmfgraph*}
 } \quad  \right]\, ,
\label{IBP2}\\
&\qquad \nonum
}
where the last two diagrams in the rhs cancel each other. Combining Eqs.~(\ref{IBP1}) and (\ref{IBP2}), we come to the equation:
\fleq{
&-(a+\ep) I(1+a) \, p^2 + (1-a-2\ep) I(a) = 
\quad
\parbox{16mm}{
\begin{fmfgraph*}(16,14)
    \fmfleft{i}
    \fmfright{o}
    \fmfleft{ve}
    \fmfright{vo}
    \fmftop{v}
    \fmffreeze
    \fmfforce{(-0.1w,0.5h)}{i}
    \fmfforce{(1.1w,0.5h)}{o}
    \fmfforce{(0w,0.5h)}{ve}
    \fmfforce{(1.0w,0.5h)}{vo}
    \fmfforce{(.5w,0h)}{v}
    \fmffreeze
    \fmf{plain}{i,ve}
    \fmf{plain,left=0.8}{ve,vo}
    \fmf{plain,left=0.9}{vo,ve}
    \fmf{plain,right=0.5}{v,ve}
	\fmf{phantom,right=0.3,label=\footnotesize{$a$},l.d=0.1h}{v,ve}
	\fmf{phantom,right=0.3,label=\footnotesize{$2$},l.d=0.1h}{ve,v}
    \fmf{plain}{vo,o}
    \fmffreeze
    \fmfdot{ve,v,vo}
  \end{fmfgraph*}
 } \quad
- ~ p^2 \quad
\parbox{16mm}{
\begin{fmfgraph*}(16,14)
    \fmfleft{i}
    \fmfright{o}
    \fmfleft{ve}
    \fmfright{vo}
    \fmftop{v}
    \fmffreeze
    \fmfforce{(-0.1w,0.5h)}{i}
    \fmfforce{(1.1w,0.5h)}{o}
    \fmfforce{(0w,0.5h)}{ve}
    \fmfforce{(1.0w,0.5h)}{vo}
    \fmfforce{(.5w,0h)}{v}
    \fmffreeze
    \fmf{plain}{i,ve}
	\fmf{plain,left=0.8,label=\footnotesize{$2$},l.d=0.1h}{ve,vo}
    \fmf{plain,left=0.9}{vo,ve}
    \fmf{plain,right=0.5}{v,ve}
	\fmf{phantom,right=0.3,label=\scriptsize{$1+a$},l.d=0.1h}{v,ve}
    \fmf{plain}{vo,o}
    \fmffreeze
    \fmfdot{ve,v,vo}
  \end{fmfgraph*}
 } \quad
\nonumber \\
&=\frac{p^{-2(a+2\ep)}}{(4\pi)^D} \left [G(D,2,a)G(D,1,1+a+\ep)-G(D,1,a+1)G(D,2,1+a+\ep)\right] \, .
\label{IBP3}
}
After some little algebra, we have for the coefficient function:
\fleq{
\text{C}_D[I(1+a)] = \frac{1-a-2\ep}{a+\ep} \text{C}_D[I(a)] + \frac{2(2a-1+3\ep)\Gamma(-a-\ep)\Gamma(a-1+2\ep)\Gamma^2(1-\ep)}{(a+\ep)\Gamma(a+1)
\Gamma(2-3\ep-a)} .
\label{IBP4}
}
Eqs.~(\ref{Fa}) and (\ref{IBP4}) are the desired functional equations for $I(1+a)$.

\subsubsection{Solution of the functional equations}
\label{sec:FunctionalSo}

To simplify the inhomogeneous part of Eq.~(\ref{IBP4}), we make the substitution:
\bea
\text{C}_D[I(1+a)] = 2\frac{\Gamma^2(1-\ep)\Gamma(-a-\ep)\Gamma(a+2\ep)}{\Gamma(a+1)\Gamma(1-3\ep-a)} \, F(1+a) \, ,
\label{IBP5}
\eea
where the function $F(1+a)$ obeys the following equations:~\footnote{We would like to note that the inhomogeneous terms in 
  Eq.~(11) of \cite{Kazakov:1983pk} and in Eq.~(2.14) of \cite{Kazakov:1984bw}
  have wrong signs. Moreover, the r.h.s. of Eqs.~(14) and (15) of \cite{Kazakov:1983pk} and also the r.h.s of Eqs.~(2.17) and (2.18)
  of \cite{Kazakov:1984bw} should have the additional sign ``$-$''.}
\begin{subequations}
\label{Ga}
\bea
&&F(1+a) = F(1-a-3\ep) \, , \label{Gaa} \\
	&&F(1+a) = -\frac{a}{a-1+3\ep} \, F(a) - \frac{1}{a-1+3\ep} \left[ \frac{1}{a+\ep} +   \frac{1}{a-1+2\ep} \right]\, .
\label{Gab}
\eea
\end{subequations}
The solution of Eqs.~(\ref{Gaa}) and (\ref{Gab}) has been found by D.~Kazakov in Refs.~\cite{Kazakov:1984km,Kazakov:1984bw}.
In order to do so, he used the analytic properties of $I(1+a)$ which can be obtained from, \eg, its $\alpha$-representation (see Eq.~(12) in \cite{Kazakov:1984km}), and are such that  
the function $I(1+a)$ is a meromorphic function, regular at $a=0$ with simple poles at $a= \pm n -2\ep$ and $a= \pm n -\ep$, where $n=1,2, ...$. 
The same conclusion follows from the inhomogeneous term in Eq.~(\ref{IBP4}). The function $F(1+a)$ obtains additional poles due to the $\Gamma$-functions in the 
denominator of Eq.~(\ref{IBP5}). Therefore, the solution to Eq.~(\ref{Ga}) can be sought for in the form of an infinite series of poles:
\bea
F(1+a) = \sum_{n=1}^{\infty} f^{(1)}_n \left(\frac{1}{n+a+\ep} +   \frac{1}{n-a-2\ep} \right) +  
\sum_{n=1}^{\infty} f^{(2)}_n \left(\frac{1}{n+a} +   \frac{1}{n-a-3\ep} \right)\, .
\label{Ga1}
\eea
Notice that Eq.~(\ref{Ga1}) satisfies Eq.~(\ref{Gaa}). 
We may now substitute Eq.~(\ref{Ga1}) into the following equation coming from  Eq.~(\ref{Gab}):
\be
(a-1+3\ep) F(1+a) = - a \, F(a) - \left[ \frac{1}{a+\ep} +   \frac{1}{a-1+2\ep} \right]\, .
\ee
Comparing the terms $1/(n+a+\ep)$ and $1/(n+a)$, we obtain the following equations for the functions $f^{(i)}_n$ $(i=1,2)$:
\begin{subequations}
\label{Ga1a}
\bea
	&&-(n+1-2\ep)(1-\delta^0_n) f^{(1)}_n = (n+\ep) f^{(1)}_{n+1} - \delta^0_n\, ,
	\\
	&&-(n+1-3\ep)(1-\delta^0_n) f^{(2)}_n = n f^{(2)}_{n+1} \, .
\eea
\end{subequations}
In the case $n\geq 1$, these equations simplify as:
\bea
f^{(1)}_n = - f^{(1)}_{n+1} \, \frac{n+\ep}{n+1-2\ep}, \qquad f^{(2)}_n = - f^{(2)}_{n+1} \, \frac{n}{n+1-3\ep}\, ,
\label{Ga2}
\eea
and their solution reads: 
\bea
f^{(1)}_n = (-1)^n c_1(\ep) \, \frac{\Gamma(n+1-2\ep)}{\Gamma(n + \ep)},~~ f^{(2)}_n = (-1)^n c_2(\ep) \, \frac{\Gamma(n+1-3\ep)}{\Gamma(n)}\, .
\label{Ga3}
\eea
In the case $n=0$, Eqs.~(\ref{Ga1a}) only fix the function $c_1(\ep)$:
\bea
 c_1(\ep)= - \frac{\Gamma(\ep)}{\Gamma(2-2\ep)}\, .
 \label{Ga4}
\eea
In order to find $c_2$, we compare the obtained solution with the known one for a particular value of $a$, \eg, $a=0$.  
 On the one hand, Eqs.~(\ref{IBP5}) and (\ref{G(1,1,1,1,1)-G4d}) yield:
\fleq{
F(1)= \frac{1}{\ep(1-2\ep)} - \frac{1}{\ep(1-2\ep)}\frac{\Gamma^2(1+\ep)\Gamma(1-\ep)\Gamma(1-3\ep)}{\Gamma^2(1-2\ep)
\Gamma(1+2\ep)}\, .
\label{Ga4a}
}
On the other hand, from (\ref{Ga1}) for $a=0$  and  (\ref{Ga4}) we have:
\fleq{
F(1) &= - \frac{\Gamma(\ep)}{\Gamma(2-2\ep)}\,\sum_{n=1}^{\infty} \, (-1)^n \frac{\Gamma(n+1-2\ep)}{\Gamma(n+\ep)} \,
\left(\frac{1}{n+\ep} +   \frac{1}{n-2\ep} \right) + 
\nonum \\
&+c_2(\ep) \sum_{n=1}^{\infty} \, (-1)^n \frac{\Gamma(n+1-3\ep)}{\Gamma(n)}\, \left(\frac{1}{n} +   \frac{1}{n-3\ep} \right)\, .
\label{Ga1.1}
}
Since
\fleq{
\sum_{n=1}^{\infty} \, (-1)^n \frac{\Gamma(n+1-2\ep)}{\Gamma(n+\ep)} \,
\left(\frac{1}{n+\ep} +   \frac{1}{n-2\ep} \right) &= \sum_{n=1}^{\infty} \, (-1)^n \frac{\Gamma(n+1-2\ep)}{\Gamma(n +1+\ep)}
+ \sum_{n=1}^{\infty} \, (-1)^n \frac{\Gamma(n-2\ep)}{\Gamma(n+\ep)} 
\nonum \\
&= - \sum_{n=2}^{\infty} \, (-1)^n \frac{\Gamma(n-2\ep)}{\Gamma(n +\ep)}
+ \sum_{n=1}^{\infty} \, (-1)^n \frac{\Gamma(n-2\ep)}{\Gamma(n+\ep)} 
\nonum \\
&= -  \frac{\Gamma(1-2\ep)}{\Gamma(1+\ep)}\, ,
}
and
\fleq{
\sum_{n=1}^{\infty} \, (-1)^n \frac{\Gamma(n+1-3\ep)}{\Gamma(n)} \,
\left(\frac{1}{n} +   \frac{1}{n-3\ep} \right) &= \sum_{n=1}^{\infty} \, (-1)^n \frac{\Gamma(n+1-3\ep)}{\Gamma(n +1)}
+ \sum_{n=1}^{\infty} \, (-1)^n \frac{\Gamma(n-3\ep)}{\Gamma(n)} 
\nonum \\
&= - \sum_{n=2}^{\infty} \, (-1)^n \frac{\Gamma(n-3\ep)}{\Gamma(n)}
+ \sum_{n=1}^{\infty} \, (-1)^n \frac{\Gamma(n-3\ep)}{\Gamma(n)} 
\nonum \\
&= - \Gamma(1-3\ep)\, ,
}
then Eq.~(\ref{Ga1.1}) may be written as:
\be
F(1)= \frac{1}{\ep(1-2\ep)} - c_2(\ep) \Gamma(1-3\ep)\, .
\label{Ga1.2}
\ee
Comparing (\ref{Ga4a}) and (\ref{Ga1.2}) then yields:
\bea
 c_2(\ep)= + \frac{\Gamma(\ep)\Gamma(1-\ep)\Gamma(1+\ep)}{\Gamma(2-2\ep)\Gamma(1-2\ep)\Gamma(1+2\ep)}\, .
 \label{Ga5}
\eea
As a result, we have:
%
\fleq{
&\text{C}_D[I(1+a)] = - 2\frac{\Gamma^2(1-\ep)\Gamma(-a-\ep)\Gamma(a+2\ep)\Gamma(\ep)}{\Gamma(a+1)\Gamma(1-3\ep-a)\Gamma(2-2\ep)} 
\Biggl[\sum_{n=1}^{\infty} (-1)^n\,\frac{\Gamma(n+1-2\ep)}{\Gamma(n+\ep)} \biggl(\frac{1}{n+a+\ep}
\nonumber \\
& 
+   \frac{1}{n-a-2\ep} \biggr) -
\frac{\Gamma(1-\ep)\Gamma(1+\ep)}{\Gamma(1-2\ep)\Gamma(1+2\ep)} \, 
\sum_{n=1}^{\infty} (-1)^n\,\frac{\Gamma(n+1-3\ep)}{\Gamma(n)} \, \left(\frac{1}{n+a} +   \frac{1}{n-a-3\ep} \right) \Biggr]\, .
\label{Fa1}
}
Ultimately, in order to ascertain the validity of the solution (\ref{Fa1}), one has to be convinced that it is impossible to add an arbitrary solution 
of the homogeneous equation. Indeed, such a solution vanishes at integer points, is an analytic function and exponentially bounded on the imaginary axis. 
Hence, due to the Carlson theorem \cite{Carlson:1914}, it is identically zero.

The last sum in (\ref{Fa1}) is equal to $-\Gamma(1+a)\Gamma(1-a-3\ep)$.  Thus, Eq.~(\ref{Fa1}) can be also rewritten as:
%
\fleq{
&\text{C}_D[F(1+a)] = - 2\frac{\Gamma^2(1-\ep)\Gamma(\ep)}{\Gamma(2-2\ep)} \,  
\Biggl[\frac{\Gamma(-a-\ep)\Gamma(a+2\ep)}{\Gamma(a+1)\Gamma(1-3\ep-a)}
\sum_{n=1}^{\infty} (-1)^n\,\frac{\Gamma(n+1-2\ep)}{\Gamma(n+\ep)} \, \nonumber \\
&\times \, \left(\frac{1}{n+a+\ep} +   \frac{1}{n-a-2\ep} \right) +
  \frac{\Gamma(1-\ep)\Gamma(1+\ep)\Gamma(-a-2\ep)\Gamma(a+2\ep)}{\Gamma(1-2\ep)\Gamma(1+2\ep)} \Biggr]\, ,
\label{Fa2}
}
%
which exactly coincides with Eq.~(\ref{res:I(al):Kazakov}) after the appropriate change of variables.
Hence, the functional equation method shows that diagram $F(1+a)$ is expressible as a combination of $\Gamma$-functions together with 
two hypergeometric functions of argument ``$-1$'' thereby proving the result advertised in Eq.~(\ref{res:I(al):Kazakov}). 

The result of Eq.~(\ref{Fa2}) for $F(1+a_5\ep)$ can be expanded in series in $\ep$. This provides additional information
and gives a possibility to construct the $\ep$-expansion for the general case  $\al_i=1+a_i \ep$ $(i=1,2,3,4,5)$.

\subsubsection{Expansion of $I(1+a_5\ep)$}
\label{FunctionalSoI}

The important master integral $I(1+a_5 \ep)$ can be calculated using Eq.~(\ref{Fa2}). 
After a little algebra, we have for its coefficient function:
%
\fleq{
C_D[I(1+a_5\ep)] &= \frac{\hat{K}_2}{1-2\ep} \, \biggl[A_0 \zeta_2 + A_1 \zeta_4 \ep+ A_2 \zeta_5 \ep^2 +
\Bigl[A_3 \zeta_6 - A_4 \zeta_3^2 \Bigr] \ep^3 \nonumber \\
&+ \Bigl[A_5 \zeta_7 - A_6 \zeta_3 \zeta_4  \Bigr] \ep^4 + O(\ep^5)\biggr]\, ,
\label{IBPex3-2}
}
with
\bea
&&A_0=6 \, ,
\nonum \\  
&&A_1=9 \, , 
\nonum \\
&&A_2= 42 + 45 a_5 + 15 a_5^2 \, ,
\nonum \\
&&A_3= \frac{5}{2} \Bigl(A_2 - 6\Bigr)  \, , 
\label{IBPex4N} \\
&&A_4= 46 + 45 a_5 + 15 a_5^2 \, ,
\nonum \\
&&A_5= 294 + \frac{2223}{4}  a_5 + \frac{3183}{8} a_5^2 + \frac{567}{4} a_5^3 +  \frac{189}{8} a_5^4,
\nonum \\
&&A_6= 3 \Bigl(A_4 - 1\Bigr)\, , 
\nonum 
\label{IBPex3-3}
\eea
where $\hat{K}_n$ was defined in Eq.~(\ref{IBPex5}).
Notice that, when $a_5=0$, we recover the result in (\ref{G(1,1,1,1,1)-exp4d}).

\subsection{Final results for $J(D,p,1+a_1\ep,1+a_2\ep,1+a_3\ep,1+a_4\ep,1+a_5\ep)$}
\label{sec:examples0}

Using the methods reviewed so far: integration by parts, the method of uniqueness, functional equations together with various transformations of the considered diagram, 
Kazakov found several terms of $\ep$-expansion of the two-loop master integral $J(D,p,1+a_1\ep,1+a_2\ep,1+a_3\ep,1+a_4\ep,1+a_5\ep)$. The result reads \cite{Kazakov:1983pk}:
%
\fleq{
&&C_D[I(1+a_1\ep,1+a_2\ep,1+a_3\ep,1+a_4\ep,1+a_5\ep)] = \frac{\hat{K}_2}{1-2\ep} \, \biggl[A_0 \zeta_2 + A_1 \zeta_4 \ep+ A_2 \zeta_5 \ep^2  \nonumber \\
&&+ \Bigl[A_3 \zeta_6 - A_4 \zeta_3^2 \Bigr] \ep^3+ \Bigl[A_5 \zeta_7 - A_6 \zeta_3 \zeta_4  \Bigr] \ep^4 + O(\ep^5)\biggr]\, ,
\label{IBPex3-4}
}
with 
\bea
&&A_0=6\, ,
\nonum \\ 
&&A_1=9\, ,
\nonum \\
&&A_2= 42 + \sum_{i=1}^2 A_{2,i} \, , 
\nonum \\
&&A_3= \frac{5}{2} \Bigl(A_2 - 6\Bigr)\, , 
\label{IBPex4M} \\
&&A_4= 46 + \sum_{i=1}^3 A_{4,i}\, , 
\nonum \\
&&A_5= 294 + \sum_{i=1}^4 A_{5,i}\, ,
\nonum \\
&&A_6= 3 \Bigl(A_4 - 1\Bigr)\, , 
\nonumber 
\eea
where $\hat{K}_n$ was defined in (\ref{IBPex5}) and
\bea
&&A_{2,1}= 30\, \overline{A}_1 + 45 a_5\, ,
\nonum \\
&&A_{4,1}= 42\,  \overline{A}_1 + 45 a_5\, ,
\nonum \\
&&A_{5,1}= 402\, \overline{A}_1 + \frac{2223}{4} a_5\, , 
\nonumber \\
&&A_{2,2}= 10\,\overline{A}_2 + 15 a_5^2 + 15\,\overline{A}_1 a_5 + 10\bigl(a_1a_2 +a_3a_4 +a_1a_4+a_2a_3\bigr)
+ 5 \bigl(a_1a_3 +a_2a_4\bigr)\, , 
\nonumber \\
&& A_{4,2}= 14\,\overline{A}_2 + 15 a_5^2 + 33 \overline{A}_1 a_5 + 50\bigl(a_1a_2 +a_3a_4\bigr) + 14\bigl(a_1a_4+a_2a_3\bigr) +
\nonumber \\
&& + 31 \bigl(a_1a_3 +a_2a_4\bigr)\, , 
\nonumber \\
&& A_{4,3}= 6 a_5\,\overline{A}_2 + 6 a_5^2\,\overline{A}_1 + 24 a_5\bigl(a_1a_2 +a_3a_4\bigr) + 12 a_5\bigl(a_1a_3+a_2a_4\bigr) + 
\nonumber \\
&& + 12 \bigl(a_1a_2a_3 +a_1a_2a_4 +a_1a_3a_4 +a_2a_3a_4\bigr) + 
12 \bigl(a_1^2a_2 +a_1 a_2^2 +a_3^2 a_4 +a_3 a_4^2\bigr) +  
\nonumber \\
&& + 6 \bigl(a_1^2a_3 +a_1 a_3^2 +a_2^2 a_4 +a_2a_4^2\bigr)\, , 
\nonumber \\
&& A_{5,2}= 260\,\overline{A}_2 + \frac{3183}{8} a_5^2 + 516\,\overline{A}_1 a_5 + 386\bigl(a_1a_2 +a_3a_4 +a_1a_4+a_2a_3\bigr) + 
\nonumber \\
&& + \frac{575}{2} \bigl(a_1a_3 +a_2a_4\bigr) \, , 
\nonumber \\
&& A_{5,3}= 84\,\overline{A}_3 +  \frac{567}{4} a_5^3+ \frac{945}{4} a_5\,\overline{A}_2 + 252 a_5^2\,\overline{A}_1 
+ \frac{693}{2} a_5\bigl(a_1a_2 +a_3 a_4 +a_1a_4+a_2a_3\bigr) + 
\nonumber \\
&& + \frac{945}{4}a_5\bigl(a_1a_3 +a_2a_4\bigr)
+ 210 \bigl(a_1a_2a_3 +a_1a_2a_4 +a_1a_3a_4 + a_2a_3a_4\bigr) + 
\nonumber \\
&& + 168 \bigl(a_1^2 a_2 +a_1 a_2^2 + a_3^2 a_4 +a_3 a_4^2 + a_1^2 a_4 +a_1 a_4^2 + a_2^2 a_3 +a_2 a_3^2
\bigr) \nonumber \\
&& + \frac{441}{4} \bigl(a_1^2 a_3 +a_1a_3^2 +a_2^2a_4 +a_2a_4^2\bigr) \, , 
\nonumber \\
&& A_{5,4}= 14\,\overline{A}_4 +  \frac{189}{8} a_5^4+ 42 a_5\,\overline{A}_3 + \frac{189}{4} a_5^3\,\overline{A}_1
+ \frac{525}{8} a_5^2  \overline{A}_2 + 
\nonumber \\
&& + \frac{357}{4}a_5^2\bigl(a_1a_2 +a_3a_4 +a_1a_4+a_2a_3\bigr) + 
\nonumber \\
&& + \frac{105}{2}a_5^2\bigl(a_1a_3 +a_2a_4\bigr)
+  \frac{357}{4} a_5\bigl(a_1a_2a_3 +a_1a_2a_4 +a_1a_3a_4 +a_2a_3a_4\bigr) + 
\nonumber \\
&& + 84 a_5\bigl(a_1^2a_2 +a_1a_2^2 +a_3^2a_4 +a_3a_4^2 + a_1^2 a_4 +a_1 a_4^2 + a_2^2 a_3 +a_2 a_3^2
\bigr) \nonumber \\
&&+ \frac{189}{4} a_5 \bigl(a_1^2a_3 +a_1a_3^2 +a_2^2a_4 +a_2a_4^2\bigr) + 
\nonumber \\
&& + 28\bigl(a_1^3a_2 +a_1a_2^3 +a_3^3a_4 +a_3a_4^3  + a_1^3 a_4 +a_1 a_4^3 + a_2^3 a_3 +a_2 a_3^3
\bigr) \nonumber \\
&& + 14 \bigl(a_1^3a_3 +a_1a_3^3 +a_2^3a_4 +a_2a_4^3\bigr) + 
\nonumber \\
&& + 42 \bigl(a_1^2a_2^2 +a_3^2a_4^2 +a_1^2a_4^2 +a_2^2a_3^2\bigr) + \frac{189}{8} \bigl(a_1^2a_3^2 +a_2^2a_4^2 \bigr) + 
\nonumber \\
&& + 42 \bigl(a_1^2a_2a_3 +a_1^2a_2a_4 +a_1^2a_3a_4 +a_2^2a_1a_4 +a_2^2a_1a_3 +a_2^2a_3a_4 + 
\nonumber \\
&& + a_3^2a_1a_4 +a_3^2a_2a_4 +a_3^2a_1a_2 + a_4^2 a_2 a_3 +a_4^2 a_1a_3 +a_4^2 a_1a_2\bigr) + 
\frac{315}{4} a_1a_2a_3a_4 \, 
\bigr)\, ,
\label{IBPex5M1}
\eea
and $ \overline{A}_n= a_1^n + a_2^n + a_3^n + a_4^n$. We would like to note that for $a_2=a_3=a_5=0$ (respectively, $a_1=a_2=a_3=a_4=0$) 
Eqs.~(\ref{IBPex3-4})-(\ref{IBPex5M1}) transform to the ones of (\ref{IBPex3})-(\ref{IBPex5}) (respectively, to the ones 
of (\ref{IBPex3-2})-(\ref{IBPex4N})).

\subsection{Three-loop master integrals}
\label{sec:three-loop}

\begin{figure}
  \begin{center}
\includegraphics{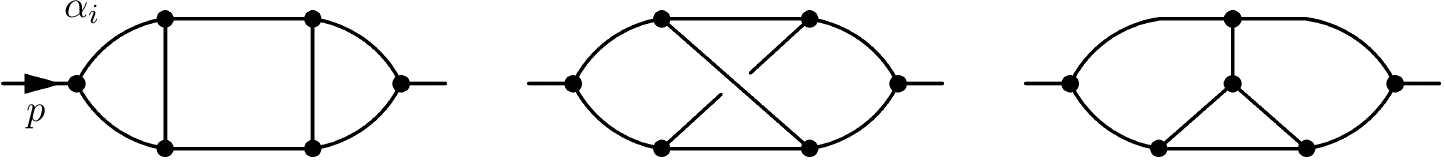}
	  \caption{\label{fig:three-loop}
  Three-loop scalar massless propagator-type diagrams.}
  \end{center}
\end{figure}

At three loops there are three generic topologies of massless propagator-type diagrams, see Fig.~\ref{fig:three-loop}. For some special values of the indices, these diagrams are primitively two-loop
and can therefore be computed with the help of the results derived so far. Let's consider for instance the following two basic massless integrals:
\begin{subequations}
\bea
&&J_1(D,p,\al_1,\al_2,\al_3,\al_4,\al_5,\al_6,\al_7) = 
\qquad \quad
\parbox{16mm}{
    \begin{fmfgraph*}(16,14)
      \fmfleft{i}
      \fmfright{o}
      \fmfleft{ve}
      \fmfright{vo}
      \fmftop{vn}
      \fmftop{vs}
      \fmffreeze
      \fmfforce{(-0.6w,0.5h)}{i}
      \fmfforce{(1.6w,0.5h)}{o}
      \fmfforce{(-0.3w,0.5h)}{ve}
      \fmfforce{(1.3w,0.5h)}{vo}
      \fmfforce{(.5w,1.05h)}{vn}
      \fmfforce{(.5w,-0.05h)}{vs}
      \fmfforce{(.5w,0.5h)}{vc}	    
      \fmffreeze
      \fmf{plain}{i,ve}
      \fmf{plain,left=0.6}{ve,vo}
	    \fmf{phantom,left=0.4,label=\footnotesize{$\al_1$},l.d=-0.01w}{ve,vn}
	    \fmf{phantom,right=0.4,label=\footnotesize{$\al_5$},l.d=-0.01w}{vo,vn}
      \fmf{plain,left=0.6}{vo,ve}
	    \fmf{phantom,left=0.4,label=\footnotesize{$\al_4$},l.d=-0.01w}{vo,vs}
	    \fmf{phantom,right=0.4,label=\footnotesize{$\al_3$},l.d=-0.01w}{ve,vs}
      \fmf{plain}{vs,vn}
	    \fmf{phantom,label=\footnotesize{$\al_6$},l.d=0.05w,l.s=left}{vn,vc}
	    \fmf{phantom,label=\footnotesize{$\al_7$},l.d=0.05w,l.s=left}{vc,vs}
	    \fmf{plain,label=\footnotesize{$\al_2$},l.d=0.05w}{ve,vc}
      \fmf{plain}{vo,o}
      \fmffreeze
      \fmfdot{ve,vn,vo,vs,vc}
    \end{fmfgraph*}
}  \qquad \quad \, ,\\
&&\nonum \qquad \\
&&\nonum \qquad \\
&&J_2(D,p,\al_1,\al_2,\al_3,\al_4,\al_5,\al_6,\al_7) = 
\qquad \quad
\parbox{16mm}{
    \begin{fmfgraph*}(16,14)
      \fmfleft{i}
      \fmfright{o}
      \fmfleft{ve}
      \fmfright{vo}
      \fmftop{vn1}
      \fmftop{vn2}
      \fmftop{vs}
      \fmffreeze
      \fmfforce{(-0.6w,0.5h)}{i}
      \fmfforce{(1.6w,0.5h)}{o}
      \fmfforce{(-0.3w,0.5h)}{ve}
      \fmfforce{(1.3w,0.5h)}{vo}
      \fmfforce{(.15w,1.05h)}{vn1}
      \fmfforce{(.85w,1.05h)}{vn2}
      \fmfforce{(.5w,-0.05h)}{vs}
      \fmffreeze
      \fmf{plain}{i,ve}
	    \fmf{plain,left=0.25,label=\footnotesize{$\al_1$},l.d=0.02w}{ve,vn1}
	    \fmf{plain,left=0.15,label=\footnotesize{$\al_2$},l.d=0.05h}{vn1,vn2}
	    \fmf{plain,left=0.25,label=\footnotesize{$\al_3$},l.d=0.02w}{vn2,vo}
      \fmf{plain,left=0.6}{vo,ve}
	    \fmf{phantom,left=0.4,label=\footnotesize{$\al_4$},l.d=-0.02w}{vo,vs}
	    \fmf{phantom,right=0.4,label=\footnotesize{$\al_5$},l.d=-0.02w}{ve,vs}
	    \fmf{plain,label=\footnotesize{$\al_6$},l.d=0.04w,l.s=left}{vs,vn1}
	    \fmf{plain,label=\footnotesize{$\al_7$},l.d=0.04w}{vs,vn2}
      \fmf{plain}{vo,o}
      \fmffreeze
      \fmfdot{ve,vn1,vn2,vo,vs}
    \end{fmfgraph*}
} \qquad \quad \, ,
\eea
\end{subequations}
which can be obtained for some special values of the indices of the last (benz) diagram in Fig.~\ref{fig:three-loop} and which are actually related to each other (see below).

When all line indices are equal to $1$, the first three-loop master, $J_1(D,p,1,1,1,1,1,1,1)$,
can be expressed in the following form (with the help of the IBP procedure for the upper triangle with the vertical line being distinguished):
%
\fleq{
\parbox{16mm}{
    \begin{fmfgraph*}(16,14)
      \fmfleft{i}
      \fmfright{o}
      \fmfleft{ve}
      \fmfright{vo}
      \fmftop{vn}
      \fmftop{vs}
      \fmffreeze
      \fmfforce{(-0.1w,0.5h)}{i}
      \fmfforce{(1.1w,0.5h)}{o}
      \fmfforce{(-0w,0.5h)}{ve}
      \fmfforce{(1.0w,0.5h)}{vo}
      \fmfforce{(.5w,0.95h)}{vn}
      \fmfforce{(.5w,0.05h)}{vs}
      \fmfforce{(.5w,0.5h)}{vc}
      \fmffreeze
      \fmf{plain}{i,ve}
      \fmf{plain,left=0.8}{ve,vo}
      \fmf{plain,left=0.8}{vo,ve}
      \fmf{plain}{vs,vn}
     \fmf{plain}{ve,vc}
	    \fmf{plain}{vo,o}
      \fmffreeze
      \fmfdot{ve,vn,vo,vs,vc}
    \end{fmfgraph*}
}  \quad  (D-4) = 2 \quad
\parbox{16mm}{
    \begin{fmfgraph*}(16,14)
      \fmfleft{i}
      \fmfright{o}
      \fmfleft{ve}
      \fmfright{vo}
      \fmftop{vn}
      \fmftop{vs}
      \fmffreeze
      \fmfforce{(-0.1w,0.5h)}{i}
      \fmfforce{(1.1w,0.5h)}{o}
      \fmfforce{(0w,0.5h)}{ve}
      \fmfforce{(1.0w,0.5h)}{vo}
      \fmfforce{(.5w,0.95h)}{vn}
      \fmfforce{(.5w,0.05h)}{vs}
      \fmfforce{(.5w,0.5h)}{vc}
      \fmffreeze
      \fmf{plain}{i,ve}
      \fmf{plain,left=0.8}{ve,vo}
      \fmf{plain,left=0.8}{vo,ve}
      \fmf{plain}{vs,vn}
      \fmf{plain}{vo,o}
      \fmf{plain,right=0.5}{ve,vn}
      \fmf{phantom,label={\scriptsize{$2$}},l.d=0.05w,l.s=left}{vc,ve}
      \fmffreeze
      \fmfdot{ve,vn,vo,vs}
    \end{fmfgraph*}
} \quad - \quad
\parbox{16mm}{
    \begin{fmfgraph*}(16,14)
      \fmfleft{i}
      \fmfright{o}
      \fmfleft{ve}
      \fmfright{vo}
      \fmftop{vn}
      \fmftop{vs}
      \fmffreeze
      \fmfforce{(-0.1w,0.5h)}{i}
      \fmfforce{(1.1w,0.5h)}{o}
      \fmfforce{(0w,0.5h)}{ve}
      \fmfforce{(1.0w,0.5h)}{vo}
      \fmfforce{(.5w,0.95h)}{vn}
      \fmfforce{(.5w,0.05h)}{vs}
      \fmfforce{(.5w,0.5h)}{vc}
      \fmffreeze
      \fmf{plain}{i,ve}
      \fmf{plain,left=0.8}{ve,vo}
      \fmf{plain,left=0.8}{vo,ve}
      \fmf{plain}{vs,vn}
      \fmf{plain}{vo,o}
      \fmf{plain,left=0.5}{ve,vs}
      \fmf{phantom,label={\scriptsize{$2$}},l.d=0.05w,l.s=left}{ve,vc}
      \fmffreeze
      \fmfdot{ve,vn,vo,vs}
    \end{fmfgraph*}
} \quad -
\quad
\parbox{16mm}{
    \begin{fmfgraph*}(16,14)
      \fmfleft{i}
      \fmfright{o}
      \fmfleft{ve}
      \fmfright{vo}
      \fmftop{vn}
      \fmftop{vs}
      \fmffreeze
      \fmfforce{(-0.1w,0.5h)}{i}
      \fmfforce{(1.1w,0.5h)}{o}
      \fmfforce{(0w,0.5h)}{ve}
      \fmfforce{(1.0w,0.5h)}{vo}
      \fmfforce{(.5w,0.95h)}{vn}
      \fmfforce{(.5w,0.05h)}{vs}
      \fmffreeze
      \fmf{plain}{i,ve}
      \fmf{plain,left=0.8}{ve,vo}
      \fmf{plain,left=0.8}{vo,ve}
      \fmf{plain}{vs,vn}
      \fmf{plain}{vo,o}
      \fmf{phantom,left=0.8,label=\scriptsize{$2$},l.d=0.2h}{ve,vo}
      \fmffreeze
      \fmfdot{ve,vn,vo,vs}
      \fmfposition
\fmfi{plain}{vloc(__vo) ..controls vloc(__o) and (xpart(vloc(__o)),1.4h) ..(xpart(vloc(__vn)),1.4h)}
\fmfi{plain}{vloc(__ve) ..controls vloc(__i) and (xpart(vloc(__i)),1.4h) ..(xpart(vloc(__vn)),1.4h)}
    \end{fmfgraph*}
} \quad \, .
\label{3loop}
}
The last term in the rhs contains the two-loop internal diagram $I(1)$, which has the $p$-dependence given by $1/p^{2(1+2\ep)}$.
So, the last term can be represented as $C_D[I(1)] G(2,1+2\ep)/p^{2(1+3\ep)}$.  Thus, the rhs of eq. (\ref{3loop}) can be transformed into the form:
\bea
G(2,1) \quad 
\parbox{16mm}{
    \begin{fmfgraph*}(16,14)
      \fmfleft{i}
      \fmfright{o}
      \fmfleft{ve}
      \fmfright{vo}
      \fmftop{vn}
      \fmftop{vs}
      \fmffreeze
      \fmfforce{(-0.1w,0.5h)}{i}
      \fmfforce{(1.1w,0.5h)}{o}
      \fmfforce{(0w,0.5h)}{ve}
      \fmfforce{(1.0w,0.5h)}{vo}
      \fmfforce{(.5w,0.95h)}{vn}
      \fmfforce{(.5w,0.05h)}{vs}
      \fmffreeze
      \fmf{plain}{i,ve}
      \fmf{plain,left=0.8}{ve,vo}
      \fmf{phantom,left=0.5,label=\footnotesize{$1+\ep$},l.d=-0.01w}{ve,vn}
      \fmf{plain,left=0.8}{vo,ve}
      \fmf{plain}{vs,vn}
      \fmf{plain}{vo,o}
      \fmffreeze
      \fmfdot{ve,vn,vo,vs}
    \end{fmfgraph*}
} \quad
- \quad G(2,1+2\ep) \, 
\quad
\parbox{16mm}{
    \begin{fmfgraph*}(16,14)
      \fmfleft{i}
      \fmfright{o}
      \fmfleft{ve}
      \fmfright{vo}
      \fmftop{vn}
      \fmftop{vs}
      \fmffreeze
      \fmfforce{(-0.1w,0.5h)}{i}
      \fmfforce{(1.5w,0.5h)}{o1}	    
      \fmfforce{(1.6w,0.5h)}{o}
      \fmfforce{(0w,0.5h)}{ve}
      \fmfforce{(1.0w,0.5h)}{vo}
      \fmfforce{(.5w,0.95h)}{vn}
      \fmfforce{(.5w,0.05h)}{vs}
      \fmffreeze
      \fmf{plain}{i,ve}
      \fmf{plain,left=0.8}{ve,vo}
      \fmf{plain,left=0.8}{vo,ve}
      \fmf{plain}{vs,vn}
	    \fmf{plain,label=\footnotesize{$\ep$}}{vo,o1}
      \fmf{plain}{o1,o}	    
      \fmffreeze
      \fmfdot{ve,vn,vo,vs,o1}
    \end{fmfgraph*}
} \qquad \quad \, ,
\nonumber
\eea
\ie, the first three-loop master integral can be expressed in terms of the two-loop master ones.
Performing various transformations on the integral, Kazakov managed to derive its $\ep$-expansion 
which takes the form \cite{Kazakov:1983pk}:~\footnote{Similar results have been recently published in Ref.~\cite{Panzer:2013cha}.}
\bea
&&C_D[J_1(D,p,1+a_1\ep,1+a_2\ep,1+a_3\ep,1+a_4\ep,1+a_5,\ep1+a_6\ep,1+a_7\ep)]   \nonumber \\
&&= \frac{\hat{K}_3}{1-2\ep} \, \biggl[B_0 \zeta_5 
+ \Bigl[B_1 \zeta_6 + B_2 \zeta_3^2 \Bigr] \ep+ \Bigl[B_3 \zeta_7 + B_4 \zeta_3 \zeta_4  \Bigr] \ep^2 + O(\ep^3)\biggr]\, ,
\label{3loop2}
\eea
with
\bea
B_0=20,~~ B_1=50,~~ B_2= 20 + 6 \overline{B} ,~~B_4= 3B_2,~~ B_3= 380 +  7\,\sum_{i=1}^2 B_{3,i}, \label{Iloop1} 
\eea
where $\overline{B}=a_4+a_5+a_6+a_7$ and 
\bea
&&B_{3,1}= \frac{1}{4} \overline{B} + 24 \bigl(a_1+a_3\bigr) + 32 a_2 + 17 \bigl(a_4+a_5\bigr)+ 33 \bigl(a_6+a_7\bigr) , \nonumber \\
&&B_{3,2}=  \frac{1}{8} \overline{B}^2  + 6 \bigl(a_1^2+a_3^2\bigr) + 8 a_2^2 + 4 \bigl(a_4^2+a_5^2\bigr)+ 8 \bigl(a_6^2+a_7^2\bigr) 
+ 8 \bigl(a_1+a_3\bigr)a_2 \nonumber \\
&&  + 2 \bigl(a_1a_4+a_3a_5\bigr) + 6 \bigl(a_1a_5+a_3a_4\bigr)+ 10 \bigl(a_1a_6+a_3a_7\bigr)
+ 6 \bigl(a_1a_7+a_3a_6\bigr)  \nonumber \\
&& +4 a_1a_3 + 4 \bigl(a_4+a_5\bigr)a_2 + 12 \bigl(a_6+a_7\bigr)a_2 + 2 a_4a_5
+4  \bigl(a_4a_6+a_5a_7\bigr) \nonumber \\
&& + 6 \bigl(a_4a_7+a_5a_6\bigr) +10 a_6a_7  \, .
\label{3loop3}
\eea

The results for the second three-loop master integral $J_2(D,p,\al_1,\al_2,\al_3,\al_4,\al_5,\al_6,\al_7)$
can be obtained in the following way:
%
\fleq{
&\qquad \nonum \\
\parbox{16mm}{
    \begin{fmfgraph*}(16,14)
      \fmfleft{i}
      \fmfright{o}
      \fmfleft{ve}
      \fmfright{vo}
      \fmftop{vn1}
      \fmftop{vn2}
      \fmftop{vs}
      \fmffreeze
      \fmfforce{(-0.6w,0.5h)}{i}
      \fmfforce{(1.6w,0.5h)}{o}
      \fmfforce{(-0.3w,0.5h)}{ve}
      \fmfforce{(1.3w,0.5h)}{vo}
      \fmfforce{(.15w,1.05h)}{vn1}
      \fmfforce{(.85w,1.05h)}{vn2}
      \fmfforce{(.5w,-0.05h)}{vs}
      \fmffreeze
	    \fmf{fermion,label=\footnotesize{$p$}}{i,ve}
	    \fmf{plain,left=0.25,label=\footnotesize{$\al_1$},l.d=0.02w}{ve,vn1}
	    \fmf{plain,left=0.15,label=\footnotesize{$\al_2$},l.d=0.05h}{vn1,vn2}
	    \fmf{plain,left=0.25,label=\footnotesize{$\al_3$},l.d=0.02w}{vn2,vo}
      \fmf{plain,left=0.6}{vo,ve}
	    \fmf{phantom,left=0.4,label=\footnotesize{$\al_4$},l.d=-0.02w}{vo,vs}
	    \fmf{phantom,right=0.4,label=\footnotesize{$\al_5$},l.d=-0.02w}{ve,vs}
	    \fmf{plain,label=\footnotesize{$\al_6$},l.d=0.04w,l.s=left}{vs,vn1}
	    \fmf{plain,label=\footnotesize{$\al_7$},l.d=0.04w}{vs,vn2}
      \fmf{plain}{vo,o}
      \fmffreeze
      \fmfdot{ve,vn1,vn2,vo,vs}
    \end{fmfgraph*}
} \qquad \quad &\underset{(\text{FT})}{=} \quad \frac{\prod_{k=1}^7 a(\al_k)}{a \left( \sum_{k=1}^7 \al_k - 3D/2 \right)}\qquad \quad
\parbox{16mm}{
    \begin{fmfgraph*}(16,14)
      \fmfleft{i}
      \fmfright{o}
      \fmfleft{ve}
      \fmfright{vo}
      \fmftop{vn1}
      \fmftop{vn2}
      \fmftop{vs}
      \fmffreeze
      \fmfforce{(-0.6w,0.5h)}{i}
      \fmfforce{(1.6w,0.5h)}{o}
      \fmfforce{(-0.3w,0.5h)}{ve}
      \fmfforce{(1.3w,0.5h)}{vo}
      \fmfforce{(.15w,1.05h)}{vn1}
      \fmfforce{(.85w,1.05h)}{vn2}
      \fmfforce{(.5w,-0.05h)}{vs}
      \fmffreeze
	    \fmf{plain,label=\footnotesize{$0$},l.s=left}{i,ve}
	    \fmf{plain,left=0.25,label=\footnotesize{$\tilde{\al}_1$},l.d=0.02w}{ve,vn1}
	    \fmf{plain,left=0.15,label=\footnotesize{$\tilde{\al}_2$},l.d=0.05h}{vn1,vn2}
	    \fmf{plain,left=0.25,label=\footnotesize{$\tilde{\al}_3$},l.d=0.02w}{vn2,vo}
      \fmf{plain,left=0.6}{vo,ve}
	    \fmf{phantom,left=0.4,label=\footnotesize{$\tilde{\al}_4$},l.d=-0.02w}{vo,vs}
	    \fmf{phantom,right=0.4,label=\footnotesize{$\tilde{\al}_5$},l.d=-0.02w}{ve,vs}
	    \fmf{plain,label=\footnotesize{$\tilde{\al}_6$},l.d=0.04w,l.s=left}{vs,vn1}
	    \fmf{plain,label=\footnotesize{$\tilde{\al}_7$},l.d=0.04w}{vs,vn2}
	    \fmf{plain,label=\footnotesize{$z$},l.s=left}{vo,o}
      \fmffreeze
      \fmfdot{ve,vn1,vn2,vo,vs}
    \end{fmfgraph*}
} \qquad \quad \nonum \\
&\quad \nonum \\
&\underset{(\text{D})}{=} \quad \frac{\prod_{k=1}^7 a(\al_k)}{a\left( \sum_{k=1}^7 \al_k - 3D/2 \right)}\qquad \quad
\parbox{16mm}{
    \begin{fmfgraph*}(16,14)
      \fmfleft{i}
      \fmfright{o}
      \fmfleft{ve}
      \fmfright{vo}
      \fmftop{vn}
      \fmftop{vs}
      \fmffreeze
      \fmfforce{(-0.6w,0.5h)}{i}
      \fmfforce{(1.6w,0.5h)}{o}
      \fmfforce{(-0.3w,0.5h)}{ve}
      \fmfforce{(1.3w,0.5h)}{vo}
      \fmfforce{(.5w,1.05h)}{vn}
      \fmfforce{(.5w,-0.05h)}{vs}
      \fmfforce{(.5w,0.5h)}{vc}
      \fmffreeze
	    \fmf{fermion,label=\footnotesize{$p$}}{i,ve}
      \fmf{plain,left=0.6}{ve,vo}
	    \fmf{phantom,left=0.4,label=\footnotesize{$\tilde{\al}_1$},l.d=-0.01w}{ve,vn}
	    \fmf{phantom,right=0.4,label=\footnotesize{$\tilde{\al}_5$},l.d=-0.01w}{vo,vn}
      \fmf{plain,left=0.6}{vo,ve}
	    \fmf{phantom,left=0.4,label=\footnotesize{$\tilde{\al}_4$},l.d=-0.01w}{vo,vs}
	    \fmf{phantom,right=0.4,label=\footnotesize{$\tilde{\al}_3$},l.d=-0.01w}{ve,vs}
      \fmf{plain}{vs,vn}
	    \fmf{phantom,label=\footnotesize{$\tilde{\al}_6$},l.d=0.05w,l.s=left}{vn,vc}
	    \fmf{phantom,label=\footnotesize{$\tilde{\al}_7$},l.d=0.05w,l.s=left}{vc,vs}
	    \fmf{plain,label=\footnotesize{$\tilde{\al}_2$},l.d=0.05w}{ve,vc}
      \fmf{plain}{vo,o}
      \fmffreeze
      \fmfdot{ve,vn,vo,vs,vc}
    \end{fmfgraph*}
}  \qquad \quad \, ,
\nonumber \\
\nonum
}
%
\ie, by first transforming the considered diagram to $x$-space with the help of a Fourier transform and later returning to $p$-space using the dual
transformation. So, we can express the coefficient function for the second three-loop master integral trough the one of the first three-loop master integral
$J_1(D,p,\al_1,\al_2,\al_3,\al_4,\al_5,\al_6,\al_7)$, \ie,
%
\fleq{
C_D[J_2(D,p,\al_1,\al_2,\al_3,\al_4,\al_5,\al_6,\al_7)] = \frac{\prod_{i=1}^7 a(\al_i)}{a \left( \sum_{i=1}^7\al_i-3D/2 \right)} \,
C_D[J_1(D,p,\tilde{\al}_1,\tilde{\al}_2,\tilde{\al}_3,\tilde{\al}_4,\tilde{\al}_5,
\tilde{\al}_6,\tilde{\al}_7]\, .
\label{3loop4}
}
%
The expansion for $C_D[J_2(D,p,1+a_1\ep,1+a_2\ep,1+a_3\ep,1+a_4\ep,1+a_5,\ep1+a_6\ep,1+a_7\ep)]$ can therefore be obtained from Eqs.~(\ref{3loop2}) and (\ref{3loop3})
with the help of the replacement $a_i \to -(a_i+1)$ $(i=1,...,7)$. 
Of course, the coefficient $\prod_{i=1}^7 a(1+a_i\ep)/a \left( 1 + \left( \sum_{i=1}^7 a_i+3 \right) \ep \right)$ should also be taken into account.

\end{fmffile}

\begin{fmffile}{fmf-review3}

	\section{More examples of calculations}
\label{sec:examples}


We shall now consider the $\Phi^4$-model and give some additional examples of concrete calculations. 
Using a combination of the methods presented in the last section, and following Kazakov, we will show that, 
up to five loops, the singular part of several complicated diagrams may be calculated exactly by reducing them 
to the two-loop massless propagator-type diagram.
	
\subsection{Three and four loop calculations in the $\Phi^4$-model}

In order to compute the three- and four-loop $\beta$-function of the $\Phi^4$-model one has, in particular, to calculate the singular parts of the following vertex diagrams:
\ms
\be
\parbox{16mm}{
    \begin{fmfgraph*}(16,14)
      \fmfleft{i1}
      \fmfright{i2}
      \fmftop{i3}
      \fmftop{i4}
      \fmfleft{v1}
      \fmfright{v2}
      \fmftop{v3}
      \fmftop{v4}
      \fmffreeze
      \fmfforce{(-0.2w,-0.2h)}{i1}
      \fmfforce{(-0.2w,1.2h)}{i2}
      \fmfforce{(1.2w,1.2h)}{i3}
      \fmfforce{(1.2w,-0.2h)}{i4}
      \fmfforce{(0w,0h)}{v1}
      \fmfforce{(0w,1h)}{v2}
      \fmfforce{(1w,1h)}{v3}
      \fmfforce{(1w,0h)}{v4}
      \fmfforce{(0.45w,0.45h)}{v1c}
      \fmfforce{(0.55w,0.55h)}{vc3}
      \fmffreeze
      \fmf{plain}{i1,v1}
      \fmf{plain}{i2,v2}
      \fmf{plain}{i3,v3}
      \fmf{plain}{i4,v4}
      \fmf{plain}{v1,v2,v3,v4,v1}
      \fmf{plain}{v1,v1c}
      \fmf{plain}{vc3,v3}
      \fmf{plain}{v2,v4}
      \fmffreeze
      \fmfdot{v1,v2,v3,v4}
    \end{fmfgraph*}
}
\qquad \qquad , \qquad \qquad
\parbox{16mm}{
    \begin{fmfgraph*}(16,14)
      \fmfleft{i1}
      \fmfright{i2}
      \fmftop{i3}
      \fmftop{i4}
      \fmfleft{v1}
      \fmfright{v2}
      \fmftop{v3}
      \fmftop{v4}
      \fmffreeze
      \fmfforce{(-0.2w,-0.2h)}{i1}
      \fmfforce{(-0.2w,1.2h)}{i2}
      \fmfforce{(1.2w,1.2h)}{i3}
      \fmfforce{(1.2w,-0.2h)}{i4}
      \fmfforce{(0w,0h)}{v1}
      \fmfforce{(0w,1h)}{v2}
      \fmfforce{(1w,1h)}{v3}
      \fmfforce{(1w,0h)}{v4}
      \fmfforce{(0.5w,0.5h)}{v5}
      \fmffreeze
      \fmf{plain}{i1,v1}
      \fmf{plain}{i2,v2}
      \fmf{plain}{i3,v3}
      \fmf{plain}{i4,v4}
      \fmf{plain}{v1,v2,v3,v4,v1}
      \fmf{plain}{v1,v3}
      \fmf{plain}{v2,v4}
      \fmffreeze
      \fmfdot{v1,v2,v3,v4,v5}
    \end{fmfgraph*}
} \qquad \, ,
\ee
\bs

\ni where all lines have index $1$. It is well known that the singular (UV) contributions of a diagram do not depend on masses and external momenta (and, in the $\overline{\text{MS}}$ scheme,
on $\gamma_E$ and $\zeta_2$ as well) \cite{Vladimirov:1979zm}. This observation is at the basis of the so-called infrared rearrangement (IRR) method \cite{Chetyrkin:1980pr} which, simply stated, allows
to set some external momenta to zero in order to extract $\ep$-poles. For the above diagrams, one may proceed in the following way:
%
\ms
\begin{subequations}
  \fleq{
&\mathrm{Sing} \left[ \quad
\parbox{16mm}{
    \begin{fmfgraph*}(16,14)
      \fmfleft{i1}
      \fmfright{i2}
      \fmftop{i3}
      \fmftop{i4}
      \fmfleft{v1}
      \fmfright{v2}
      \fmftop{v3}
      \fmftop{v4}
      \fmffreeze
      \fmfforce{(-0.2w,-0.2h)}{i1}
      \fmfforce{(-0.2w,1.2h)}{i2}
      \fmfforce{(1.2w,1.2h)}{i3}
      \fmfforce{(1.2w,-0.2h)}{i4}
      \fmfforce{(0w,0h)}{v1}
      \fmfforce{(0w,1h)}{v2}
      \fmfforce{(1w,1h)}{v3}
      \fmfforce{(1w,0h)}{v4}
      \fmfforce{(0.45w,0.45h)}{v1c}
      \fmfforce{(0.55w,0.55h)}{vc3}
      \fmffreeze
      \fmf{plain}{i1,v1}
      \fmf{plain}{i3,v3}
      \fmf{plain}{v1,v2,v3,v4,v1}
      \fmf{plain}{v1,v1c}
      \fmf{plain}{vc3,v3}
      \fmf{plain,fore=red}{v2,v4}
      \fmffreeze
      \fmfdot{v1,v2,v3,v4}
    \end{fmfgraph*}
} 
\quad \right] ~ = ~
\mathrm{Sing} \left[ \quad
\parbox{16mm}{
    \begin{fmfgraph*}(16,14)
      \fmftop{dt}
      \fmftop{db}
      \fmfleft{vde}
      \fmfright{vdo}
      \fmftop{vdn}
      \fmftop{vds}
      \fmffreeze
      \fmfforce{(-.2w,.5h)}{dl}
      \fmfforce{(1.2w,.5h)}{dr}
      \fmfforce{(.0w,0.5h)}{vde}
      \fmfforce{(1.0w,0.5h)}{vdo}
      \fmfforce{(.5w,0.5h)}{vdc}
      \fmfforce{(.5w,1.0h)}{vdn}
      \fmfforce{(.5w,0.0h)}{vds}
      \fmffreeze
      \fmf{plain}{dl,vde}
      \fmf{plain}{vdo,dr}
      \fmf{plain}{vde,vdc}
      \fmf{plain}{vdo,vdc}
      \fmf{plain}{vdn,vdc}
      \fmf{plain,left=0.9}{vde,vdo}
      \fmf{plain,right=0.9}{vde,vdo}
      \fmffreeze
      \fmfdot{vde,vdo,vdc,vdn}
    \end{fmfgraph*}
}
\quad \right] ~ = ~
\frac{1}{3\veps} ~
\mathrm{Finite}  \left[ \quad
\parbox{16mm}{
    \begin{fmfgraph*}(16,14)
      \fmftop{dt}
      \fmftop{db}
      \fmfleft{vde}
      \fmfright{vdo}
      \fmftop{vdn}
      \fmftop{vds}
      \fmffreeze
      \fmfforce{(-.2w,.5h)}{dl}
      \fmfforce{(1.2w,.5h)}{dr}
      \fmfforce{(.0w,0.5h)}{vde}
      \fmfforce{(1.0w,0.5h)}{vdo}
      \fmfforce{(.5w,0.5h)}{vdc}
      \fmfforce{(.5w,1.0h)}{vdn}
      \fmfforce{(.5w,0.0h)}{vds}
      \fmffreeze
      \fmf{plain}{dl,vde}
      \fmf{plain}{vdo,dr}
      \fmf{plain}{vdn,vds}
      \fmf{plain,left=0.9}{vde,vdo}
      \fmf{plain,right=0.9}{vde,vdo}
      \fmffreeze
      \fmfdot{vde,vdo,vdn,vds}
    \end{fmfgraph*}
} \quad \right] ~ = ~\frac{2}{\ep} \, \zeta_3 \, .
\\
&\mathrm{Sing} \left[\quad
\parbox{16mm}{
    \begin{fmfgraph*}(16,14)
      \fmfleft{i1}
      \fmfright{i2}
      \fmftop{i3}
      \fmftop{i4}
      \fmfleft{v1}
      \fmfright{v2}
      \fmftop{v3}
      \fmftop{v4}
      \fmffreeze
      \fmfforce{(-0.2w,-0.2h)}{i1}
      \fmfforce{(-0.2w,1.2h)}{i2}
      \fmfforce{(1.2w,1.2h)}{i3}
      \fmfforce{(1.2w,-0.2h)}{i4}
      \fmfforce{(0w,0h)}{v1}
      \fmfforce{(0w,1h)}{v2}
      \fmfforce{(1w,1h)}{v3}
      \fmfforce{(1w,0h)}{v4}
      \fmfforce{(0.5w,0.5h)}{v5}
      \fmffreeze
      \fmf{plain}{i1,v1}
      \fmf{plain}{i4,v4}
      \fmf{plain}{v1,v2,v3,v4,v1}
      \fmf{plain}{v1,v3}
      \fmf{plain}{v2,v4}
      \fmffreeze
      \fmfdot{v1,v2,v3,v4,v5}
    \end{fmfgraph*}
}
\quad \right] ~ = ~
\mathrm{Sing} \left[ \quad
\parbox{16mm}{
    \begin{fmfgraph*}(16,14)
      \fmftop{dt}
      \fmftop{db}
      \fmfleft{vde}
      \fmfright{vdo}
      \fmftop{vdn1}
      \fmftop{vdn2}
      \fmftop{vds}
      \fmffreeze
      \fmfforce{(-.2w,.5h)}{dl}
      \fmfforce{(1.2w,.5h)}{dr}
      \fmfforce{(.0w,0.5h)}{vde}
      \fmfforce{(1.0w,0.5h)}{vdo}
      \fmfforce{(.5w,0.5h)}{vdc}
      \fmfforce{(.25w,.95h)}{vdn1}
      \fmfforce{(.75w,.95h)}{vdn2}
      \fmfforce{(.5w,0.0h)}{vds}
      \fmffreeze
      \fmf{plain}{dl,vde}
      \fmf{plain}{vdo,dr}
      \fmf{plain}{vde,vdc}
      \fmf{plain}{vdo,vdc}
      \fmf{plain}{vdn1,vdc}
      \fmf{plain}{vdn2,vdc}
      \fmf{plain,left=0.9}{vde,vdo}
      \fmf{plain,right=0.9}{vde,vdo}
      \fmffreeze
      \fmfdot{vde,vdo,vdc,vdn1,vdn2}
    \end{fmfgraph*}
}
\quad \right] ~ = ~
\frac{1}{4\veps} ~
\mathrm{Finite}  \left[ \quad
\parbox{16mm}{
    \begin{fmfgraph*}(16,14)
      \fmfleft{i}
      \fmfright{o}
      \fmfleft{ve}
      \fmfright{vo}
      \fmftop{vn}
      \fmftop{vs}
      \fmffreeze
      \fmfforce{(-0.2w,0.5h)}{i}
      \fmfforce{(1.2w,0.5h)}{o}
      \fmfforce{(0w,0.5h)}{ve}
      \fmfforce{(1.0w,0.5h)}{vo}
      \fmfforce{(.3w,0.9h)}{v1}
      \fmfforce{(.7w,0.9h)}{v2}
      \fmfforce{(.5w,0.05h)}{vs}
      \fmffreeze
      \fmf{plain}{i,ve}
      \fmf{plain}{vo,o}
      \fmf{plain,left=0.8}{ve,vo}
      \fmf{plain,left=0.8}{vo,ve}
      \fmf{plain}{vs,v1}
      \fmf{plain}{vs,v2}
      \fmffreeze
      \fmfdot{ve,v1,v2,vo,vs}
    \end{fmfgraph*}
} \quad \right]  ~  = ~ \frac{5}{\ep} \,\zeta_5 \, ,
}
\end{subequations}
where the vertex diagrams were reduced to p-type ones and the results of the previous section were used to compute the finite parts of the massless two-loop and three-loop V-shaped p-type diagrams. 

\subsection{Five loop calculations in the $\Phi^4$-model}
\label{sec:examples5}

\begin{figure}
  \begin{center}
\includegraphics{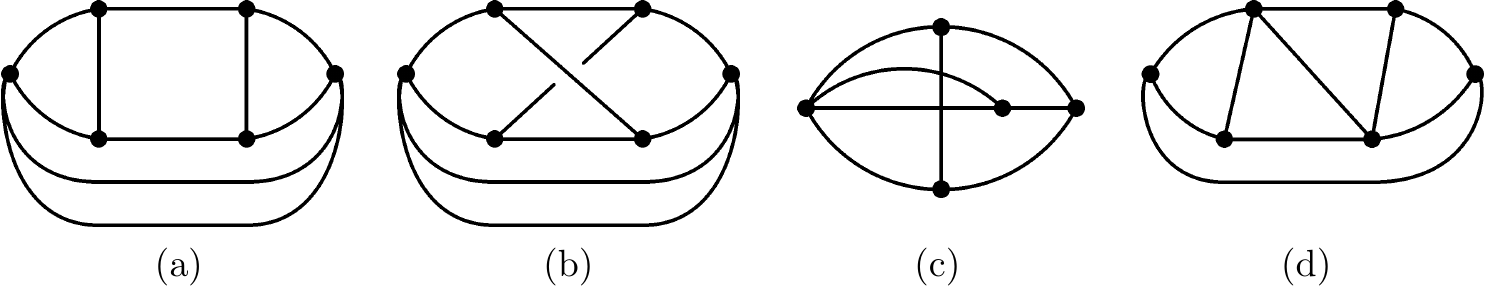}
\caption{\label{fig:Phi4}
  The four most complicated $5$-loop diagrams of the $\Phi^4$-model.}
  \end{center}
\end{figure}

Following Kazakov, we may even go further and apply the described techniques to five-loop renormalization-group calculations in the $\Phi^4$-model. 
In this connection, let's recall that the five-loop anomalous dimensions and beta-function in this
model in the MS-scheme were calculated long time ago in \cite{Gorishnii:1983gp}.
 The full computation required the evaluation of about 120 diagrams. All but four were calculated analytically using integration by parts. For the
remaining four most complicated diagrams, see Fig.~\ref{fig:Phi4}, an expansion in Gegenbauer polynomials provides
the result in the form of a triple infinite convergent series. Computer summation of the series made it
possible to achieve good accuracy. However, it is understandable that one would like to obtain an analytic
 evaluation of these contributions. In order to test the possibilities of the considered above methods,
we will present the calculation some of
these complicated diagrams. 
 As it was shown in \cite{Kazakov:1983ns}, already in \cite{Gorishnii:1983gp},
it proved to be possible to sum the series for
diagram ``a'' analytically and express the obtained result in terms of $\zeta_5$ and $\zeta_6$.
The results for the other diagrams have also been obtained by Kazakov in \cite{Kazakov:1983ns,Kazakov:1984km,Kazakov:1983pk}. 

In the following, we will focus on a detailed computation of diagrams ``a'' and ``d''.

\subsubsection{Diagram ``a''}
\label{sec:examples5a}

For diagram ``a'' (where all indices are equal to $1$) the problem is equivalent to calculating the diagram:
%
\bea
 I_a ~~ = ~\quad 
 \parbox{16mm}{
\begin{fmfgraph*}(16,14)
      \fmfleft{vl}
      \fmfright{vr}
      \fmffreeze
      \fmfforce{(-0.2w,0.5h)}{vl}
      \fmfforce{(-0.1w,0.5h)}{vl1}
      \fmfforce{(1.2w,0.5h)}{vr}
      \fmfforce{(1.1w,0.5h)}{vr4}
      \fmfforce{(-0.05w,0.5h)}{v1}
      \fmfforce{(1.05w,0.5h)}{v4}
      \fmfforce{(.25w,0.8h)}{vn1}
      \fmfforce{(.75w,0.8h)}{vn2}
      \fmfforce{(.25w,0.2h)}{vs1}
      \fmfforce{(.75w,0.2h)}{vs2}
      \fmffreeze
      \fmf{plain}{vl,v1}
      \fmf{plain,left=0.25}{v1,vn1}
      \fmf{plain}{vn1,vn2}
      \fmf{plain,left=0.25}{vn2,v4}
      \fmf{plain,left=0.25}{v4,vs2}
      \fmf{plain}{vs2,vs1}
      \fmf{plain,left=0.25}{vs1,v1}
      \fmf{plain}{vs1,vn1}
      \fmf{plain}{vs2,vn2}
      \fmf{plain}{v4,vr}
\fmfdot{v1,v4,vn1,vn2,vs1,vs2}
\fmffreeze
\end{fmfgraph*}
} \qquad \, ,
\nonumber 
\eea
with $O(\ep)$ accuracy.

Applying the IBP procedure to the right triangle with the vertical distinguished line, we first obtain the following relation:
\bea
 \parbox{16mm}{
\begin{fmfgraph*}(16,14)
      \fmfleft{vl}
      \fmfright{vr}
      \fmffreeze
      \fmfforce{(-0.15w,0.5h)}{vl}
      \fmfforce{(-0.1w,0.5h)}{vl1}
      \fmfforce{(1.15w,0.5h)}{vr}
      \fmfforce{(1.1w,0.5h)}{vr4}
      \fmfforce{(-0.05w,0.5h)}{v1}
      \fmfforce{(1.05w,0.5h)}{v4}
      \fmfforce{(.25w,0.8h)}{vn1}
      \fmfforce{(.75w,0.8h)}{vn2}
      \fmfforce{(.25w,0.2h)}{vs1}
      \fmfforce{(.75w,0.2h)}{vs2}
      \fmffreeze
      \fmf{plain}{vl,v1}
      \fmf{plain,left=0.25}{v1,vn1}
      \fmf{plain}{vn1,vn2}
      \fmf{plain,left=0.25}{vn2,v4}
      \fmf{plain,left=0.25}{v4,vs2}
      \fmf{plain}{vs2,vs1}
      \fmf{plain,left=0.25}{vs1,v1}
      \fmf{plain}{vs1,vn1}
      \fmf{plain}{vs2,vn2}
      \fmf{plain}{v4,vr}
\fmfdot{v1,v4,vn1,vn2,vs1,vs2}
\fmffreeze
\end{fmfgraph*}
} \quad  (D-4) ~~ = ~~ 2 \left[ \quad 
 \parbox{16mm}{
\begin{fmfgraph*}(16,14)
      \fmfleft{vl}
      \fmfright{vr}
      \fmffreeze
      \fmfforce{(-0.15w,0.5h)}{vl}
      \fmfforce{(-0.1w,0.5h)}{vl1}
      \fmfforce{(1.6w,0.5h)}{vr}
      \fmfforce{(1.55w,0.5h)}{vr4}
      \fmfforce{(-0.05w,0.5h)}{v1}
      \fmfforce{(1.5w,0.5h)}{v4}
      \fmfforce{(.5w,0.8h)}{vn}
      \fmfforce{(1.0w,0.5h)}{v3}
      \fmfforce{(.5w,0.2h)}{vs}
      \fmffreeze
      \fmf{plain}{vl,v1}
      \fmf{plain,left=0.5}{v1,v3}
      \fmf{plain,left=0.5}{v3,v1}
      \fmf{plain,left=0.5}{v3,v4}
	\fmf{plain,left=0.5,label=\footnotesize{$2$},l.d=0.05h}{v4,v3}
      \fmf{plain}{vn,vs}
      \fmf{plain}{v4,vr}
\fmfdot{v1,v4,vn,vs,v3}
\fmffreeze
\end{fmfgraph*}
} \qquad \qquad
- \qquad
\parbox{16mm}{
\begin{fmfgraph*}(16,14)
      \fmfleft{vl}
      \fmfright{vr}
      \fmffreeze
      \fmfforce{(-0.15w,0.5h)}{vl}
      \fmfforce{(-0.1w,0.5h)}{vl1}
      \fmfforce{(1.15w,0.5h)}{vr}
      \fmfforce{(1.1w,0.5h)}{vr4}
      \fmfforce{(-0.05w,0.5h)}{v1}
      \fmfforce{(1.05w,0.5h)}{v4}
      \fmfforce{(.25w,0.8h)}{vn1}
      \fmfforce{(.75w,0.8h)}{vn2}
      \fmfforce{(.25w,0.2h)}{vs1}
      \fmfforce{(.75w,0.2h)}{vs2}
      \fmfforce{(.5w,0.2h)}{vs}
      \fmffreeze
      \fmf{plain}{vl,v1}
      \fmf{plain,left=0.25}{v1,vn1}
      \fmf{plain}{vn1,vn2}
      \fmf{plain,left=0.25}{vn2,v4}
	\fmf{plain,left=0.25,label=\footnotesize{$2$},l.d=0.04w}{v4,vs2}
      \fmf{plain}{vs2,vs1}
      \fmf{plain,left=0.25}{vs1,v1}
      \fmf{plain}{vs,vn2}
      \fmf{plain}{vs,vn1}
      \fmf{plain}{v4,vr}
\fmfdot{v1,v4,vn1,vn2,vs}
\fmffreeze
\end{fmfgraph*}
}
\quad
\right] \quad \, .
\label{DiA1}
\eea
To evaluate the second diagram in the rhs of Eq.~(\ref{DiA1}), it is convenient to consider the second three loop master integral 
$J_2(D,p,\al_1,\al_2,\al_3,\al_4,\al_5,\al_6,\al_7)$ with $\al_i=1$ $(i=1,...,7)$. We may then apply the IBP procedure to this diagram twice: 
for the right triangle with the upper distinguished line and for the middle triangle with the upper distinguished line. The corresponding results read:
\begin{subequations}
  \fleq{
&\parbox{16mm}{
\begin{fmfgraph*}(16,14)
      \fmfleft{vl}
      \fmfright{vr}
      \fmffreeze
      \fmfforce{(-0.15w,0.5h)}{vl}
      \fmfforce{(-0.1w,0.5h)}{vl1}
      \fmfforce{(1.15w,0.5h)}{vr}
      \fmfforce{(1.1w,0.5h)}{vr4}
      \fmfforce{(-0.05w,0.5h)}{v1}
      \fmfforce{(1.05w,0.5h)}{v4}
      \fmfforce{(.25w,0.8h)}{vn1}
      \fmfforce{(.75w,0.8h)}{vn2}
      \fmfforce{(.25w,0.2h)}{vs1}
      \fmfforce{(.75w,0.2h)}{vs2}
      \fmfforce{(.5w,0.2h)}{vs}
      \fmffreeze
      \fmf{plain,label=$p$}{vl,v1}
      \fmf{plain,left=0.25}{v1,vn1}
      \fmf{plain}{vn1,vn2}
      \fmf{plain,left=0.25}{vn2,v4}
      \fmf{plain,left=0.25}{v4,vs2}
      \fmf{plain}{vs2,vs1}
      \fmf{plain,left=0.25}{vs1,v1}
      \fmf{plain}{vs,vn2}
      \fmf{plain}{vs,vn1}
      \fmf{plain}{v4,vr}
\fmfdot{v1,v4,vn1,vn2,vs}
\fmffreeze
\end{fmfgraph*}
}
\quad (D-4) ~~ = ~~ 2 \quad
\parbox{16mm}{
    \begin{fmfgraph*}(16,14)
      \fmfleft{i}
      \fmfright{o}
      \fmfleft{ve}
      \fmfright{vo}
      \fmftop{vn}
      \fmftop{vs}
      \fmffreeze
      \fmfforce{(-0.1w,0.5h)}{i}
      \fmfforce{(1.1w,0.5h)}{o}
      \fmfforce{(0w,0.5h)}{ve}
      \fmfforce{(1.0w,0.5h)}{vo}
      \fmfforce{(.5w,0.95h)}{vn}
      \fmfforce{(.5w,0.05h)}{vs}
      \fmfforce{(.5w,0.5h)}{vc}
      \fmffreeze
      \fmf{plain}{i,ve}
      \fmf{plain,left=0.8}{ve,vo}
      \fmf{plain,left=0.8}{vo,ve}
      \fmf{plain}{vs,vn}
      \fmf{plain}{vo,o}
      \fmf{plain,right=0.5}{vo,vs}
      \fmf{phantom,label={\scriptsize{$2$}},l.d=0.05w,l.s=left}{vc,vo}
      \fmffreeze
      \fmfdot{ve,vn,vo,vs}
    \end{fmfgraph*}
} \quad - \quad
\parbox{16mm}{
    \begin{fmfgraph*}(16,14)
      \fmfleft{i}
      \fmfright{o}
      \fmfleft{ve}
      \fmfright{vo}
      \fmftop{vn}
      \fmftop{vs}
      \fmffreeze
      \fmfforce{(-0.1w,0.5h)}{i}
      \fmfforce{(1.1w,0.5h)}{o}
      \fmfforce{(0w,0.5h)}{ve}
      \fmfforce{(1.0w,0.5h)}{vo}
      \fmfforce{(.5w,0.95h)}{vn}
      \fmfforce{(.5w,0.05h)}{vs}
      \fmffreeze
      \fmf{plain}{i,ve}
      \fmf{plain,left=0.8}{ve,vo}
      \fmf{plain,left=0.8}{vo,ve}
      \fmf{plain,left=0.3}{vs,vn}
	    \fmf{plain,left=0.3,label=\scriptsize{$2$},l.d=0.05w}{vn,vs}
      \fmf{plain}{vo,o}
      \fmffreeze
      \fmfdot{ve,vn,vo,vs}
    \end{fmfgraph*}
} \quad - \quad p^2 \quad
\parbox{16mm}{
\begin{fmfgraph*}(16,14)
      \fmfleft{vl}
      \fmfright{vr}
      \fmffreeze
      \fmfforce{(-0.15w,0.5h)}{vl}
      \fmfforce{(-0.1w,0.5h)}{vl1}
      \fmfforce{(1.15w,0.5h)}{vr}
      \fmfforce{(1.1w,0.5h)}{vr4}
      \fmfforce{(-0.05w,0.5h)}{v1}
      \fmfforce{(1.05w,0.5h)}{v4}
      \fmfforce{(.25w,0.8h)}{vn1}
      \fmfforce{(.75w,0.8h)}{vn2}
      \fmfforce{(.25w,0.2h)}{vs1}
      \fmfforce{(.75w,0.2h)}{vs2}
      \fmfforce{(.5w,0.2h)}{vs}
      \fmffreeze
      \fmf{plain}{vl,v1}
      \fmf{plain,left=0.25}{v1,vn1}
      \fmf{plain}{vn1,vn2}
      \fmf{plain,left=0.25}{vn2,v4}
	\fmf{plain,left=0.25,label=\footnotesize{$2$},l.d=0.04w}{v4,vs2}
      \fmf{plain}{vs2,vs1}
      \fmf{plain,left=0.25}{vs1,v1}
      \fmf{plain}{vs,vn2}
      \fmf{plain}{vs,vn1}
      \fmf{plain}{v4,vr}
\fmfdot{v1,v4,vn1,vn2,vs}
\fmffreeze
\end{fmfgraph*}
}
\quad \, ,  \\
&\nonum \\
&\parbox{16mm}{
\begin{fmfgraph*}(16,14)
      \fmfleft{vl}
      \fmfright{vr}
      \fmffreeze
      \fmfforce{(-0.15w,0.5h)}{vl}
      \fmfforce{(-0.1w,0.5h)}{vl1}
      \fmfforce{(1.15w,0.5h)}{vr}
      \fmfforce{(1.1w,0.5h)}{vr4}
      \fmfforce{(-0.05w,0.5h)}{v1}
      \fmfforce{(1.05w,0.5h)}{v4}
      \fmfforce{(.25w,0.8h)}{vn1}
      \fmfforce{(.75w,0.8h)}{vn2}
      \fmfforce{(.25w,0.2h)}{vs1}
      \fmfforce{(.75w,0.2h)}{vs2}
      \fmfforce{(.5w,0.2h)}{vs}
      \fmffreeze
      \fmf{plain}{vl,v1}
      \fmf{plain,left=0.25}{v1,vn1}
      \fmf{plain}{vn1,vn2}
      \fmf{plain,left=0.25}{vn2,v4}
      \fmf{plain,left=0.25}{v4,vs2}
      \fmf{plain}{vs2,vs1}
      \fmf{plain,left=0.25}{vs1,v1}
      \fmf{plain}{vs,vn2}
      \fmf{plain}{vs,vn1}
      \fmf{plain}{v4,vr}
\fmfdot{v1,v4,vn1,vn2,vs}
\fmffreeze
\end{fmfgraph*}
}
\quad (D-4) ~~ = ~~ 2~~ \left[
\quad
\parbox{16mm}{
    \begin{fmfgraph*}(16,14)
      \fmfleft{i}
      \fmfright{o}
      \fmfleft{ve}
      \fmfright{vo}
      \fmftop{vn}
      \fmftop{vs}
      \fmffreeze
      \fmfforce{(-0.1w,0.5h)}{i}
      \fmfforce{(1.1w,0.5h)}{o}
      \fmfforce{(0w,0.5h)}{ve}
      \fmfforce{(1.0w,0.5h)}{vo}
      \fmfforce{(.5w,0.95h)}{vn}
      \fmfforce{(.5w,0.05h)}{vs}
      \fmffreeze
      \fmf{plain}{i,ve}
      \fmf{plain,left=0.8}{ve,vo}
      \fmf{plain,left=0.8}{vo,ve}
      \fmf{plain,left=0.3}{vs,vn}
	    \fmf{plain,left=0.3,label=\scriptsize{$2$},l.d=0.05w}{vn,vs}
      \fmf{plain}{vo,o}
      \fmffreeze
      \fmfdot{ve,vn,vo,vs}
    \end{fmfgraph*}
} \quad - \quad 
\parbox{16mm}{
    \begin{fmfgraph*}(16,14)
      \fmfleft{i}
      \fmfright{o}
      \fmfleft{ve}
      \fmfright{vo}
      \fmftop{vn}
      \fmftop{vs}
      \fmffreeze
      \fmfforce{(-0.1w,0.5h)}{i}
      \fmfforce{(1.1w,0.5h)}{o}
      \fmfforce{(0w,0.5h)}{ve}
      \fmfforce{(1.0w,0.5h)}{vo}
      \fmfforce{(.5w,0.95h)}{vn}
      \fmfforce{(.5w,0.05h)}{vs}
      \fmffreeze
      \fmf{plain}{i,ve}
      \fmf{plain,left=0.8}{ve,vo}
      \fmf{plain,left=0.8}{vo,ve}
	    \fmf{phantom,left=0.5,label=\footnotesize{$2$},l.d=-0.01w}{vo,vs}
      \fmf{plain}{vs,vn}
      \fmf{plain}{vo,o}
      \fmf{plain,right=0.5}{vo,vs}
      \fmffreeze
      \fmfdot{ve,vn,vo,vs}
    \end{fmfgraph*}
} \quad
\right] \quad \, .
}
\end{subequations}
Taking the combination of these two equations, we obtain, for the second diagram in the rhs of (\ref{DiA1}), the following result:
\bea
\parbox{16mm}{
\begin{fmfgraph*}(16,14)
      \fmfleft{vl}
      \fmfright{vr}
      \fmffreeze
      \fmfforce{(-0.15w,0.5h)}{vl}
      \fmfforce{(-0.1w,0.5h)}{vl1}
      \fmfforce{(1.15w,0.5h)}{vr}
      \fmfforce{(1.1w,0.5h)}{vr4}
      \fmfforce{(-0.05w,0.5h)}{v1}
      \fmfforce{(1.05w,0.5h)}{v4}
      \fmfforce{(.25w,0.8h)}{vn1}
      \fmfforce{(.75w,0.8h)}{vn2}
      \fmfforce{(.25w,0.2h)}{vs1}
      \fmfforce{(.75w,0.2h)}{vs2}
      \fmfforce{(.5w,0.2h)}{vs}
      \fmffreeze
      \fmf{plain}{vl,v1}
      \fmf{plain,left=0.25}{v1,vn1}
      \fmf{plain}{vn1,vn2}
      \fmf{plain,left=0.25}{vn2,v4}
	\fmf{plain,left=0.25,label=\footnotesize{$2$},l.d=0.04w}{v4,vs2}
      \fmf{plain}{vs2,vs1}
      \fmf{plain,left=0.25}{vs1,v1}
      \fmf{plain}{vs,vn2}
      \fmf{plain}{vs,vn1}
      \fmf{plain}{v4,vr}
\fmfdot{v1,v4,vn1,vn2,vs}
\fmffreeze
\end{fmfgraph*}
}
\quad = \quad \frac{1}{p^2}~ \left[ \quad 4 \quad
\parbox{16mm}{
    \begin{fmfgraph*}(16,14)
      \fmfleft{i}
      \fmfright{o}
      \fmfleft{ve}
      \fmfright{vo}
      \fmftop{vn}
      \fmftop{vs}
      \fmffreeze
      \fmfforce{(-0.1w,0.5h)}{i}
      \fmfforce{(1.1w,0.5h)}{o}
      \fmfforce{(0w,0.5h)}{ve}
      \fmfforce{(1.0w,0.5h)}{vo}
      \fmfforce{(.5w,0.95h)}{vn}
      \fmfforce{(.5w,0.05h)}{vs}
      \fmfforce{(.5w,0.5h)}{vc}
      \fmffreeze
      \fmf{plain}{i,ve}
      \fmf{plain,left=0.8}{ve,vo}
      \fmf{plain,left=0.8}{vo,ve}
      \fmf{plain}{vs,vn}
      \fmf{plain}{vo,o}
      \fmf{plain,right=0.5}{vo,vs}
      \fmf{phantom,label={\scriptsize{$2$}},l.d=0.05w,l.s=left}{vc,vo}
      \fmffreeze
      \fmfdot{ve,vn,vo,vs}
    \end{fmfgraph*}
} \quad - \quad 3 \quad
\parbox{16mm}{
    \begin{fmfgraph*}(16,14)
      \fmfleft{i}
      \fmfright{o}
      \fmfleft{ve}
      \fmfright{vo}
      \fmftop{vn}
      \fmftop{vs}
      \fmffreeze
      \fmfforce{(-0.1w,0.5h)}{i}
      \fmfforce{(1.1w,0.5h)}{o}
      \fmfforce{(0w,0.5h)}{ve}
      \fmfforce{(1.0w,0.5h)}{vo}
      \fmfforce{(.5w,0.95h)}{vn}
      \fmfforce{(.5w,0.05h)}{vs}
      \fmffreeze
      \fmf{plain}{i,ve}
      \fmf{plain,left=0.8}{ve,vo}
      \fmf{plain,left=0.8}{vo,ve}
      \fmf{plain,left=0.3}{vs,vn}
	    \fmf{plain,left=0.3,label=\scriptsize{$2$},l.d=0.05w}{vn,vs}
      \fmf{plain}{vo,o}
      \fmffreeze
      \fmfdot{ve,vn,vo,vs}
    \end{fmfgraph*}
} \quad
\right] \quad \, .
\label{DiA1a}
\eea
Then, substituting back the result of Eq.~(\ref{DiA1a}) in the rhs of Eq.~(\ref{DiA1}) and calculating the remaining one-loop parts, yields, for rhs of (\ref{DiA1}):
%
\fleq{
\frac{2G(D,2,1)}{p^2} \left[ 
\quad
\parbox{16mm}{
    \begin{fmfgraph*}(16,14)
      \fmfleft{i}
      \fmfright{o}
      \fmfleft{ve}
      \fmfright{vo}
      \fmftop{vn}
      \fmftop{vs}
      \fmffreeze
      \fmfforce{(-0.1w,0.5h)}{i}
      \fmfforce{(1.1w,0.5h)}{o}
      \fmfforce{(0w,0.5h)}{ve}
      \fmfforce{(1.0w,0.5h)}{vo}
      \fmfforce{(.5w,0.95h)}{vn}
      \fmfforce{(.5w,0.05h)}{vs}
      \fmffreeze
      \fmf{plain}{i,ve}
      \fmf{plain,left=0.8}{ve,vo}
      \fmf{plain,left=0.8}{vo,ve}
      \fmf{plain}{vs,vn}
      \fmf{plain}{vo,o}
      \fmffreeze
      \fmfdot{ve,vn,vo,vs}
    \end{fmfgraph*}
} \quad p^{2\ep} -  \quad 4 \quad
\parbox{16mm}{
    \begin{fmfgraph*}(16,14)
      \fmfleft{i}
      \fmfright{o}
      \fmfleft{ve}
      \fmfright{vo}
      \fmftop{vn}
      \fmftop{vs}
      \fmffreeze
      \fmfforce{(-0.1w,0.5h)}{i}
      \fmfforce{(1.1w,0.5h)}{o}
      \fmfforce{(0w,0.5h)}{ve}
      \fmfforce{(1.0w,0.5h)}{vo}
      \fmfforce{(.5w,0.95h)}{vn}
      \fmfforce{(.5w,0.05h)}{vs}
      \fmffreeze
      \fmf{plain}{i,ve}
      \fmf{plain,left=0.8}{ve,vo}
      \fmf{plain,left=0.8}{vo,ve}
	    \fmf{phantom,left=0.5,label=\footnotesize{$1+\ep$},l.d=-0.01w}{vo,vs}
      \fmf{plain}{vs,vn}
      \fmf{plain}{vo,o}
      \fmffreeze
      \fmfdot{ve,vn,vo,vs}
    \end{fmfgraph*}
} \quad
+ \quad 3 \quad 
\parbox{16mm}{
    \begin{fmfgraph*}(16,14)
      \fmfleft{i}
      \fmfright{o}
      \fmfleft{ve}
      \fmfright{vo}
      \fmftop{vn}
      \fmftop{vs}
      \fmffreeze
      \fmfforce{(-0.1w,0.5h)}{i}
      \fmfforce{(1.1w,0.5h)}{o}
      \fmfforce{(0w,0.5h)}{ve}
      \fmfforce{(1.0w,0.5h)}{vo}
      \fmfforce{(.5w,0.95h)}{vn}
      \fmfforce{(.5w,0.05h)}{vs}
      \fmffreeze
      \fmf{plain}{i,ve}
      \fmf{plain,left=0.8}{ve,vo}
      \fmf{plain,left=0.8}{vo,ve}
      \fmf{plain,label=\tiny{$1+\ep$},l.d=0.01w}{vs,vn}
      \fmf{plain}{vo,o}
      \fmffreeze
      \fmfdot{ve,vn,vo,vs}
    \end{fmfgraph*}
}
\quad \right] \quad \, .
\nonumber
}
Evaluating the two-loop master integrals appearing in this equation using the expansions provided by Eqs.~(\ref{IBPex3}) and
(\ref{IBPex3-2}),
we obtain 
the expression of the diagram $I_a$ with an accuracy even reaching $O(\ep^2)$:
%
\fleq{
\parbox{16mm}{
\begin{fmfgraph*}(16,14)
      \fmfleft{vl}
      \fmfright{vr}
      \fmffreeze
      \fmfforce{(-0.2w,0.5h)}{vl}
      \fmfforce{(-0.1w,0.5h)}{vl1}
      \fmfforce{(1.2w,0.5h)}{vr}
      \fmfforce{(1.1w,0.5h)}{vr4}
      \fmfforce{(-0.05w,0.5h)}{v1}
      \fmfforce{(1.05w,0.5h)}{v4}
      \fmfforce{(.25w,0.8h)}{vn1}
      \fmfforce{(.75w,0.8h)}{vn2}
      \fmfforce{(.25w,0.2h)}{vs1}
      \fmfforce{(.75w,0.2h)}{vs2}
      \fmffreeze
      \fmf{plain}{vl,v1}
      \fmf{plain,left=0.25}{v1,vn1}
      \fmf{plain}{vn1,vn2}
      \fmf{plain,left=0.25}{vn2,v4}
      \fmf{plain,left=0.25}{v4,vs2}
      \fmf{plain}{vs2,vs1}
      \fmf{plain,left=0.25}{vs1,v1}
      \fmf{plain}{vs1,vn1}
      \fmf{plain}{vs2,vn2}
      \fmf{plain}{v4,vr}
\fmfdot{v1,v4,vn1,vn2,vs1,vs2}
\fmffreeze
\end{fmfgraph*}
} \quad  = 
\frac{\hat{K}_3}{{1-2\ep}}
\, \biggl[20 \zeta_5  
+ \Bigl[50 \zeta_6 + 44 \zeta_3^2 \Bigr] \ep+ \Bigl[317 \zeta_7 + 132 \zeta_3 \zeta_4  \Bigr] \ep^2 + O(\ep^3)\biggr] \, \frac{1}{p^{{2}(2+3)\ep}}\, .
\label{DiA2}
}
Taking into account of the two additional loops produces an additional factor:
%
\fleq{
G(D,1,2+3\ep)G(D,1,1+4\ep) = - \frac{1}{20\ep^2} \, \frac{1}{(1+3\ep)(1-6\ep)} \, 
\frac{\Gamma^2(1-\ep)\Gamma(1-4\ep)\Gamma(1+5\ep)}{\Gamma(1+{3}\ep)\Gamma(1-6\ep)}\, .
  \label{Dia3}
}
  %
Hence, the final result for 
diagram ``a'' has the following form:
%
\bea
\mathrm{Sing} \left[ \quad  
\parbox{15mm}{
\begin{fmfgraph*}(15,11)
      \fmfleft{vl}
      \fmfright{vr}
      \fmffreeze
      \fmfforce{(-0.2w,0.5h)}{vl}
      \fmfforce{(-0.1w,0.5h)}{vl1}
      \fmfforce{(1.2w,0.5h)}{vr}
      \fmfforce{(1.1w,0.5h)}{vr4}
      \fmfforce{(-0.05w,0.5h)}{v1}
      \fmfforce{(1.05w,0.5h)}{v4}
      \fmfforce{(.25w,0.8h)}{vn1}
      \fmfforce{(.75w,0.8h)}{vn2}
      \fmfforce{(.25w,0.2h)}{vs1}
      \fmfforce{(.75w,0.2h)}{vs2}
      \fmffreeze
      \fmf{phantom}{vl,v1}
      \fmf{plain,left=0.25}{v1,vn1}
      \fmf{plain}{vn1,vn2}
      \fmf{plain,left=0.25}{vn2,v4}
      \fmf{plain,left=0.25}{v4,vs2}
      \fmf{plain}{vs2,vs1}
      \fmf{plain,left=0.25}{vs1,v1}
      \fmf{plain}{vs1,vn1}
      \fmf{plain}{vs2,vn2}
      \fmf{phantom}{v4,vr}
\fmfdot{v1,v4,vn1,vn2,vs1,vs2}
\fmffreeze
\fmfposition
\fmfi{plain}{vloc(__v4) ..controls vloc(__vr4) and (xpart(vloc(__vr4)),-0.0h) ..(xpart(vloc(__vs2)),-0.0h)}
\fmfi{plain}{(xpart(vloc(__vs1)),-0.0h) ..controls (xpart(vloc(__vl1)),-0.0h) and vloc(__vl1) ..vloc(__v1)}
\fmfi{plain}{(xpart(vloc(__vs1)),-0.0h)..(xpart(vloc(__vs2)),-0.0h)}
\fmfi{plain}{vloc(__v4) ..controls vloc(__vr4) and (xpart(vloc(__vr4)),-0.2h) ..(xpart(vloc(__vs2)),-0.2h)}
\fmfi{plain}{(xpart(vloc(__vs1)),-0.2h) ..controls (xpart(vloc(__vl1)),-0.2h) and vloc(__vl1) ..vloc(__v1)}
\fmfi{plain}{(xpart(vloc(__vs1)),-0.2h)..(xpart(vloc(__vs2)),-0.2h)}
\end{fmfgraph*}
}
\quad \right] &=& - \frac{\hat{K}_5}{20(1-2\ep)(1+3\ep)(1-6\ep)} \, \biggl[\frac{20}{\ep^2} \zeta_5  \nonumber \\
&&+ \Bigl[50 \zeta_6 + 44 \zeta_3^2 \Bigr] \frac{1}{\ep}+ 317 \zeta_7 + 132 \zeta_3 \zeta_4  + O(\ep)\biggr] \, \frac{1}{p^{{10}\ep}}\, .
\label{DiA4}
\eea

\subsubsection{Diagram ``d''}
\label{sec:examples5d}

The evaluation of the singular part of diagram ``d'' (where all indices are equal to $1$) 
is equivalent to calculating the diagram:
\be
I_d ~~ = ~\quad \parbox{16mm}{
\begin{fmfgraph*}(16,14)
      \fmfleft{vl}
      \fmfright{vr}
      \fmffreeze
      \fmfforce{(-0.2w,0.5h)}{vl}
      \fmfforce{(-0.1w,0.5h)}{vl1}
      \fmfforce{(1.2w,0.5h)}{vr}
      \fmfforce{(1.1w,0.5h)}{vr4}
      \fmfforce{(-0.05w,0.5h)}{v1}
      \fmfforce{(1.05w,0.5h)}{v4}
      \fmfforce{(.3w,0.8h)}{vn1}
      \fmfforce{(.78w,0.8h)}{vn2}
      \fmfforce{(.2w,0.2h)}{vs1}
      \fmfforce{(.7w,0.2h)}{vs2}
      \fmffreeze
      \fmf{plain}{vl,v1}
      \fmf{plain,left=0.25}{v1,vn1}
      \fmf{plain}{vn1,vn2}
      \fmf{plain,left=0.25}{vn2,v4}
      \fmf{plain,left=0.25}{v4,vs2}
      \fmf{plain}{vs2,vs1}
      \fmf{plain,left=0.25}{vs1,v1}
      \fmf{plain}{vs1,vn1}
      \fmf{plain}{vs2,vn2}
      \fmf{plain}{vn1,vs2}
      \fmf{plain}{v4,vr}
\fmfdot{v1,v4,vn1,vn2,vs1,vs2}
\fmffreeze
\end{fmfgraph*}
} \quad \, ,
\ee
with $O(\ep^0)$ accuracy. Indeed, the additional loop provides the factor: $G(1,1+4\ep) = \Gamma(1+\ep)/(5\ep) + O(\ep^0)$.

In order to proceed with the computation, it is firstly convenient to transform the considered diagram to $x$-space by Fourier transform and then to return to $p$-space using the dual
transformation:
\bea
\parbox{16mm}{
\begin{fmfgraph*}(16,14)
      \fmfleft{vl}
      \fmfright{vr}
      \fmffreeze
      \fmfforce{(-0.3w,0.5h)}{vl}
      \fmfforce{(-0.1w,0.5h)}{vl1}
      \fmfforce{(1.2w,0.5h)}{vr}
      \fmfforce{(1.1w,0.5h)}{vr4}
      \fmfforce{(-0.05w,0.5h)}{v1}
      \fmfforce{(1.05w,0.5h)}{v4}
      \fmfforce{(.3w,0.8h)}{vn1}
      \fmfforce{(.78w,0.8h)}{vn2}
      \fmfforce{(.2w,0.2h)}{vs1}
      \fmfforce{(.7w,0.2h)}{vs2}
      \fmffreeze
	\fmf{fermion,label=\footnotesize{$p$}}{vl,v1}
      \fmf{plain,left=0.25}{v1,vn1}
      \fmf{plain}{vn1,vn2}
      \fmf{plain,left=0.25}{vn2,v4}
      \fmf{plain,left=0.25}{v4,vs2}
      \fmf{plain}{vs2,vs1}
      \fmf{plain,left=0.25}{vs1,v1}
      \fmf{plain}{vs1,vn1}
      \fmf{plain}{vs2,vn2}
      \fmf{plain}{vn1,vs2}
      \fmf{plain}{v4,vr}
\fmfdot{v1,v4,vn1,vn2,vs1,vs2}
\fmffreeze
\end{fmfgraph*}
} \qquad \underset{(\text{Du})}{=} \qquad
\parbox{16mm}{
\begin{fmfgraph*}(16,14)
      \fmfleft{vl}
      \fmfright{vr}
      \fmffreeze
      \fmfforce{(-0.2w,0.5h)}{vl}
      \fmfforce{(-0.1w,0.5h)}{vl1}
      \fmfforce{(1.2w,0.25h)}{vr}
      \fmfforce{(1.1w,0.25h)}{vr4}
      \fmfforce{(-0.05w,0.5h)}{v1}
      \fmfforce{(1.05w,0.25h)}{v4}
      \fmfforce{(.5w,0.95h)}{vn}
      \fmfforce{(.5w,0.5h)}{vcc}	
      \fmfforce{(.5w,0.25h)}{vcb}
      \fmfforce{(.5w,-0.05h)}{vs}
      \fmffreeze
	\fmf{phantom,label=\footnotesize{$0$}}{vl,v1}
      \fmf{plain}{v1,vn}
      \fmf{plain}{vn,v4}
      \fmf{plain}{v4,vs}
      \fmf{plain}{vs,v1}
      \fmf{plain}{v1,vcc}
      \fmf{plain}{vcb,v4}
      \fmf{plain}{vn,vs}
	\fmf{phantom,label=\footnotesize{$x$}}{v4,vr}
\fmfdot{v1,v4,vn,vs,vcc,vcb}
\fmffreeze
\end{fmfgraph*}
} \qquad \underset{(\text{F})}{=} \qquad
\parbox{16mm}{
\begin{fmfgraph*}(16,14)
      \fmfleft{vl}
      \fmfright{vr}
      \fmffreeze
      \fmfforce{(-0.3w,0.5h)}{vl}
      \fmfforce{(-0.1w,0.5h)}{vl1}
      \fmfforce{(1.2w,0.25h)}{vr}
      \fmfforce{(1.1w,0.25h)}{vr4}
      \fmfforce{(-0.05w,0.5h)}{v1}
      \fmfforce{(1.05w,0.25h)}{v4}
      \fmfforce{(.5w,0.95h)}{vn}
      \fmfforce{(.5w,0.5h)}{vcc}
      \fmfforce{(.5w,0.25h)}{vcb}
      \fmfforce{(.5w,-0.05h)}{vs}
      \fmffreeze
	\fmf{fermion,label=\footnotesize{$p$}}{vl,v1}
      \fmf{plain}{v1,vn}
      \fmf{plain}{vn,v4}
      \fmf{plain}{v4,vs}
      \fmf{plain}{vs,v1}
      \fmf{plain}{v1,vcc}
      \fmf{plain}{vcb,v4}
      \fmf{plain}{vn,vs}
      \fmf{plain}{v4,vr}
\fmfdot{v1,v4,vn,vs,vcc,vcb}
\fmffreeze
\end{fmfgraph*}
} \quad  + \Ord(\ep) \, .
\eea
Since the considered accuracy is $O(\ep^0)$, we may as well consider the following diagram:
\bea
\parbox{16mm}{
\begin{fmfgraph*}(16,14)
      \fmfleft{vl}
      \fmfright{vr}
      \fmffreeze
      \fmfforce{(-0.3w,0.5h)}{vl}
      \fmfforce{(-0.1w,0.5h)}{vl1}
      \fmfforce{(1.2w,0.25h)}{vr}
      \fmfforce{(1.1w,0.25h)}{vr4}
      \fmfforce{(-0.05w,0.5h)}{v1}
      \fmfforce{(1.05w,0.25h)}{v4}
      \fmfforce{(.5w,0.95h)}{vn}
      \fmfforce{(.5w,0.5h)}{vcc}
      \fmfforce{(.5w,0.25h)}{vcb}
      \fmfforce{(.5w,-0.05h)}{vs}
      \fmffreeze
	\fmf{fermion,label=\footnotesize{$p$}}{vl,v1}
      \fmf{plain}{v1,vn}
	\fmf{plain,label=\footnotesize{$1-2\ep$},l.d=0.1h,l.s=left}{vn,v4}
      \fmf{plain}{v4,vs}
      \fmf{plain}{vs,v1}
      \fmf{plain}{v1,vcc}
      \fmf{plain}{vcb,v4}
      \fmf{plain}{vn,vs}
      \fmf{plain}{v4,vr}
\fmfdot{v1,v4,vn,vs,vcc,vcb}
\fmffreeze
\end{fmfgraph*}
} \qquad \qquad \, .
\eea
Applying the IBP procedure for the right triangle with the vertical distinguished line, we have the following relation:
%
\fleq{
\parbox{16mm}{
\begin{fmfgraph*}(16,14)
      \fmfleft{vl}
      \fmfright{vr}
      \fmffreeze
      \fmfforce{(-0.3w,0.5h)}{vl}
      \fmfforce{(-0.1w,0.5h)}{vl1}
      \fmfforce{(1.2w,0.25h)}{vr}
      \fmfforce{(1.1w,0.25h)}{vr4}
      \fmfforce{(-0.05w,0.5h)}{v1}
      \fmfforce{(1.05w,0.25h)}{v4}
      \fmfforce{(.5w,0.95h)}{vn}
      \fmfforce{(.5w,0.5h)}{vcc}
      \fmfforce{(.5w,0.25h)}{vcb}
      \fmfforce{(.5w,-0.05h)}{vs}
      \fmffreeze
	\fmf{fermion,label=\footnotesize{$p$}}{vl,v1}
      \fmf{plain}{v1,vn}
	\fmf{plain,label=\footnotesize{$1-2\ep$},l.d=0.1h,l.s=left}{vn,v4}
      \fmf{plain}{v4,vs}
      \fmf{plain}{vs,v1}
      \fmf{plain}{v1,vcc}
      \fmf{plain}{vcb,v4}
      \fmf{plain}{vn,vs}
      \fmf{plain}{v4,vr}
\fmfdot{v1,v4,vn,vs,vcc,vcb}
\fmffreeze
\end{fmfgraph*}
} \qquad ~(D-4) \quad = \quad 2 \quad
\parbox{16mm}{
    \begin{fmfgraph*}(16,14)
      \fmfleft{i}
      \fmfright{o}
      \fmfleft{ve}
      \fmfright{vo}
      \fmftop{vn}
      \fmftop{vs}
      \fmffreeze
      \fmfforce{(-0.1w,0.5h)}{i}
      \fmfforce{(1.1w,0.5h)}{o}
      \fmfforce{(0w,0.5h)}{ve}
      \fmfforce{(1.0w,0.5h)}{vo}
      \fmfforce{(.5w,0.95h)}{vn}
      \fmfforce{(.5w,0.05h)}{vs}
      \fmfforce{(.5w,0.5h)}{vc}
      \fmffreeze
      \fmf{plain}{i,ve}
      \fmf{plain,left=0.8}{ve,vo}
	    \fmf{phantom,right=0.5,label=\footnotesize{$1-2\ep$},l.d=-0.01w}{vo,vn}
      \fmf{plain,left=0.8}{vo,ve}
      \fmf{plain}{vs,vn}
      \fmf{plain}{vo,o}
      \fmf{plain,right=0.5}{vo,vs}
      \fmf{phantom,label=\scriptsize{$2$},l.d=0.05h,l.s=left}{vc,vo}
      \fmf{plain}{ve,vc}
      \fmffreeze
      \fmfdot{ve,vn,vo,vs,vc}
    \end{fmfgraph*}
} \qquad -
\quad
\parbox{16mm}{
    \begin{fmfgraph*}(16,14)
      \fmfleft{i}
      \fmfright{o}
      \fmfleft{ve}
      \fmfright{vo}
      \fmftop{vn}
      \fmftop{vs}
      \fmffreeze
      \fmfforce{(-0.1w,0.5h)}{i}
      \fmfforce{(1.1w,0.5h)}{o}
      \fmfforce{(0w,0.5h)}{ve}
      \fmfforce{(1.0w,0.5h)}{vo}
      \fmfforce{(.5w,0.95h)}{vn}
      \fmfforce{(.5w,0.05h)}{vs}
      \fmfforce{(.5w,0.5h)}{vc}
      \fmffreeze
      \fmf{plain}{i,ve}
      \fmf{plain,left=0.8}{ve,vo}
	    \fmf{phantom,right=0.5,label=\footnotesize{$1-2\ep$},l.d=-0.01w}{vo,vn}
      \fmf{plain,left=0.8}{vo,ve}
      \fmf{plain}{vs,vn}
      \fmf{plain}{vo,o}
      \fmf{phantom,label=\scriptsize{$2$},l.d=0.05h,l.s=left}{vc,vo}
       \fmf{plain}{ve,vo}
      \fmffreeze
      \fmfdot{ve,vn,vo,vs,vc}
    \end{fmfgraph*}
}
\quad
 - \qquad
 \parbox{16mm}{
\begin{fmfgraph*}(16,14)
      \fmfleft{vl}
      \fmfright{vr}
      \fmffreeze
      \fmfforce{(-0.2w,0.5h)}{vl}
      \fmfforce{(-0.1w,0.5h)}{vl1}
      \fmfforce{(1.2w,0.5h)}{vr}
      \fmfforce{(1.1w,0.5h)}{vr4}
      \fmfforce{(-0.05w,0.5h)}{v1}
      \fmfforce{(1.05w,0.5h)}{v4}
      \fmfforce{(.5w,0.9h)}{vn}
      \fmfforce{(.5w,0.1h)}{vs}
      \fmfforce{(.5w,0.5h)}{vc}
      \fmfforce{(.2w,0.2h)}{vs1}
      \fmfforce{(.7w,0.2h)}{vs2}
	\fmffreeze
      \fmf{plain}{vl,v1}
      \fmf{plain,left=0.66}{v1,v4}
      \fmf{plain,left=0.66}{v4,v1}
	\fmf{phantom,left=0.9,label=\scriptsize{$2$},l.d.=0.1h}{v4,v1}	
	\fmf{phantom,right=0.5,label=\footnotesize{$1-2\ep$},l.d=-0.01w}{v4,vn}
      \fmf{plain}{vn,vs}
      \fmf{plain}{v1,vc}
      \fmf{plain}{v4,vr}
\fmfdot{v1,v4,vn,vs,vc}
\fmffreeze
\fmfposition
\fmfi{plain}{vloc(__v4) ..controls vloc(__vr4) and (xpart(vloc(__vr4)),-0.1h) ..(xpart(vloc(__vs2)),-0.1h)}
\fmfi{plain}{(xpart(vloc(__vs1)),-0.1h) ..controls (xpart(vloc(__vl1)),-0.1h) and vloc(__vl1) ..vloc(__v1)}
\fmfi{plain}{(xpart(vloc(__vs1)),-0.1h)..(xpart(vloc(__vs2)),-0.1h)} \, .
\end{fmfgraph*}
}
 \label{DiD1}
}
 %
The first diagram in the rhs of the above equation can be simplified as follows:
\bea
\parbox{16mm}{
    \begin{fmfgraph*}(16,14)
      \fmfleft{i}
      \fmfright{o}
      \fmfleft{ve}
      \fmfright{vo}
      \fmftop{vn}
      \fmftop{vs}
      \fmffreeze
      \fmfforce{(-0.1w,0.5h)}{i}
      \fmfforce{(1.1w,0.5h)}{o}
      \fmfforce{(0w,0.5h)}{ve}
      \fmfforce{(1.0w,0.5h)}{vo}
      \fmfforce{(.5w,0.95h)}{vn}
      \fmfforce{(.5w,0.05h)}{vs}
      \fmfforce{(.5w,0.5h)}{vc}
      \fmffreeze
      \fmf{plain}{i,ve}
      \fmf{plain,left=0.8}{ve,vo}
	    \fmf{phantom,right=0.5,label=\footnotesize{$1-2\ep$},l.d=-0.01w}{vo,vn}
      \fmf{plain,left=0.8}{vo,ve}
      \fmf{plain}{vs,vn}
      \fmf{plain}{vo,o}
      \fmf{plain,right=0.5}{vo,vs}
      \fmf{phantom,label=\scriptsize{$2$},l.d=0.05h,l.s=left}{vc,vo}
      \fmf{plain}{ve,vc}
      \fmffreeze
      \fmfdot{ve,vn,vo,vs,vc}
    \end{fmfgraph*}
} \qquad =
\quad
     G(D,2,1)\quad 
\parbox{16mm}{
    \begin{fmfgraph*}(16,14)
      \fmfleft{i}
      \fmfright{o}
      \fmfleft{ve}
      \fmfright{vo}
      \fmftop{vn}
      \fmftop{vs}
      \fmffreeze
      \fmfforce{(-0.1w,0.5h)}{i}
      \fmfforce{(1.1w,0.5h)}{o}
      \fmfforce{(0w,0.5h)}{ve}
      \fmfforce{(1.0w,0.5h)}{vo}
      \fmfforce{(.5w,0.95h)}{vn}
      \fmfforce{(.5w,0.05h)}{vs}
      \fmfforce{(.5w,0.5h)}{vc}
      \fmffreeze
      \fmf{plain}{i,ve}
      \fmf{plain,left=0.8}{ve,vo}
	    \fmf{phantom,right=0.5,label=\footnotesize{$1-2\ep$},l.d=-0.01w}{vo,vn}
      \fmf{plain,left=0.8}{vo,ve}
	    \fmf{phantom,left=0.5,label=\footnotesize{$1+\ep$},l.d=-0.01w}{vo,vs}
      \fmf{plain}{vs,vn}
      \fmf{plain}{vo,o}
      \fmf{plain}{ve,vc}
      \fmffreeze  
      \fmfdot{ve,vn,vo,vs,vc}
    \end{fmfgraph*}
} \qquad \, .   
 \label{DiD2}
\eea
On the other hand, the last diagram can be represented as:
\bea
&&\parbox{16mm}{
\begin{fmfgraph*}(16,14)
      \fmfleft{vl}
      \fmfright{vr}
      \fmffreeze
      \fmfforce{(-0.2w,0.5h)}{vl}
      \fmfforce{(-0.1w,0.5h)}{vl1}
      \fmfforce{(1.2w,0.5h)}{vr}
      \fmfforce{(1.1w,0.5h)}{vr4}
      \fmfforce{(-0.05w,0.5h)}{v1}
      \fmfforce{(1.05w,0.5h)}{v4}
      \fmfforce{(.5w,0.9h)}{vn}
      \fmfforce{(.5w,0.1h)}{vs}
      \fmfforce{(.5w,0.5h)}{vc}
      \fmfforce{(.2w,0.2h)}{vs1}
      \fmfforce{(.7w,0.2h)}{vs2}
        \fmffreeze
      \fmf{plain}{vl,v1}
      \fmf{plain,left=0.66}{v1,v4}
      \fmf{plain,left=0.66}{v4,v1}
	\fmf{phantom,left=0.9,label=\footnotesize{$2$},l.d.=0.1h}{v4,v1}
	\fmf{phantom,right=0.5,label=\footnotesize{$1-2\ep$},l.d=-0.01w}{v4,vn}
      \fmf{plain}{vn,vs}
      \fmf{plain}{v1,vc}
      \fmf{plain}{v4,vr}
\fmfdot{v1,v4,vn,vs,vc}
\fmffreeze
\fmfposition
\fmfi{plain}{vloc(__v4) ..controls vloc(__vr4) and (xpart(vloc(__vr4)),-0.1h) ..(xpart(vloc(__vs2)),-0.1h)}
\fmfi{plain}{(xpart(vloc(__vs1)),-0.1h) ..controls (xpart(vloc(__vl1)),-0.1h) and vloc(__vl1) ..vloc(__v1)}
\fmfi{plain}{(xpart(vloc(__vs1)),-0.1h)..(xpart(vloc(__vs2)),-0.1h)}
\end{fmfgraph*}
}
\quad   = \quad  G(D,2,1+\ep) \quad
\parbox{16mm}{
    \begin{fmfgraph*}(16,14)
      \fmfleft{i}
      \fmfright{o}
      \fmfleft{ve}
      \fmfright{vo}
      \fmftop{vn}
      \fmftop{vs}
      \fmffreeze
      \fmfforce{(-0.1w,0.5h)}{i}
      \fmfforce{(1.1w,0.5h)}{o}
      \fmfforce{(0w,0.5h)}{ve}
      \fmfforce{(1.0w,0.5h)}{vo}
      \fmfforce{(.5w,0.95h)}{vn}
      \fmfforce{(.5w,0.05h)}{vs}
      \fmfforce{(.5w,0.5h)}{vc}
      \fmffreeze
      \fmf{plain}{i,ve}
      \fmf{plain,left=0.8}{ve,vo}
	    \fmf{phantom,right=0.5,label=\footnotesize{$1-2\ep$},l.d=-0.01w}{vo,vn}
      \fmf{plain,left=0.8}{vo,ve}
      \fmf{plain}{vs,vn}
      \fmf{plain}{vo,o}
      \fmf{plain}{ve,vc}
      \fmffreeze
      \fmfdot{ve,vn,vo,vs,vc}
    \end{fmfgraph*}
} \qquad \, ,
\label{DiD3} \\
&&\nonum 
\eea
because the diagram contains an internal three-loop one with $p$-dependence of the form:
$\sim 1/p^{(2(1+\ep))}$. Finally, the right upper triangle of the second diagram in the rhs of Eq.~(\ref{DiD1}) is unique; so, we have:
\bea
\parbox{16mm}{
    \begin{fmfgraph*}(16,14)
      \fmfleft{i}
      \fmfright{o}
      \fmfleft{ve}
      \fmfright{vo}
      \fmftop{vn}
      \fmftop{vs}
      \fmffreeze
      \fmfforce{(-0.1w,0.5h)}{i}
      \fmfforce{(1.1w,0.5h)}{o}
      \fmfforce{(0w,0.5h)}{ve}
      \fmfforce{(1.0w,0.5h)}{vo}
      \fmfforce{(.5w,0.95h)}{vn}
      \fmfforce{(.5w,0.05h)}{vs}
      \fmfforce{(.5w,0.5h)}{vc}
      \fmffreeze
      \fmf{plain}{i,ve}
      \fmf{plain,left=0.8}{ve,vo}
	    \fmf{phantom,right=0.5,label=\footnotesize{$1-2\ep$},l.d=-0.01w}{vo,vn}
      \fmf{plain,left=0.8}{vo,ve}
      \fmf{plain}{vs,vn}
      \fmf{plain}{vo,o}
      \fmf{phantom,label=\scriptsize{$2$},l.d=0.05h,l.s=left}{vc,vo}
       \fmf{plain}{ve,vo}
      \fmffreeze
      \fmfdot{ve,vn,vo,vs,vc}
    \end{fmfgraph*}
}
\quad   = \quad G(D,2,1) \qquad
\parbox{16mm}{
    \begin{fmfgraph*}(16,14)
      \fmfleft{i}
      \fmfright{o}
      \fmfleft{ve}
      \fmfright{vo}
      \fmftop{vn}
      \fmftop{vs}
      \fmffreeze
      \fmfforce{(-0.3w,0.5h)}{i}
      \fmfforce{(1.3w,0.5h)}{o}
      \fmfforce{(-0.2w,0.5h)}{ve}
      \fmfforce{(1.2w,0.5h)}{vo}
      \fmfforce{(.5w,1.05h)}{vn}
      \fmfforce{(.5w,-0.05h)}{vs}
      \fmfforce{(.5w,0.5h)}{vc}
      \fmffreeze
      \fmf{plain}{i,ve}
      \fmf{plain,left=0.7}{ve,vo}
	    \fmf{phantom,left=0.5,label=\footnotesize{$1-\ep$},l.d=-0.01w}{ve,vn}
	    \fmf{phantom,right=0.5,label=\footnotesize{$1-\ep$},l.d=-0.01w}{vo,vn}
      \fmf{plain,left=0.7}{vo,ve}
      \fmf{plain}{vs,vn}
      \fmf{plain}{vo,o}
	    \fmf{phantom,label={\tiny{$1+\ep$}},l.d=0.05w,l.s=left}{vn,vc}
      \fmf{plain}{ve,vc}
      \fmffreeze
      \fmfdot{ve,vn,vo,vs,vc}
    \end{fmfgraph*}
} \qquad \, .
\label{DiD4}
\eea
Combining the above results yields the following expression for the diagram of Eq.~(\ref{DiD1}):
%
\fleq{
\parbox{16mm}{
\begin{fmfgraph*}(16,14)
      \fmfleft{vl}
      \fmfright{vr}
      \fmffreeze
      \fmfforce{(-0.3w,0.5h)}{vl}
      \fmfforce{(-0.1w,0.5h)}{vl1}
      \fmfforce{(1.2w,0.25h)}{vr}
      \fmfforce{(1.1w,0.25h)}{vr4}
      \fmfforce{(-0.05w,0.5h)}{v1}
      \fmfforce{(1.05w,0.25h)}{v4}
      \fmfforce{(.5w,0.95h)}{vn}
      \fmfforce{(.5w,0.5h)}{vcc}
      \fmfforce{(.5w,0.25h)}{vcb}
      \fmfforce{(.5w,-0.05h)}{vs}
      \fmffreeze
      \fmf{fermion,label=$p$}{vl,v1}
      \fmf{plain}{v1,vn}
	\fmf{plain,label=\footnotesize{$1-2\ep$},l.d=0.1h,l.s=left}{vn,v4}
      \fmf{plain}{v4,vs}
      \fmf{plain}{vs,v1}
      \fmf{plain}{v1,vcc}
      \fmf{plain}{vcb,v4}
      \fmf{plain}{vn,vs}
      \fmf{plain}{v4,vr}
\fmfdot{v1,v4,vn,vs,vcc,vcb}
\fmffreeze
\end{fmfgraph*}
} \qquad & ~~\times ~(D-4)  \quad = \quad G(D,2,1) \left[ ~~2\quad 
\parbox{16mm}{
    \begin{fmfgraph*}(16,14)
      \fmfleft{i}
      \fmfright{o}
      \fmfleft{ve}
      \fmfright{vo}
      \fmftop{vn}
      \fmftop{vs}
      \fmffreeze
      \fmfforce{(-0.1w,0.5h)}{i}
      \fmfforce{(1.1w,0.5h)}{o}
      \fmfforce{(0w,0.5h)}{ve}
      \fmfforce{(1.0w,0.5h)}{vo}
      \fmfforce{(.5w,0.95h)}{vn}
      \fmfforce{(.5w,0.05h)}{vs}
      \fmfforce{(.5w,0.5h)}{vc}
      \fmffreeze
      \fmf{plain}{i,ve}
      \fmf{plain,left=0.8}{ve,vo}
	    \fmf{phantom,right=0.5,label=\footnotesize{$1-2\ep$},l.d=-0.01w}{vo,vn}
      \fmf{plain,left=0.8}{vo,ve}
	    \fmf{phantom,left=0.5,label=\footnotesize{$1+\ep$},l.d=-0.01w}{vo,vs}
      \fmf{plain}{vs,vn}
      \fmf{plain}{vo,o}
      \fmf{plain}{ve,vc}
      \fmffreeze
      \fmfdot{ve,vn,vo,vs,vc}
    \end{fmfgraph*}
} \qquad
- \quad \qquad 
\parbox{16mm}{
    \begin{fmfgraph*}(16,14)
      \fmfleft{i}
      \fmfright{o}
      \fmfleft{ve}
      \fmfright{vo}
      \fmftop{vn}
      \fmftop{vs}
      \fmffreeze
      \fmfforce{(-0.3w,0.5h)}{i}
      \fmfforce{(1.3w,0.5h)}{o}
      \fmfforce{(-0.2w,0.5h)}{ve}
      \fmfforce{(1.2w,0.5h)}{vo}
      \fmfforce{(.5w,1.05h)}{vn}
      \fmfforce{(.5w,-0.05h)}{vs}
      \fmfforce{(.5w,0.5h)}{vc}
      \fmffreeze
      \fmf{plain}{i,ve}
      \fmf{plain,left=0.7}{ve,vo}
	    \fmf{phantom,left=0.5,label=\footnotesize{$1-\ep$},l.d=-0.01w}{ve,vn}
	    \fmf{phantom,right=0.5,label=\footnotesize{$1-\ep$},l.d=-0.01w}{vo,vn}
      \fmf{plain,left=0.7}{vo,ve}
      \fmf{plain}{vs,vn}
      \fmf{plain}{vo,o}
            \fmf{phantom,label={\tiny{$1+\ep$}},l.d=0.05w,l.s=left}{vn,vc}
      \fmf{plain}{ve,vc}
      \fmffreeze
      \fmfdot{ve,vn,vo,vs,vc}
    \end{fmfgraph*}
}
\qquad
- \right . \nonum \\
&\left . - \quad \frac{G(D,2,1+\ep)}{G(D,2,1)}\quad 
\parbox{16mm}{
    \begin{fmfgraph*}(16,14)
      \fmfleft{i}
      \fmfright{o}
      \fmfleft{ve}
      \fmfright{vo}
      \fmftop{vn}
      \fmftop{vs}
      \fmffreeze
      \fmfforce{(-0.1w,0.5h)}{i}
      \fmfforce{(1.1w,0.5h)}{o}
      \fmfforce{(0w,0.5h)}{ve}
      \fmfforce{(1.0w,0.5h)}{vo}
      \fmfforce{(.5w,0.95h)}{vn}
      \fmfforce{(.5w,0.05h)}{vs}
      \fmfforce{(.5w,0.5h)}{vc}
      \fmffreeze
      \fmf{plain}{i,ve}
      \fmf{plain,left=0.8}{ve,vo}
	    \fmf{phantom,right=0.5,label=\footnotesize{$1-2\ep$},l.d=-0.01w}{vo,vn}
      \fmf{plain,left=0.8}{vo,ve}
      \fmf{plain}{vs,vn}
      \fmf{plain}{vo,o}
      \fmf{plain}{ve,vc}
      \fmffreeze
      \fmfdot{ve,vn,vo,vs,vc}
    \end{fmfgraph*}
} \qquad \quad
\right] \quad \, .
\label{DiD5}
}
%
Notice that the ratio $G(D,2,1+\ep)/G(D,2,1)= 1+ O(\ep^3)$ so it can be replaced by 1 and Eq.~(\ref{DiD5}) may be further simplified as:
%
\fleq{
\parbox{16mm}{
\begin{fmfgraph*}(16,14)
      \fmfleft{vl}
      \fmfright{vr}
      \fmffreeze
      \fmfforce{(-0.3w,0.5h)}{vl}
      \fmfforce{(-0.1w,0.5h)}{vl1}
      \fmfforce{(1.2w,0.25h)}{vr}
      \fmfforce{(1.1w,0.25h)}{vr4}
      \fmfforce{(-0.05w,0.5h)}{v1}
      \fmfforce{(1.05w,0.25h)}{v4}
      \fmfforce{(.5w,0.95h)}{vn}
      \fmfforce{(.5w,0.5h)}{vcc}
      \fmfforce{(.5w,0.25h)}{vcb}
      \fmfforce{(.5w,-0.05h)}{vs}
      \fmffreeze
      \fmf{fermion,label=$p$}{vl,v1}
      \fmf{plain}{v1,vn}
	\fmf{plain,label=\footnotesize{$1-2\ep$},l.d=0.1h,l.s=left}{vn,v4}
      \fmf{plain}{v4,vs}
      \fmf{plain}{vs,v1}
      \fmf{plain}{v1,vcc}
      \fmf{plain}{vcb,v4}
      \fmf{plain}{vn,vs}
      \fmf{plain}{v4,vr}
\fmfdot{v1,v4,vn,vs,vcc,vcb}
\fmffreeze
\end{fmfgraph*}
} \qquad & ~~\times ~(D-4)  \quad = \quad G(D,2,1) \left[ ~~2\quad
\parbox{16mm}{
    \begin{fmfgraph*}(16,14)
      \fmfleft{i}
      \fmfright{o}
      \fmfleft{ve}
      \fmfright{vo}
      \fmftop{vn}
      \fmftop{vs}
      \fmffreeze
      \fmfforce{(-0.1w,0.5h)}{i}
      \fmfforce{(1.1w,0.5h)}{o}
      \fmfforce{(0w,0.5h)}{ve}
      \fmfforce{(1.0w,0.5h)}{vo}
      \fmfforce{(.5w,0.95h)}{vn}
      \fmfforce{(.5w,0.05h)}{vs}
      \fmfforce{(.5w,0.5h)}{vc}
      \fmffreeze
      \fmf{plain}{i,ve}
      \fmf{plain,left=0.8}{ve,vo}
	    \fmf{phantom,right=0.5,label=\footnotesize{$1-2\ep$},l.d=-0.01w}{vo,vn}
      \fmf{plain,left=0.8}{vo,ve}
	    \fmf{phantom,left=0.5,label=\footnotesize{$1+\ep$},l.d=-0.01w}{vo,vs}
      \fmf{plain}{vs,vn}
      \fmf{plain}{vo,o}
      \fmf{plain}{ve,vc}
      \fmffreeze
      \fmfdot{ve,vn,vo,vs,vc}
    \end{fmfgraph*}
} \qquad
- \quad \qquad
\parbox{16mm}{
    \begin{fmfgraph*}(16,14)
      \fmfleft{i}
      \fmfright{o}
      \fmfleft{ve}
      \fmfright{vo}
      \fmftop{vn}
      \fmftop{vs}
      \fmffreeze
      \fmfforce{(-0.3w,0.5h)}{i}
      \fmfforce{(1.3w,0.5h)}{o}
      \fmfforce{(-0.2w,0.5h)}{ve}
      \fmfforce{(1.2w,0.5h)}{vo}
      \fmfforce{(.5w,1.05h)}{vn}
      \fmfforce{(.5w,-0.05h)}{vs}
      \fmfforce{(.5w,0.5h)}{vc}
      \fmffreeze
      \fmf{plain}{i,ve}
      \fmf{plain,left=0.7}{ve,vo}
	    \fmf{phantom,left=0.5,label=\footnotesize{$1-\ep$},l.d=-0.01w}{ve,vn}
	    \fmf{phantom,right=0.5,label=\footnotesize{$1-\ep$},l.d=-0.01w}{vo,vn}
      \fmf{plain,left=0.7}{vo,ve}
      \fmf{plain}{vs,vn}
      \fmf{plain}{vo,o}
            \fmf{phantom,label={\tiny{$1+\ep$}},l.d=0.05w,l.s=left}{vn,vc}
      \fmf{plain}{ve,vc}
      \fmffreeze
      \fmfdot{ve,vn,vo,vs,vc}
    \end{fmfgraph*}
}
\qquad
- \right . \nonum \\
&\left . - \quad
\parbox{16mm}{
    \begin{fmfgraph*}(16,14)
      \fmfleft{i}
      \fmfright{o}
      \fmfleft{ve}
      \fmfright{vo}
      \fmftop{vn}
      \fmftop{vs}
      \fmffreeze
      \fmfforce{(-0.1w,0.5h)}{i}
      \fmfforce{(1.3w,0.5h)}{o1}
      \fmfforce{(1.4w,0.5h)}{o}
      \fmfforce{(0w,0.5h)}{ve}
      \fmfforce{(1.0w,0.5h)}{vo}
      \fmfforce{(.5w,0.95h)}{vn}
      \fmfforce{(.5w,0.05h)}{vs}
      \fmfforce{(.5w,0.5h)}{vc}
      \fmffreeze
      \fmf{plain}{i,ve}
      \fmf{plain,left=0.8}{ve,vo}
	    \fmf{phantom,right=0.5,label=\footnotesize{$1-2\ep$},l.d=-0.01w}{vo,vn}
      \fmf{plain,left=0.8}{vo,ve}
      \fmf{plain}{vs,vn}
	    \fmf{plain,label=\footnotesize{$\ep$}}{vo,o1}
      \fmf{plain}{o1,o}
      \fmf{plain}{ve,vc}
      \fmffreeze
      \fmfdot{ve,vn,vo,vs,vc,o1}
    \end{fmfgraph*}
} \qquad \quad
\right] \quad \, .
\label{DiD6}
}
%
Evaluating the three-loop master integrals in the rhs of Eq.~(\ref{DiD6}) with the help of Eqs.~(\ref {3loop2}) and (\ref{3loop3}),
yields the following result for the considered diagram:
\bea
\parbox{16mm}{
\begin{fmfgraph*}(16,14)
      \fmfleft{vl}
      \fmfright{vr}
      \fmffreeze
      \fmfforce{(-0.2w,0.5h)}{vl}
      \fmfforce{(-0.1w,0.5h)}{vl1}
      \fmfforce{(1.2w,0.5h)}{vr}
      \fmfforce{(1.1w,0.5h)}{vr4}
      \fmfforce{(-0.05w,0.5h)}{v1}
      \fmfforce{(1.05w,0.5h)}{v4}
      \fmfforce{(.3w,0.8h)}{vn1}
      \fmfforce{(.78w,0.8h)}{vn2}
      \fmfforce{(.2w,0.2h)}{vs1}
      \fmfforce{(.7w,0.2h)}{vs2}
      \fmffreeze
      \fmf{plain}{vl,v1}
      \fmf{plain,left=0.25}{v1,vn1}
      \fmf{plain}{vn1,vn2}
      \fmf{plain,left=0.25}{vn2,v4}
      \fmf{plain,left=0.25}{v4,vs2}
      \fmf{plain}{vs2,vs1}
      \fmf{plain,left=0.25}{vs1,v1}
      \fmf{plain}{vs1,vn1}
      \fmf{plain}{vs2,vn2}
      \fmf{plain}{vn1,vs2}
      \fmf{plain}{v4,vr}
\fmfdot{v1,v4,vn1,vn2,vs1,vs2}
\fmffreeze
\end{fmfgraph*}
} \qquad   = \qquad
\parbox{16mm}{
\begin{fmfgraph*}(16,14)
      \fmfleft{vl}
      \fmfright{vr}
      \fmffreeze
      \fmfforce{(-0.3w,0.5h)}{vl}
      \fmfforce{(-0.1w,0.5h)}{vl1}
      \fmfforce{(1.2w,0.25h)}{vr}
      \fmfforce{(1.1w,0.25h)}{vr4}
      \fmfforce{(-0.05w,0.5h)}{v1}
      \fmfforce{(1.05w,0.25h)}{v4}
      \fmfforce{(.5w,0.95h)}{vn}
      \fmfforce{(.5w,0.5h)}{vcc}
      \fmfforce{(.5w,0.25h)}{vcb}
      \fmfforce{(.5w,-0.05h)}{vs}
      \fmffreeze
      \fmf{plain}{vl,v1}
      \fmf{plain}{v1,vn}
	\fmf{plain,label=\footnotesize{$1-2\ep$},l.d=0.1h,l.s=left}{vn,v4}
      \fmf{plain}{v4,vs}
      \fmf{plain}{vs,v1}
      \fmf{plain}{v1,vcc}
      \fmf{plain}{vcb,v4}
      \fmf{plain}{vn,vs}
      \fmf{plain}{v4,vr}
\fmfdot{v1,v4,vn,vs,vcc,vcb}
\fmffreeze
\end{fmfgraph*}
} \qquad + ~ O(\ep) ~ = ~ \frac{441}{8}\, \zeta_7\, \frac{1}{p^{2(1+4\ep)}} + O(\ep)\, .
\label{DiD7a}
\eea
Hence, the singular part of diagram ``d'' reads:
%
\bea
\mathrm{Sing} \left[ \quad
\parbox{15mm}{
\begin{fmfgraph*}(15,11)
      \fmfleft{vl}
      \fmfright{vr}
      \fmffreeze
      \fmfforce{(-0.2w,0.5h)}{vl}
      \fmfforce{(-0.1w,0.5h)}{vl1}
      \fmfforce{(1.2w,0.5h)}{vr}
      \fmfforce{(1.1w,0.5h)}{vr4}
      \fmfforce{(-0.05w,0.5h)}{v1}
      \fmfforce{(1.05w,0.5h)}{v4}
      \fmfforce{(.3w,0.8h)}{vn1}
      \fmfforce{(.78w,0.8h)}{vn2}
      \fmfforce{(.2w,0.2h)}{vs1}
      \fmfforce{(.7w,0.2h)}{vs2}
      \fmffreeze
      \fmf{phantom}{vl,v1}
      \fmf{plain,left=0.25}{v1,vn1}
      \fmf{plain}{vn1,vn2}
      \fmf{plain,left=0.25}{vn2,v4}
      \fmf{plain,left=0.25}{v4,vs2}
      \fmf{plain}{vs2,vs1}
      \fmf{plain,left=0.25}{vs1,v1}
      \fmf{plain}{vs1,vn1}
      \fmf{plain}{vs2,vn2}
      \fmf{plain}{vn1,vs2}
      \fmf{phantom}{v4,vr}
\fmfdot{v1,v4,vn1,vn2,vs1,vs2}
\fmffreeze
\fmfposition
\fmfi{plain}{vloc(__v4) ..controls vloc(__vr4) and (xpart(vloc(__vr4)),-0.0h) ..(xpart(vloc(__vs2)),-0.0h)}
\fmfi{plain}{(xpart(vloc(__vs1)),-0.0h) ..controls (xpart(vloc(__vl1)),-0.0h) and vloc(__vl1) ..vloc(__v1)}
\fmfi{plain}{(xpart(vloc(__vs1)),-0.0h)..(xpart(vloc(__vs2)),-0.0h)}
\end{fmfgraph*}
}  \quad \right] \quad = \quad   \frac{441}{40 \ep} \, \zeta_7 \, \frac{1}{p^{10\ep}} + O(\ep^0)\, .
\label{DiD7b}
\eea

\end{fmffile}

\section{Conclusion}
\label{sec:conclusion}

In the present article, we have reviewed several powerful multi-loop techniques devoted to the {\it analytical} calculations of massless Feynman diagrams: integration by parts
\cite{Tkachov:1981wb,Chetyrkin:1981qh} (and similar approach in coordinate space \cite{Vasiliev:1981dg,Vasil'evbook}),
the method of uniqueness \cite{Vasiliev:1981dg,Usyukina:1983gj,Kazakov:1983ns,Kazakov:1984km} (see also Vassiliev's book
\cite{Vasil'evbook}), Kazakov's functional equations \cite{Kazakov:1983pk} (see also Kazakov's lectures  \cite{Kazakov:1984bw})
and the Gegenbauer polynomial technique \cite{Chetyrkin:1980pr,Kotikov:1995cw}. 
We would like to note that there is another powerful technique \cite{Tarasov:1996br} (see also an extension in \cite{Lee:2009dh})
which is based on difference equations connecting Feynman diagrams in $D$ and $(D+2)$ dimensions; this technique is very popular now (see the recent
Ref.~\cite{Lee:2017ftw} for short review of the application of the approach) but its consideration is beyond the scope our present paper.
Based on the method of uniqueness, we have shown various possibilities 
to transform the basic two-loop diagram to ones with different powers of propagators. 
The full set of such transformations can be found in Refs.~\cite{Vasiliev:1981dg,Gorishnii:1984te,Broadhurst:1986bx,Barfoot:1987kg}
and in the book \cite{Vasil'evbook}. Our review closely followed the nice lectures of Kazakov which were published in  \cite{Kazakov:1984bw}. But contrary to the lectures
\cite{Kazakov:1984bw} as well as the original papers  \cite{Vasiliev:1981dg,Usyukina:1983gj,Kazakov:1983ns,Kazakov:1984km}, we presented all our results
in momentum space, which is more popular at the present time.

As concrete examples of analytic calculations, we presented the evaluations of two-, three-, four- and even five-loop Feynman integrals of rather
complicated topologies. The five-loop diagrams considered in detail in Sec.~\ref{sec:examples5}, are part of the set of the four most complicated
Feynman integrals which were calculated firstly numerically \cite{Gorishnii:1983gp} and later analytically
\cite{Kazakov:1983ns,Kazakov:1984km,Kazakov:1983pk} in order to evaluate the $5$-loop correction to the $\beta$-function of the $\Phi^4$-model in $D=4$ dimensions.
That the presented purely analytical techniques allow to compute such complicated diagrams is quite suggestive of their power and efficiency.

Moreover, the considered methods may be applied to a wide range of models across various fields. In particular, they can be of crucial importance 
for odd-dimensional models and/or to evaluate Feynman integrals with propagators having nonzero indices, \ie, nontrivial powers of their momenta, all the more that, precisely in these case,
the most popular modern computer programs undergo strong restrictions. Let's add that diagrams with non-trivial indices generally appear upon considering high orders
of perturbation theory where such indices come from calculations of subgraphs. However, in some effective field theories, such propagators appear already at the
leading order of some expansion parameter. One example is reduced QED (see, for example,
\cite{Teber:2012de}-\cite{Teber:2018goo}), which can be considered as describing the ultrarelativistic limit of planar Dirac liquids.
The second one \cite{Kotikov:1989nm}-\cite{Kotikov:2016prf} is the $1/N$-expansion of three-dimensional QED with $N$-species of fermions.
Both of these models contain photon propagators of rather specific forms. Their accurate study with the help of the methods reviewed in this manuscript 
could reveal some of their duality \cite{Kotikov:2016yrn} and, as a consequence, some relation between the results for these two different models (see \cite{Teber:2017hab} for a review).
Among other recent achievements of these techniques, let's briefly mention: 
their application to models with broken-Lorentz invariance \cite{Teber:2014ita,Teber:2018qcn}, the calculation of the critical index $\eta$ of the $\Phi^{3}$-model at four-loops \cite{Pismensky:2015mba}, 
the computation of the anomalous dimension of double trace operators in hexagonal fishnet models \cite{Mamroud:2017uyz}, .... and more to come in the next years.

In closing, we hope that our review will be useful to both experts in calculations as well as to novices interested in the methods of analytical calculations of Feynman diagrams.

\section*{Acknowledgements}

We are very grateful to David Broadhurst, John Gracey, Valery Gusynin and Mikhail Kompaniets for their comments. The  work  of  A.V.K.\  was  supported  in  part  by  the Russian 
Foundation  for  Basic  Research  (Grant  No.16-02-00790-a).

\section{Appendix A. Gegenbauer polynomial technique}
\label{sec:meth:Gegenbauer}
\label{App:A}
\def\theequation{A\arabic{equation}}
\setcounter{equation}{0}

  
This Appendix is devoted to a short presentation of the Gegenbauer polynomial technique. The latter should be considered as the effective (but rather cumbersome) method
for calculating dimensionally regularized Feynman diagrams. In its modern form, it has been introduced by Chetyrkin, Kataev and Tkachov \cite{Chetyrkin:1980pr}. Later, subtle and important improvements
were brought up by Kotikov \cite{Kotikov:1995cw} and we shall follow this reference in our brief review of the technique.

Hereafter we will use the variables $x,y, ...$, which are usually used in coordinate space. But we can also think about the variables $x,y, ...$ 
as being some momenta. Thus, all formulae in this Appendix are also applicable in the momentum space.
Such type of ``duality'' has already been considered in Sec.~\ref{sec:meth:FT+Du}.

\subsection{Presentation of the method}

The basic motivation for this technique lays in the fact that, in multi-loop computations, the complicated part of the integration is often the one over the angular variables.
This task is considerably simplified by expanding some of the propagators in the integrand in terms of the Gegenbauer polynomials (the so-called multipole expansion):
\bea
\frac{1}{(x_1 - x_2)^{2\lambda}}  =
\sum_{n=0}^\infty\,C_n^\lambda(\hat{x}_1 \cdot \hat{x}_2)\,\Bigg[ \frac{(x_1^2)^{n/2}}{(x_2^2)^{n/2+\lambda}}\,\Theta(x_2^2-x_1^2) + (x_1^2 \longleftrightarrow x_2^2) \Bigg]\, ,
\label{gegenbauer:multipole}
\eea
where $C_n^\lambda$ is the Gegenbauer polynomial of degree $n$ and $\hat{x}=x/\sqrt{x^2}$, and then using the orthogonality relation of Gegenbauer polynomials on the unit $D$-dimensional sphere: 
\be
\frac{1}{\Omega_D}\, \int\,\D_D\, \hat{x}\,C_n^\lambda(\hat{z} \cdot \hat{x})C_m^\lambda(\hat{x} \cdot \hat{z})=\delta_{n,m}\,\frac{\lambda\,\Gamma(n+2\lambda)}{\Gamma(2\lambda)\,(n+\lambda)\,n!},
\qquad \lambda = \frac{D}{2} -1\, ,
\label{gegenbauer:orth}
\ee
where $\D_D\,\hat{x}$ is the surface element of the unit $D$-dimensional sphere and $\Omega_D = 2 \pi^{D/2} / \Gamma(D/2)$. The Gegenbauer polynomials can be defined from their generating function:

\be
\frac{1}{(1-2xw+w^2)^\beta} = \sum_{k=0}^\infty\,C_k^\beta(x)\,w^k \qquad C_n^\beta(1) = \frac{\Gamma(n+2\beta)}{\Gamma(2\beta)\,n!}\, ,
\label{gegenbauer:def}
\ee
with some additional particular values given by:
\be
C_0^\lambda (x) = 1, \quad C_1^\lambda(x) = 2\lambda x, \quad C_2^\lambda(x) = 2\lambda(\lambda+1)x^2-\lambda\, .
\label{gegenbauer:values}
\ee
For our purpose, it is convenient to express the Gegenbauer polynomials in terms of traceless symmetric tensors \cite{Kotikov:1995cw}:
\be
C_n^\lambda(\hat{x}\cdot \hat{z})\,(x^2\,z^2)^{n/2} = S_n(\lambda)\,x^{\mu_1 \mu_2 \cdots \mu_n}\,z^{\mu_1 \mu_2 \cdots \mu_n}, \qquad S_n(\lambda) =  \frac{2^n \Gamma(n+\lambda)}{n!\,\Gamma(\lambda)}\, .
\label{gegenbauer:TSTGegen}
\ee
From Eq.~(\ref{gegenbauer:TSTGegen}) for $x=z$ and the last equation in (\ref{gegenbauer:def}), we deduce the following equation for products of traceless tensors:
\be
S_n(\lambda)\,z^{\mu_1 \mu_2 \cdots \mu_n}\,z^{\mu_1 \mu_2 \cdots \mu_n} = \frac{\Gamma(n+2\lambda)}{\Gamma(2\lambda)\,n!}\,z^{2n}\, .
\label{gegenbauer:TSTP}
\ee
With the help of Eq.~(\ref{gegenbauer:TSTGegen}), Eq.~(\ref{gegenbauer:multipole}) can be rewritten as:
\be
\frac{1}{(x_1 - x_2)^{2\lambda}}  =
\sum_{n=0}^\infty\,S_n(\lambda)\,x_1^{\mu_1 \cdots \mu_n}\,x_2^{\mu_1 \cdots \mu_n}\,
\Bigg[ \frac{1}{(x_2^2)^{n+\lambda}}\,\Theta(x_2^2-x_1^2) + (x_1^2 \longleftrightarrow x_2^2) \Bigg]\, .
\label{gegenbauer:multipole2}
\ee
Notice that, for a propagator with arbitrary index, Eq.~(\ref{gegenbauer:multipole}) can be generalized as:
\bea
\frac{1}{(x_1 - x_2)^{2\beta}}  =
\sum_{n=0}^\infty\,C_n^\beta(\hat{x}_1 \cdot \hat{x}_2)\,\Bigg[ \frac{(x_1^2)^{n/2}}{(x_2^2)^{n/2+\beta}}\,\Theta(x_2^2-x_1^2) + (x_1^2 \longleftrightarrow x_2^2) \Bigg]\, ,
\label{gegenbauer:multipole3}
\eea
where $C_n^\beta(x)$ can then be related to $C_{n-2k}^\lambda(x)$ ($0 \leq k \leq [n/2]$) with the help of:
\be
C_n^\delta(x) =  \sum_{k=0}^{[n/2]}\,C_{n-2k}^\lambda(x)\,\frac{(n-2k+\lambda)\Gamma(\lambda)}{k!\,\Gamma(\delta)}\,
\frac{\Gamma(n+\delta-k)\Gamma(k+\delta-\lambda)}{\Gamma(n-k+\lambda+1) \Gamma(\delta-\lambda)}\, .
\label{gegenbauer:diffind}
\ee
Moreover, the series appearing upon expanding the propagators and after performing all integrations may sometimes be resummed in the form of a  generalized
hypergeometric function ${}_3F_2$ of unit argument. There is a very useful transformation property relating such hypergeometric functions.
Even though not directly connected with Gegenbauer polynomials, we mention it here:
\bea
&& {}_3F_2(a,b,c;e,f;1) = \frac{\Gamma(1-a)\Gamma(e)\Gamma(f)\Gamma(c-b)}{\Gamma(e-b)\Gamma(f-b)\Gamma(1+b-a)\Gamma(c)}
\nonum 
\\
&& \quad \times {}_3F_2(b,b-e+1,b-f+1;1+b-c,1+b-a;1) + \big( b \longleftrightarrow c \big)\, .
\label{gegenbauer:hyperg-tranf0}
\eea
Of peculiar importance is the case where $e=b+1$ in which case the ${}_3F_2$ function can be expressed in terms of another ${}_3F_2$ plus a term involving only products of Gamma functions:
\bea
&&\sum_{p=0}^\infty \frac{\Gamma(p+a) \Gamma(p+c)}{p!\,\Gamma(p+f)}\,\frac{1}{p+b} = \frac{\Gamma(a) \Gamma(1-a)\Gamma(b)\Gamma(c-b)}{\Gamma(f-b)\Gamma(1+b-a)}
\nonum 
\\
\quad && - \frac{\Gamma(1-a)\Gamma(a)}{\Gamma(f-c)\Gamma(1+c-f)} \, \sum_{p=0}^\infty \frac{\Gamma(p+c-f+1) \Gamma(p+c)}{p!\,\Gamma(p+1+c-a)}\,\frac{1}{p+c-b}\, .
\label{gegenbauer:hyperg-tranf}
\eea

\subsection{One-loop integral}

Let's consider some simple examples in order to illustrate the method. We start with the one-loop massless p-type diagram with two arbitrary indices in $x$-space (transformation rules between $x$-space and $p$-space are provided
in Sec.~\ref{sec:meth:FT+Du}):
\bea
J(D,z,\al,\beta)
= \int \frac{\D^D x}{x^{2\al} (x-z)^{2\beta}}\, , \qquad \D^D x = \frac{1}{2}\,x^{2\lambda}\,d x^2\,d_D\hat{x}\, .
\label{def:A00}
\eea
Combining Eqs.~(\ref{gegenbauer:multipole3}) and (\ref{gegenbauer:diffind}),
the integral can be separated into a radial and an angular part as follows:
\begin{flalign}
J(D,z,\al,\beta) &= \frac{1}{2}\, \sum_{n=0}^\infty \sum_{k=0}^{[n/2]} \int_0^\infty \D x^2 \, (x^2)^{\lambda - \al}\,
\bigg[\frac{(x^2)^{n/2}}{(z^2)^{n/2+\beta}}\,\Theta(z^2-x^2) + (x^2 \leftrightarrow y^2) \Bigg]
\nonum \\
&\times \underbrace{\int \D_D\, \hat{x}\, C_{n-2k}^\lambda(\hat{x} \cdot \hat{z})}_{\Omega_D\,\delta_{n,2k}}\,\frac{(n-2k+\lambda)\,\Gamma(\lambda)}{k!\,\Gamma(\beta)}\,
\frac{\Gamma(n+\beta-k)\Gamma(k+\beta-\lambda)}{\Gamma(n+\lambda+1-k)\Gamma(\beta-\lambda)}\, ,
\end{flalign}
where the orthogonality relation, Eq.~(\ref{gegenbauer:orth}) has been used to compute the angular part. It then follows that $n$ must be an even integer: $n=2p$ and $k=[n/2]=p$.
The remaining radial integrals are easily performed. The resulting expression can be conveniently written as a sum of two one-fold series:
\begin{flalign}
J(D,z,\al,\beta) &= \frac{\pi^{D/2}}{(z^2)^{\al+\beta-\lambda-1}}\,\frac{1}{\Gamma(\beta)\Gamma(\beta-\lambda)}\,
\nonum \\
&\times \, \sum_{p=0}^\infty\,\frac{\Gamma(p+\beta)\Gamma(p+\beta-\lambda)}{p!\,\Gamma(p+\lambda+1)}\,
\bigg[ \frac{1}{p+\al+\beta-1-\lambda} + \frac{1}{p-\al+\lambda+1} \bigg]\, .
\end{flalign}
This expression can be further simplified by transforming the first sum with the help of Eq.~(\ref{gegenbauer:hyperg-tranf}) with $a=\beta-\lambda$, $b=\al+\beta-1-\lambda$, $c=\beta$
and $f=\lambda+1$. Indeed, this yields:
\begin{flalign}
& \sum_{p=0}^\infty\,\frac{\Gamma(p+\beta)\Gamma(p+\beta-\lambda)}{p!\,\Gamma(p+\lambda+1)}\,\frac{1}{p+\al+\beta-1-\lambda} =
\\
&\frac{\Gamma(\beta-\lambda)\Gamma(1+\lambda-\al) \Gamma(1+\lambda-\beta) \Gamma(\al + \beta - 1 -\lambda)}{\Gamma(\al) \Gamma(2+2\lambda-\al-\beta)}
- \sum_{p=0}^\infty\,\frac{\Gamma(p+\beta)\Gamma(p+\beta-\lambda)}{p!\,\Gamma(p+\lambda+1)}\,\frac{1}{p-\al+1+\lambda}\, ,
\nonum
\end{flalign}
and the sum on the lhs is simply the opposite of the second sum in $J(D,z,\al,\beta)$. Hence, the sum of the two one-fold series reduces to a product of $\Gamma$-functions
and we recover the well-known result:
\bea
J(D,z,\al,\beta) = \frac{\pi^{D/2}}{(z^2)^{\al+\beta-\lambda-1}}\,G(D,\al,\beta), \qquad G(D,\al,\beta) = \frac{a(\al) a(\beta)}{a(\al+\beta-1-\lambda)}\, ,
\eea
where $a(\al) = \Gamma(D/2-\al)/\Gamma(\al)$ and which was given in Eq.~(\ref{def:one-loop-G-func}) in $p$-space.

\subsection{One-loop integral with traceless products}

We may next generalize this result to the case where a traceless product appears in the numerator:
\bea
J^{\mu_1 \cdots \mu_n}(D,z,\al,\beta)
= \int \D^D x\, \frac{x^{\mu_1 \cdots \mu_n}}{x^{2\al} (x-z)^{2\beta}}\, , \qquad \D^D x = \frac{1}{2}\,x^{2\lambda}\,d x^2\,d_D\hat{x}\, .
\label{def:An0}
\eea
Dimensional analysis suggests that this integral should have the form:
\be
J^{\mu_1 \cdots \mu_n}(D,z,\al,\beta) = \pi^{D/2}\,\frac{z^{\mu_1 \cdots \mu_n}}{(z^2)^{\al+\beta-\lambda-1}}\,G^{(n,0)}(\al,\beta)\, ,
\label{def:An0-form}
\ee
where the coefficient function, $G^{(n,0)}(\al,\beta)$, is yet to be determined.
In order to do so, we consider the scalar function:
\be
z^{\mu_1 \cdots \mu_n}\,J^{\mu_1 \cdots \mu_n}(D,z,\al,\beta) =\pi^{D/2}\,\frac{z^{2n}}{(z^2)^{\al+\beta-\lambda-1}}\,\frac{\Gamma(\lambda) \Gamma(n+2\lambda)}{2^n \Gamma(2\lambda) \Gamma(n+\lambda)}\,G^{(n,0)}(\al,\beta)\, ,
\ee
where Eqs.~(\ref{def:An0-form}) and (\ref{gegenbauer:TSTP}) have been used.
The corresponding integral can be evaluated by using the relation between traceless products and Gegenbauer polynomials, Eq.~(\ref{gegenbauer:TSTGegen}):
\begin{flalign}
z^{\mu_1 \cdots \mu_n}\,J^{\mu_1 \cdots \mu_n}(D,z,\al,\beta) = \int \D^D x\, \frac{z^{\mu_1 \cdots \mu_n} x^{\mu_1 \cdots \mu_n}}{x^{2\al} (x-z)^{2\beta}}
= \frac{n! \Gamma(\lambda)}{2^n \Gamma(n+\lambda)}\,\int \D^D x\, \frac{C_n(\hat{z} \cdot \hat{x})\,(x^2 z^2)^{n/2}}{x^{2\al} (x-z)^{2\beta}}\, ,
\end{flalign}
and then expanding the propagator in Gegenbauer polynomials as before. This yields:
\begin{flalign}
&z^{\mu_1 \cdots \mu_n}\,J^{\mu_1 \cdots \mu_n}(D,z,\al,\beta) = \frac{n! \Gamma(\lambda)}{2^n \Gamma(n+\lambda)}\,\frac{1}{2}\,\sum_{p=0}^\infty \sum_{k=0}^{[p/2]}\,
\int dx^2\,(x^2)^{\lambda-\al}
\bigg[\frac{(x^2)^{\frac{p+n}{2}}}{(y^2)^{\frac{p-n}{2}+\beta}}\,\Theta(z^2-x^2) + (x^2 \leftrightarrow y^2) \Bigg]
\nonum \\
& \times \int \D_D\, \hat{x}\,C_n(\hat{z} \cdot \hat{x}) C_{p-2k}^\lambda(\hat{x} \cdot \hat{z})\,\frac{(p-2k+\lambda)\,\Gamma(\lambda)}{k!\,\Gamma(\beta)}\,
\frac{\Gamma(p+\beta-k)\Gamma(k+\beta-\lambda)}{\Gamma(p+\lambda+1-k)\Gamma(\beta-\lambda)}\, .
\end{flalign}
The angular integral is non-zero for $2k=p-n$ which implies that $p$ must have the same parity as $n$ and $p \geq n$. Separate analysis of the even and odd $n$ cases
yield, after some simple manipulations:
\bea
&& z^{\mu_1 \cdots \mu_n}\,J^{\mu_1 \cdots \mu_n}(D,z,\al,\beta) =  \pi^{D/2}\,\frac{z^{2n}}{(z^2)^{\al+\beta-\lambda-1}}\,\frac{\Gamma(\lambda) \Gamma(n+2\lambda)}{2^n \Gamma(2\lambda) \Gamma(n+\lambda)}
\times \nonum \\
&& \qquad \times \, \sum_{m=0}^\infty B(m,n| \beta,\lambda) \left( \frac{1}{m+\al+\beta-1-\lambda} + \frac{1}{m+n-\al+\lambda+1} \right)\, ,
\label{def:An0-calc}
\eea
where:
\be
B(m,n| \beta,\lambda) = \frac{\Gamma(m+n+\beta)}{m! \Gamma(m+n+\lambda+1) \Gamma(\beta)}\,\frac{\Gamma(m+\beta-\lambda)}{\Gamma(\beta-\lambda)}\, .
\label{gegenbauer-B-def}
\ee
Comparing Eqs.~(\ref{def:An0-calc}) and (\ref{def:An0-form}), we see that the coefficient function
equals the sum of two one-fold series:
\be
G^{(n,0)} (D,\al,\beta) = \sum_{m=0}^\infty B(m,n| \beta,\lambda) \left( \frac{1}{m+\al+\beta-1-\lambda} + \frac{1}{m+n-\al+\lambda+1} \right)\, .
\label{master-1loop-Gn-series}
\ee
Such a series representation reduces to a product of $\Gamma$-functions upon using the transformation properties of hypergeometric functions:
\be
G^{(n,0)} (D,\al,\beta) = \frac{a_n(\al) a_0(\beta)}{a_n(\al+\beta-\lambda-1)}\, , \qquad a_n(\al) = \frac{\Gamma(n+D/2-\al)}{\Gamma(\al)}\, ,
\ee
in accordance with Eq.~(\ref{one-loop-Gn}).

\subsection{One-loop integral with Heaviside functions}

The above results yield {\it integration rules for Feynman integrals involving traceless products and Heaviside functions} which were given in Ref.~\cite{Kotikov:1995cw}. 
From Eq.~(\ref{def:An0-calc}) we indeed recover the basic results of this reference:
%
%
\bea
&&\int \D^D x\, \frac{x^{\mu_1 \cdots \mu_n}}{x^{2\al} (x-y)^{2\beta}}\,\Theta(x^2 - y^2) = \pi^{D/2}\,\frac{y^{\mu_1 \cdots \mu_n}}{(y^2)^{\al+\beta-\lambda-1}}\,
\sum_{m=0}^\infty \frac{B(m,n| \beta,\lambda)}{m+\al+\beta-1-\lambda}\, 
\nonum \\
&& \qquad \stackrel{(\beta = \lambda)}{=} \pi^{D/2}\,\frac{y^{\mu_1 \cdots \mu_n}}{(y^2)^{\al-1}}\,\frac{1}{\Gamma(\lambda)}\,\frac{1}{(\al-1) (n+\lambda)}\, ,
\eea
and
\bea
&&\int \D^D x\, \frac{x^{\mu_1 \cdots \mu_n}}{x^{2\al} (x-y)^{2\beta}}\,\Theta(y^2 - x^2) = \pi^{D/2}\,\frac{y^{\mu_1 \cdots \mu_n}}{(y^2)^{\al+\beta-\lambda-1}}\,
\sum_{m=0}^\infty \frac{B(m,n| \beta,\lambda)}{m+n-\al+1+\lambda}\, 
\nonum \\
&& \qquad \stackrel{(\beta = \lambda)}{=} \pi^{D/2}\,\frac{y^{\mu_1 \cdots \mu_n}}{(y^2)^{\al-1}}\,\frac{1}{\Gamma(\lambda)}\,\frac{1}{(n+\lambda + 1 - \al) (n+\lambda)}\, ,
\eea
where the peculiar case $\beta = \lambda$ has been explicitly displayed. The following more complicated cases are also useful
(see \cite{Kotikov:1995cw}):
\begin{flalign}
&\int \D^D x\, \frac{x^{\mu_1 \cdots \mu_n}}{x^{2\al} (x-y)^{2\beta}}\,\Theta(x^2 - z^2) = \pi^{D/2}\,y^{\mu_1 \cdots \mu_n}\, \left [ \frac{\Theta(y^2-z^2)}{(y^2)^{\al+\beta-\lambda-1}}\,G^{n,0} (\al,\beta) 
\right .
\nonum \\
& \left . \quad + \sum_{m=0}^\infty \frac{B(m,n| \beta,\lambda)}{(z^2)^{\al+\beta-\lambda-1}}\,\left( \left(\frac{y^2}{z^2} \right)^m\,\frac{\Theta(z^2-y^2)}{m+\al+\beta-1-\lambda}
- \left(\frac{z^2}{y^2} \right)^{m+n+\beta}\,\frac{\Theta(y^2-z^2)}{m-\al+n+1+\lambda}\right) \right]\, 
\nonum \\
& \quad \stackrel{(\beta = \lambda)}{=} \pi^{D/2}\,\frac{1}{\Gamma(\lambda)}\,y^{\mu_1 \cdots \mu_n}\, \left [ \frac{\Theta(y^2-z^2)}{(y^2)^{\al-1}}\,\frac{1}{(\al-1)(n+\lambda+1-\al)}
\right .
\nonum \\
& \left . \quad + \frac{1}{(z^2)^{\al-1}}\,\frac{1}{n+\lambda}\,\left( \frac{\Theta(z^2-y^2)}{\al-1}
- \left(\frac{z^2}{y^2} \right)^{n+\lambda}\,\frac{\Theta(y^2-z^2)}{n+1+\lambda-\al}\right) \right]\, ,
\end{flalign}
and
\begin{flalign}
&\int \D^D x\, \frac{x^{\mu_1 \cdots \mu_n}}{x^{2\al} (x-y)^{2\beta}}\,\Theta(z^2 - x^2) = \pi^{D/2}\,y^{\mu_1 \cdots \mu_n}\, 
\left [ \frac{\Theta(z^2-y^2)}{(y^2)^{\al+\beta-\lambda-1}}\,G^{n,0} (\al,\beta) \right .
\nonum \\
& \left . \quad - \sum_{m=0}^\infty \frac{B(m,n| \beta,\lambda)}{(z^2)^{\al+\beta-\lambda-1}}\,
\left( \left(\frac{y^2}{z^2} \right)^m\,\frac{\Theta(z^2-y^2)}{m+\al+\beta-1-\lambda}
- \left(\frac{z^2}{y^2} \right)^{m+n+\beta}\,\frac{\Theta(y^2-z^2)}{m-\al+n+1+\lambda}\right) \right]\, 
\nonum \\
& \quad \stackrel{(\beta = \lambda)}{=} \pi^{D/2}\,\frac{1}{\Gamma(\lambda)}\,y^{\mu_1 \cdots \mu_n}\, \left [ \frac{\Theta(z^2-y^2)}{(y^2)^{\al-1}}\,\frac{1}{(\al-1)(n+\lambda+1-\al)}
\right .
\nonum \\
& \left . \quad - \frac{1}{(z^2)^{\al-1}}\,\frac{1}{n+\lambda}\,\left( \frac{\Theta(z^2-y^2)}{\al-1}
- \left(\frac{z^2}{y^2} \right)^{n+\lambda}\,\frac{\Theta(y^2-z^2)}{n+1+\lambda-\al}\right) \right]\, /
\end{flalign}

%
%
%

With these rules in hand, the Gegenbauer polynomials technique allows to compute the massless p-type two-loop master integral with up
three arbitrary indices as a linear combination of up to four hypergeometric functions ${}_3F_2$ of argument $1$, a result which can be found in 
Ref.~\cite{Kotikov:1995cw}. In particular, the method provides an alternative representation for the integral $I(1+a)$ found in the previous section with functional
equations (see Eq.~(\ref{Fa2})).

\subsection{Application to $I(1+a)$}
\label{sec:meth:GegenbauerEx}

Here we reconsider the simple but very important example of: $I(1+a) = J(D,p,1,1,1,1,1+a)$ 
(see Eq.~(\ref{def:two-loop-p-int})).
Applying the rules of the previous paragraph, its coefficient function $\text{C}_D[I(1+a)]$ can be expressed as:
\bea
&&\text{C}_D[I(a+1)] = 2\frac{\Gamma^2(1-\ep)\Gamma(-\ep-a)\Gamma(a+2\ep)}{\Gamma(2-2\ep)\Gamma(1-3\ep-a)} \,  
\Biggl[\frac{\pi\Gamma(1-a-3\ep)\Gamma(1-a-2\ep)\Gamma(a+2\ep)}{\Gamma(1-\ep)\Gamma(1/2-a-2\ep)\Gamma(1/2+2\ep+a)}\, \nonumber \\
&&- \sum_{n=1}^{\infty} \frac{\Gamma(n+1-2\ep)}{\Gamma(n+1+a)} \,
\frac{1}{n+a+1+\ep} 
\Biggr] \, ,
\label{Fa2K}
\eea
which coincides with Eq.~(\ref{res:I(al):Kotikov}) after changing of variables. 

So, using the method of Gegenbauer polynomials, the results for $I(1+a)$  can be expressed as a combination of 
$\Gamma$-functions together with one hypergeometric function with the arguments ``$1$''. Such result can be successfully used for an efficient $\ep$-expansion of the diagram.
Moreover, the combination of the two results (\ref{Fa2}) and (\ref{Fa2K}) provides the advertised relation (\ref{id:F32})
between two hypergeometric functions of argument ``$-1$'' and one hypergeometric function of argument ``$1$''. Such a relation is absent in standard textbooks 
and was recently proven exactly in \cite{Kotikov:2016rgs}.


\end{document}